\documentclass[a4paper,twocolumn,10pt,accepted=2025-03-05]{quantumarticle}
\pdfoutput=1
\usepackage{dsfont}
\usepackage{amsmath}
\usepackage{bm}
\usepackage{amssymb}
\usepackage{graphicx}
\usepackage{braket}
\usepackage{amsthm}
\usepackage{comment}
\usepackage{array}
\usepackage{multirow,booktabs}
\usepackage{caption}
\usepackage{subcaption}
\usepackage{diagbox}
\usepackage{cite}
\usepackage{mathrsfs}  
\usepackage{comment}

\newtheorem{Proposition}{Proposition}
\newtheorem{Definition}{Definition}
\newtheorem{Corollary}{Corollary}

\usepackage[colorlinks=true,urlcolor=blue,citecolor=blue,linkcolor=blue]{hyperref}
\usepackage{physics}
\DeclareGraphicsExtensions{.eps,.ps,.pdf}
\newcommand{\la}{\lambda}
\newcommand{\ie}{\hbox{\em i.e.{}}}

\newcommand{\rhs}{\hbox{r.h.s.{}}}
\newcommand{\Rr}{\mathsf{R}}
\newcommand{\Ro}{\mathsf{O}}
\newcommand{\Hs}{\mathcal{H}}
\newcommand{\HSs}{\mathcal{B}}
\newcommand{\BHs}{\mathscr{B}}
\newcommand{\bcdot}{\boldsymbol{\cdot}}
\renewcommand{\vec}{\mathbf}

\newcommand{\co}{\mathfrak{C}}
\newcommand{\rot}{C}
\newcommand{\group}{\mathcal{G}}
\newcommand{\pointg}[1]{\mathrm{#1}}
\newcommand{\SetG}{\mathcal{F}}

\begin{document} 
\title{{Platonic dynamical decoupling sequences for interacting spin systems}} 
\author{Colin Read}
\email{cread@uliege.be}
\affiliation{Institut de Physique Nucléaire, Atomique et de Spectroscopie, CESAM, University of Liège
\\
B-4000 Liège, Belgium}
\author{Eduardo Serrano-Ens\'astiga}
\email{ed.ensastiga@uliege.be}
\affiliation{Institut de Physique Nucléaire, Atomique et de Spectroscopie, CESAM, University of Liège
\\
B-4000 Liège, Belgium}
\author{John Martin}
\email{jmartin@uliege.be}
\affiliation{Institut de Physique Nucléaire, Atomique et de Spectroscopie, CESAM, University of Liège
\\
B-4000 Liège, Belgium}

\maketitle

\begin{abstract}
In the NISQ era, where quantum information processing is hindered by the decoherence and dissipation of elementary quantum systems, developing new protocols to extend the lifetime of quantum states is of considerable practical and theoretical importance. A well-known technique, known as dynamical decoupling, uses a carefully designed sequence of pulses applied to a quantum system, such as a spin-$j$ (which represents a qudit with $d=2j+1$ levels), to suppress the coupling Hamiltonian between the system and its environment, thereby mitigating dissipation. While dynamical decoupling of qubit systems has been widely studied, the decoupling of qudit systems has been far less explored and often involves complex sequences and operations. In this work, we design efficient decoupling sequences composed solely of global $\mathrm{SU}(2)$ rotations and based on tetrahedral, octahedral, and icosahedral point groups, which we call Platonic sequences. We extend the Majorana representation for Hamiltonians to develop a simple framework that establishes the decoupling properties of each Platonic sequence and show its effectiveness on many examples. These sequences are universal in their ability to cancel any type of interaction with the environment for single spin-$j$ with spin quantum number $j\leqslant 5/2$, and they are capable of decoupling up to $5$-body interactions in an ensemble of interacting spin-$1/2$ with only global pulses, provided that the interaction Hamiltonian has no isotropic component, with the exception of the global identity. 
We also discuss their inherent robustness to finite pulse duration and a wide range of pulse errors, as well as their potential application as building blocks for dynamically corrected gates.
\end{abstract}
\section{Introduction}
\label{sec.Introduction}
Many quantum hardware devices are limited by the coherence time of their noisy constituents due to undesirable interactions among them or with their environment. These interactions lead to decoherence and dissipation and consequently to a deterioration of the overall quantum state that contains the useful information to be processed, stored, and extracted. In a world where quantum error correction is not yet available to correct this loss of information, the development of protocols to mitigate undesirable interactions is of major practical interest~\cite{Preskill_2018,Bharti_2022}, but also of fundamental theoretical importance. In particular, such protocols can be useful in emerging quantum technologies or for revealing minute physical effects masked by surrounding noise~\cite{Preskill_2018}.

\par A well-known technique for mitigating decoherence is called dynamical decoupling (DD) and consists of applying a periodic sequence of pulses to the system of interest to average out its unwanted interaction with the environment. Since Viola's seminal work~\cite{Viola_1999}, numerous DD sequences have been constructed —some guided by physical intuition and others by numerical methods—, firstly for a single qubit~\cite{Viola_2003, PhysRevLett.98.100504, Biercuk_2009, Lidar_2013} and then for multiqubit systems~\cite{Stollsteimer_2001,Paz-Silva_2016,Lukin_2020, Cappellaro_2022}. They have been successfully implemented on a variety of physical platforms, such as NV centers in diamond~\cite{Bradley_2019}, trapped ions~\cite{Biercuk_2009}, superconducting flux qubits~\cite{Bylander_2011}, solid-state spin qubits~\cite{du2009preserving,doi:10.1126/science.1192739,PhysRevB.90.054304}, photonic systems~\cite{PhysRevLett.103.040502,PhysRevA.85.022340} and Bose-Einstein condensates~\cite{PhysRevA.84.013606,Edri_2021}. They are routinely used in experiments such as NMR and electron spin resonance to reduce decoherence, but also to measure transverse relaxation times more accurately~\cite{BHATTACHARYYA_2020} or to finely probe the local spin environment in systems of interacting spins~\cite{Goldblatt_2024}.

\par Although qubits are the most elementary building blocks for quantum information processing, many quantum platforms naturally possess a richer, multi-level structure, in which a qubit is defined by selecting a two-level subsystem. In recent years, considerable efforts have been made to exploit these extra dimensions~\cite{Ringbauer_2022,Hrmo_2023,Liu_2023,Fischer_2023,roy_2024,Blok_2021,Omanakuttan_2024} which could offer advantages in certain quantum algorithms~\cite{Lanyon_2009,Wang_2020}, but also more efficient quantum simulation~\cite{Neeley_2009,Ringbauer_2024,Popov_2024,vezvaee_2024}, metrology~\cite{Martin_2020,serrano_2024} and error correction schemes~\cite{Gottesman_2001, Cafaro_2012, Muralidharan_2017}.

\par Dynamical decoupling of qudit ensembles has been much less explored than for qubits. However, they have gained interest in recent years. A universal DD sequence for a single qudit based on the Heisenberg-Weyl group has been recently constructed and experimentally implemented~\cite{tripathi_2024}. An elegant generalization of the CPMG sequence~\cite{Viola_1998}, which suppresses dephasing in a single qubit, has been constructed for a dephasing qudit and its efficiency was experimentally demonstrated for a qutrit on IBM quantum computers~\cite{iiyama_2024}. Some sequences have been constructed from orthogonal arrays that were shown to decouple an ensemble of interacting qudits, assuming that individual control over each qudit is possible~\cite{Wocjan_2002, Wocjan_2006, Rotteler_2006, Rötteler_Wocjan_2013}. However, the complexity of these protocols increases drastically with the number of qudits in the ensemble. This issue is avoided by allowing only global pulses, $\ie$, each subsystem undergoes exactly the same pulse at each moment of the sequence. Recent works have relied on numerical approaches to find such sequences that decouple relevant interactions between qudits~\cite{Lukin_2017,Lukin_2023}.

\par In this work, we construct novel DD sequences applicable to qudits and in various single- and multispin systems, using an innovative approach based on operator space symmetries. We use the symmetrization formalism of Refs.~\cite{Viola_1999,Zanardi_1999}, where a DD sequence is seen to perform a \textit{symmetrization operation} on the unwanted interaction Hamiltonian. In this way, the latter Hamiltonian is transformed into a Hamiltonian that possesses the symmetry enforced by the sequence. We note that choosing a sequence enforcing an \textit{inaccessible symmetry} (a concept that will be rigorously defined in the next section) will cancel the Hamiltonian, and that studying the accessible symmetries of a generic Hamiltonian provides important information about the DD sequences that cancel it. By representing an operator as a geometric object (a set of \textit{constellations}) in three-dimensional space and studying its rotational symmetries, using the so-called Majorana representation~\cite{Maj:32,ser.mar:23}, we show how to find inaccessible symmetries in various operator spaces, leading us to the construction of three novel sequences relevant in a wide range of quantum systems. We demonstrate their decoupling properties in single and multispin systems for numerous relevant decoherence models.

\par We call these three new sequences \textit{Platonic} DD sequences, as they are associated with the symmetry group of a Platonic solid, namely the tetrahedron, octahedron and icosahedron. They are composed solely of \textit{global $\mathrm{SU(2)}$ operations} and are naturally robust to finite-duration pulses as well as many systematic control errors (in particular, flip-angle and axis misspecification~\cite{Lidar_2023_survey}). They provide decoupling for a wide range of Hamiltonians in single- and multispin systems. For instance, we find a Platonic sequence that decouples up to five-body interactions in a spin-$1/2$ ensemble (and more generally up to five-body multilinear interactions in any spin-$j$ ensemble) and provides universal decoupling of an arbitrary spin with quantum number $j<3$ (spin with $d<7$ levels). This work aims at presenting the framework that leads to the construction of Platonic sequences and providing an extensive analysis of their decoupling and robustness properties, in order to identify relevant experimental systems where they might outperform other state-of-the-art sequences.

\par This paper is organized as follows. In Sec.~\ref{Sec.Dyn.decoupling}, we recall the necessary notions of dynamical decoupling, explain and define the concept of inaccessible symmetries, and present a pedagogical summary of the general framework used in this paper. In Sec.~\ref{Sec.Mathematical.tools}, we describe the necessary mathematical tools for studying the $\mathrm{SU(2)}$ symmetries of a given operator space, introducing the different point groups (finite groups of rotational symmetries) and the decomposition of Hilbert spaces into $\pointg{SU}(2)$ irreps\footnote{We use the usual convention of calling $\pointg{SU}(2)$ $j$-irrep both the respective set of $(2j+1)\times(2j+1)$ matrices defining the action of the group elements, and the $(2j+1)$-dimensional vector space where the action of the group is defined through the same matrices.}. In Sec.~\ref{Sec.Point}, we generalize Majorana's representation to arbitrary operators and systematically study the accessible rotational symmetries of operators by analyzing those of their associated constellations. In particular, we define and list the set of largest point groups for different types of Hamiltonians with different irreps decomposition. We then use these results to develop a simple framework for selecting the relevant decoupling groups for a single spin-$j$ system (Sec.~\ref{Sec. single spin}) and a multispin system (Sec.~\ref{Sec. multi spin}). We study the robustness of these sequences with respect to various pulse errors in Sec.~\ref{Sec.robustness}. 
Finally, we discuss their potential application for dynamically corrected gates in Sec.~\ref{Sec.DCG} and conclude with a summary and outlook in Sec.~\ref{Sec.Conclusions}.
\section{Dynamical decoupling}
\label{Sec.Dyn.decoupling}
In this section, we first recall in Sec.~\ref{introDDsec} the useful notions of dynamical decoupling at first order of the Magnus series and then introduce the concept of decoupling group in the context of dynamical decoupling. In Sec.~\ref{Sec.In.Sym}, we define inaccessible symmetries and present the general idea used in this work to find new decoupling groups. We also present in Sec.~\ref{Sec.Gen.Frame} a pedagogical summary of the more mathematical material presented in Secs.~\ref{Sec.Mathematical.tools} and \ref{Sec.Point}, with the aim of helping the reader understand the important results and the framework developed without going rigorously through the technical parts of the paper. We then briefly explain the concept of Cayley graph, which is used to construct a DD sequence.

\subsection{Basics of dynamical decoupling}
\label{introDDsec}
Consider a quantum system ($S$) suffering from decoherence arising from unwanted interaction with its environment (bath $B$). The "system--bath" interaction Hamiltonian is written in Schmidt decomposition as
\begin{equation}\label{Schmidtoperator}
    H_{SB}=\sum_{\alpha} S_{\alpha} \otimes  B_{\alpha},
\end{equation}
where $S_{\alpha}$ (resp.\ $B_{\alpha}$) are operators, not necessarily Hermitian, acting on the Hilbert space of the system (resp.\ of the bath). From this decomposition, we usually define the \textit{interaction subspace} $\mathcal{I}_S = \mathrm{span}\qty(\qty{S_{\alpha}})$ as the vector space spanned by the system operators. The free evolution from $t_0$ to $t$ under such Hamiltonian leads to unwanted dynamics through the propagator $U_{SB}(t,t_0) = \exp(-i\Phi)$ where $\Phi \equiv (t-t_0) H_{SB}$ is called the error phase operator (EPO)~\cite{Khodjasteh_2009}. 
\par To reduce this error, we can send a dynamical decoupling sequence ($\mbox{---}P_1\mbox{---}P_2\dots \mbox{---}P_N$) that acts only on the system, which is made up of $N$ infinitely short and strong pulses, where $P_k$ is the unitary operator corresponding to the action of the $k$th pulse, and where each dash ($\mbox{---}$) corresponds to a free evolution of duration $\tau_0$, the time interval between two successive pulses. When subjected to this DD sequence, the error phase operator can be considered as undergoing the following series of unitary transformations~\cite{Viola_1999,Khodjasteh_2009} in the toggling frame with respect to the DD pulses\footnote{The toggling frame corresponds here to the interaction picture with respect to the Hamiltonian implementing the pulse sequence.},
\begin{equation}
    \Phi \xrightarrow[\mbox{---}P_1]{} P_1^{\dagger} \Phi P_1 \xrightarrow[\mbox{---}P_2]{} P_1^{\dagger}  P_2^{\dagger} \Phi P_2 P_1 \xrightarrow[\mbox{---}P_3]{}\cdots 
\end{equation}
where the last pulse $P_N$ is chosen to satisfy the cyclic condition $\prod_{i=1}^NP_i=\mathds{1}_S$. An average EPO over the whole sequence can then be calculated by performing a Magnus expansion in the toggling frame. This leads to $\Phi_{\text{av}} = \sum_{n=1}^{\infty} \Phi_{\text{av}}^{[n]}$ with $\Phi_{\text{av}}^{[n]}$ the $n$th-order term of the Magnus series, scaling as\footnote{$\norm{A}$ denotes here the supremum operator norm of $A$ defined as $\norm{A}= \sup_{\ket{\psi}\in\mathcal{H}_S}\frac{\norm{A\ket{\psi}}}{\norm{\ket{\psi}}}$.} $\norm{\Phi_{\text{av}}^{[n]}} \leq \pi \qty(\frac{T \norm{H_{SB}}}{\xi})^n$ with $T\equiv  N\tau_0$ the total duration of the sequence and $\xi \approx 1.0868$ a convergence radius~\cite{Lidar_2013,moan_2002,Blanes_2009}. If decoherence is small enough (\ie, $T \norm{H_{SB}} \ll 1$), then the average EPO is well approximated by its first-order term, 
\begin{equation}
    \Phi_{\text{av}} \approx \Phi_{\text{av}}^{[1]} = \sum_{k=1}^N (g_k^{\dagger}\otimes \mathds{1}_B) \Phi (g_k\otimes \mathds{1}_B),
\end{equation}
where we defined the propagator acting on $S$ at each step $k$ of the sequence as 
\begin{equation}
    g_1 \equiv \mathds{1}_S,\quad g_k \equiv \prod_{i=k-1}^{1} P_i \quad(\mathrm{for}~1<k\leq N).\label{eq:DD propag}
\end{equation}
The products of unitary operators $P_i$ in \eqref{eq:DD propag} are ordered chronologically, so that $P_i$ stands to the right of $P_{i+1}$. Using the Schmidt decomposition \eqref{Schmidtoperator} for $\Phi$, we see that the DD sequence implements the following operation on each system operator $S_\alpha$~\cite{Viola_1999,Zanardi_1999} 
\begin{equation}
    \Pi_{\group}: \HSs(\Hs_S) \to \HSs(\Hs_S): S \mapsto \Pi_{\group}(S) = \frac{1}{N}\sum_{k=1}^{N} g_k^{\dagger} S g_k 
    \label{eq: symmetrization}
\end{equation}
where $\group=\qty{g_k}_{k=1}^{N}$ denotes the set of propagators defined in Eq.~\eqref{eq:DD propag}.
\par In the case where $\group$ forms a group of unitary operators, we have that
\begin{equation}
    g_k^{\dagger}\Pi_{\group}(S_{\alpha})g_k= \Pi_{\group}(S_{\alpha}) \quad \forall \,\alpha,k
\end{equation}
such that the quantum operation $\Pi_{\group}$ projects each operator $S_{\alpha}\in\mathcal{I}_S$ onto a $\group$-invariant subspace of the space of operators; each $S_{\alpha}$ is \textit{symmetrized}~\cite{Zanardi_1999} to include $\group$ as its symmetry group. Interestingly, if the group $\group$ is chosen such that the vector space $\Pi_{\group}(\mathcal{I}_S)$ only includes operators proportional to the identity, then we ensure that 
\begin{equation}
    \Phi_{\text{av}}^{[1]} \propto \mathds{1}_S\otimes B,
\end{equation}
where $B$ is some operator acting on the bath Hilbert space. Thereby, the undesirable dynamics caused by the interaction with the bath is eliminated at the dominant order of the Magnus series. In this case, $\group$ is called a \textit{decoupling group}~\cite{Viola_1999}. Each decoupling group has an associated \emph{correctable subspace}~\cite{Khodjasteh_2009_PRA} $\mathcal{C}_{\group} \subseteq \HSs\qty(\Hs_S)$ which contains all operators of $\HSs(\Hs_S)$ whose image under $\Pi_{\group}$ is proportional to the identity\footnote{Formally speaking, the correctable subspace is defined as the pre-image $\mathcal{C}_{\group}=\Pi_{\group}^{-1}\qty(\mathrm{span}\left( \left\{ \mathds{1}_S \right\} \right))$, which is a vector space.}. The decoupling group should then be chosen such that $\mathcal{I}_S \subseteq \mathcal{C}_{\group}$. In particular, we call a group $\mathcal{G}_{\mathrm{uni}}$ a \textit{universal decoupling group} if its correctable subspace is equal to the entire space of operators $\mathcal{I}_S=\HSs(\Hs_S)$.

\subsection{Inaccessible symmetries and decoupling groups}\label{Sec.In.Sym}
Consider a subspace $V$ of the total space of operators acting on the state of a quantum system, $\HSs(\Hs_S)$. We say that a group $\group$ is an inaccessible symmetry of $V$ if the only elements of $V$ having $\group$ as a symmetry group are operators proportional to the identity (including the zero operator). The rigorous definition of inaccessible symmetry is given below.

\begin{Definition}
\label{Def.1}
Consider a quantum system with a Hilbert space $\Hs_S$ and a subspace $V\subseteq \HSs(\Hs_S)$ of operators acting on $\Hs_S$. The group of unitary operators $\group = \qty{g_k}_{k=1}^N$ is an inaccessible symmetry for $V$ if the $\group$-invariant subspace of $V$ contains only operators proportional to the identity.
\end{Definition}

The following proposition using this definition highlights the link between inaccessible symmetries and decoupling groups.

\begin{Proposition}
\label{Result.0} 
Consider a group of unitary operators $\group = \qty{g_k}_{k=1}^N$ and a subspace $V$ closed under unitary transformations by the elements of $\group$ which satisfies $\mathcal{I}_S\subseteq V \subseteq \HSs\qty(\Hs_S)$. If $\group$ is an inaccessible symmetry for $V$, then it is a decoupling group for $\mathcal{I}_S$.
\end{Proposition}
\textit{Proof.} Because $V$ is closed under unitary transformations by any element $g\in \group$, we have that $\Pi_{\group}(\mathcal{I}_S)\subseteq V$ where, by construction, the subspace $\Pi_{\group}(\mathcal{I}_S)$ includes only $\group$-invariant operators. If $\group$ is an inaccessible symmetry for $V$, the only $\group$-invariant operators in the vector space are those proportional to the identity. Hence, the subspace $\Pi_{\group}(\mathcal{I}_S)$ can only include operators proportional to the identity, and, consequently, $\group$ is a decoupling group for $\mathcal{I}_S$.
\par 
By finding inaccessible symmetries for certain subspaces, it is then possible to find decoupling groups for relevant interaction subspaces. In this work, we focus on finding symmetries of a generic Hamiltonian of an ensemble of spin-$j$ under global $\mathrm{SU(2)}$ unitary transformations. By global $\mathrm{SU(2)}$, we mean that each spin in the ensemble undergoes the same unitary evolution $R(\mathbf{n},\theta) = e^{-i\theta \mathbf{J}\boldsymbol{\cdot}\vec{n}}$ where $\vec{J}=\qty(J_x,J_y,J_z)$ are the angular momentum operators of a single spin-$j$ and $R(\mathbf{n},\theta)$ represents a rotation $\mathrm{R}\in \mathrm{SO(3)}$ of angle $\theta$ of the spin around some axis $\mathbf{n}$. The most common experimental implementation of the $\mathrm{SU(2)}$ group is through physical rotations which can be induced, for instance, by applying a magnetic field to an electronic or nuclear spin, as in NMR~\cite{RevModPhys.76.1037}. However, there are different implementations of $\pointg{SU}(2)$ in other physical platforms such as multiphoton systems~\cite{PhysRevLett.103.040502, Ferretti:24} or two-component Bose-Einstein condensates~\cite{PhysRevLett.132.173401}. Thus, our results will also apply to a generic Hamiltonian of a multiqudit system where the $\mathrm{SU(2)}$ unitaries are defined by associating to each qudit a fictitious spin-$j$ and angular momentum operators $(J_x,J_y,J_z)$ with $j = \frac{d-1}{2}$.

\subsection{Conceptual framework}
\label{Sec.Gen.Frame}
\begin{figure*}[t]
    \centering
    \includegraphics[width=\linewidth]{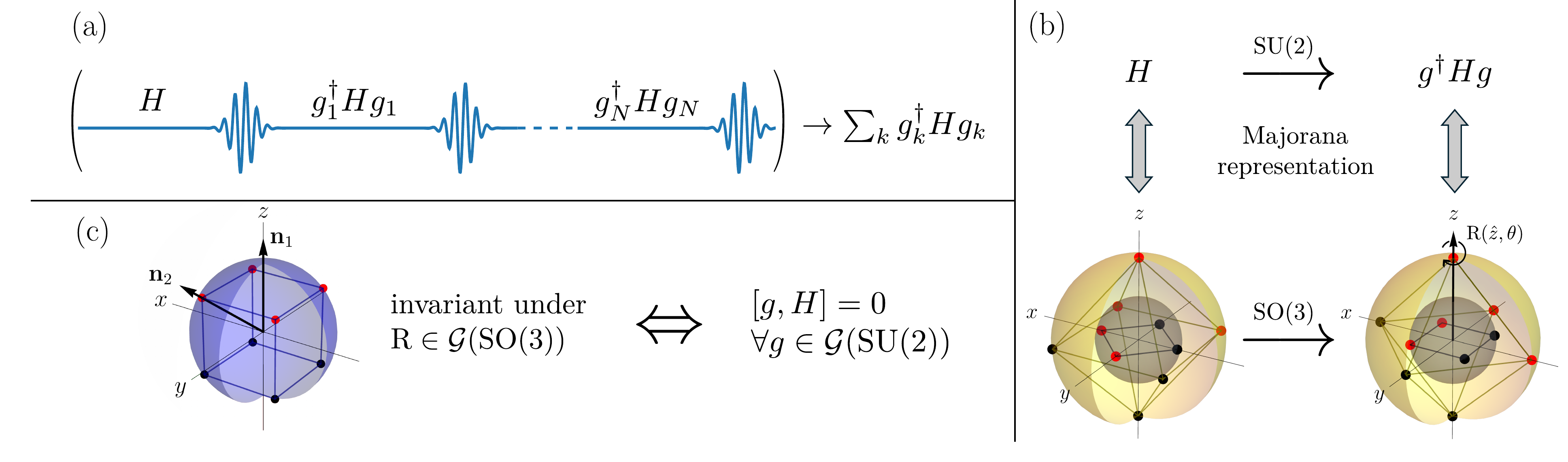}
    \caption{(a) Representation of the symmetrization operation provided by a DD sequence. (b) Representation of the equivalence between a $\mathrm{SU(2)}$ transformation of an operator and the $\mathrm{SO(3)}$ rotation of its associated constellation. (c) Representation of the equivalence between the rotation symmetries of a constellation and the symmetries of its corresponding operator under unitary transformation. }
    \label{fig:fig10}
\end{figure*}

In order to study $\mathrm{SU(2)}$ symmetries, the natural strategy is to decompose a generic subspace of operators $V$ into a direct sum of subspaces that transform independently under $\mathrm{SU(2)}$ transformations.
The (unique) decomposition leaves us with
\begin{equation}
    V = \bigoplus_{(L,\alpha)} \mathscr{B}^{(L,\alpha)},
\end{equation}
where each of these subspaces $\mathscr{B}^{(L,\alpha)}$ is called an \textit{irrep}. The index $L$ indicates the dimension of the subspace (which is equal to $2L+1$), while the index $\alpha$ may be necessary to distinguish between different irreps of the same dimension. An introduction to the irrep decomposition of operator spaces is provided in Sec.~\ref{Sec.Mathematical.tools}. 
\\
\indent
In Sec.~\ref{Sec.Point}, we will show that the operators that live in each subspace of the decomposition ($\ie$, each irrep $\mathscr{B}^{(L,\alpha)}$) can be associated with a (two-color) constellation on a 2-sphere, where the number of points in the constellation is equal to $2L$ and therefore depends only on the dimension of the irreps~\cite{Maj:32, Ser.Bra:20}. Crucially, an $\mathrm{SU(2)}$ transformation of an operator $O\in \mathscr{B}^{(L,\alpha)}$ is equivalent to the corresponding $\mathrm{SO(3)}$ rotation of its constellation, and we understand that it is possible to associate the (rather abstract) symmetries of an operator under $\mathrm{SU(2)}$ transformations with the rotational symmetries of a geometric object in physical space. We give a visual representation of the ideas explained above in Fig.~\ref{fig:fig10}.
\\ \indent 
Using this approach, we systematically identify the $\mathrm{SU(2)}$ symmetries accessible for each irrep by listing the rotational symmetries in the associated constellations of different sizes. The results are displayed in Table~\ref{Table.1}, where we list the largest rotational symmetry groups (or so-called \textit{point groups}) for irreps $\mathscr{B}^{(L)}$ of different dimensions $2L+1$ as well as an example of operator that possesses each symmetry. We now have all the tools we need to find inaccessible rotational symmetries in an interaction subspace $\mathcal{I}_S$ using a  two-step procedure: (i) decompose $\mathcal{I}_S$ into its irreps and (ii) check Table~\ref{Table.1} to identify possible symmetries absent from the irreps involved in the decomposition. More precisely, (i) we search for a subspace defined by a direct sum of irreps $V=\bigoplus_{(L,\alpha)} \mathscr{B}^{(L,\alpha)}$ such that $\mathcal{I}_S \subseteq V$, and (ii) we find an inaccessible symmetry in the irreps of $V$. By Proposition~\ref{Result.0}, $\group$ is then guaranteed to be a decoupling group. As a result, we obtain that the point groups associated to the Platonic solids (see Appendix~\ref{App.Three.exceptional} for more details about these groups) serve as inaccessible symmetries for several interaction spaces.
\subsection{From decoupling group to DD sequence}
Once a decoupling group has been identified, the remaining task is to find a sequence of pulses implementing the symmetrization operation~\eqref{eq: symmetrization} for this group. This can be done systematically using the Cayley graph formalism~\cite{Viola_2003, Viola_2013, Bollobás1998}, which is briefly explained in this subsection for the sake of  completeness. 
\\
\indent 
Starting from a decoupling group $\group$, we can pick a generating set $\Gamma \subseteq \group$, such that every element of the group can be written as a product of the elements of $\Gamma$ (called \textit{generators}). The Cayley graph of $\group$ with respect to $\Gamma$ is constructed by assigning a vertex to each element of $\group$ and connecting a vertex $g_1$ to another vertex $g_2$ by an edge if there exists a generator $\gamma\in\Gamma$ such that $g_2=\gamma g_1$. The edges are then colored to distinguish the generators connecting the group elements. As explained in Ref.~\cite{Viola_2003}, any Eulerian path on the graph corresponds to a sequence of pulses that guarantees~\eqref{eq: symmetrization}, \ie, a DD sequence. Further information on Cayley graphs and their construction can be found in Appendix~\ref{Ap.CayleyDiagram}. 

\section{Mathematical tools}
\label{Sec.Mathematical.tools}
In this section, we present some general reminders about point groups and representation theory of Hilbert spaces which are useful for this work. Readers who are primarily interested in the results and applications of the Platonic sequences, rather than their derivation and the proof of their decoupling properties, can skip Sections 3 and 4 and go straight to the following Sections.
\subsection{Point groups}
Point groups in three dimensions are classified into different families, made explicit by the Schönflies notation~\cite{dresselhaus2007group,Bra.Cra:10}. In this work, we only consider proper point groups, \ie, those that do not contain reflections. Any of these point groups can be generated by at most two rotations, also called generators (see Appendix~\ref{App.Three.exceptional} for more details). We denote by $C_n^k$  a $2\pi k/n$ rotation about an axis. We now briefly describe every proper point group:

i) \emph{Cyclic groups} $\pointg{C}_n$: such groups have $n$ elements consisting of $2\pi q/n$ rotations along the same axis of rotation, where $q=0,1,\dots,n-1$. The regular polygons on the sphere that do not lie in a great circle\footnote{The circles on the sphere whose center coincides with the center of the sphere are called \emph{great circles}.} have this symmetry. In addition, there is one group of infinite order, denoted $\pointg{C}_{\infty}$, consisting of the subgroup of rotations about a fixed axis by any angle (the symmetry of a point on the sphere). Each $\pointg{C}_n$ has only one generator $\rot_n$.

ii) \emph{Dihedral groups} $\pointg{D}_n$ : For $n\geqslant 2$, they contain $2n$ elements: $n$ rotations $C_n$ over the principal axis, and $n$ additional $C_2$ rotations about axes perpendicular to the principal axis. This is the proper point group of a regular prism with $n>2$ sides, antiprisms\footnote{An antiprism with $n$ sides has a point group $\pointg{D}_n$ unless it is a tetrahedron or octahedron (Platonic solids for $n=2,3$), respectively.} (with $n>3$) and regular $n$-gons on a great circle of the sphere for $n>2$. The group $\pointg{D}_{ \infty }$ has as subgroup $\pointg{C}_{\infty}$ on a principal axis of rotation, as well as any rotation of $\pi$ about an axis perpendicular to it. It is equivalent to the point group associated with a pair of antipodal points. Each $\pointg{D}_n$ has two generators: $\rot_2$ and $\rot_n$.

iii) \emph{Tetrahedral group} $\pointg{T}$: It consists of 12 elements $\pointg{T}=\{E , 8\,\rot_3 , 3\,\rot_2 \}$ corresponding to the rotations which leave a regular tetrahedron invariant. A possible set of generators of $\pointg{T}$ consists of two $\rot_3$ rotations about different axes.

iv) \emph{Octahedral group} $\pointg{O}$: It consists of 24 elements $\pointg{O}=  \{ E , 8\,\rot_3 , 9\,\rot_2 , 6\,\rot_4\}$ that transform a regular octahedron (or cube) into itself. It can be generated by a $\rot_3$ and a $\rot_4$ rotation.

v) \emph{Icosahedral group} $\pointg{I}$: It consists of 60 elements $\{ E , 12\,\rot_5 , 12\,\rot_5^2 , 20\,\rot_3 , 15\,\rot_2\}$ and is equivalent to the symmetry point group of a regular icosahedron (or dodecahedron). It has two generators: $\rot_3$ and $\rot_5$.

The point group notation is abstract in the sense that it does not specify the orientation of the axes. A procedure for constructing the three exceptional groups ($\pointg{T}$, $\pointg{O}$ and $\pointg{I}$) from their generators is presented in Appendix~\ref{App.Three.exceptional}. Their corresponding Cayley graphs and DD sequences are presented in Appendix~\ref{Ap.CayleyDiagram}.
\subsection{Representation theory of Hilbert spaces}
Consider a general quantum state $\ket{\Psi}\in \Hs$ in a finite-dimensional Hilbert space $\Hs$ where the elements of the rotation group $\Rr \in \mathrm{SO}(3)$ are represented by the unitary operators $R(\mathbf{n} , \theta  ) = e^{-i\theta \, \mathbf{J} \boldsymbol{\cdot} \mathbf{n}}$, with $\mathbf{n} = (n_x , n_y , n_z) $ and $\theta$ the axis-angle parameters, and $\mathbf{J} = (J_x , J_y , J_z)$ the angular momentum operator of the quantum system. From representation theory, $\Hs$ decomposes into irreps (see subsection~\ref{Sec.Gen.Frame}),
\begin{equation}
    \Hs = \bigoplus_{(j,\alpha_j)} \Hs^{(j,\alpha_j)} 
\end{equation}
where the superindex $(j,\alpha_j)$ is used to label the spin-$j$ irrep. An additional integer $\alpha_j$ may be used to distinguish between different subspaces of the same dimension.  
For example, for a two-qubit system, we have $\Hs = \Hs^{(0)}\oplus\Hs^{(1)}$ because there is no degenerate irrep, whereas for a three-qubit system, we have $\Hs = \Hs^{(1/2,1)}\oplus\Hs^{(1/2,2)}\oplus \Hs^{(3/2)}$. In the following, we explain our approach for a single spin-$j$ system with $\Hs = \Hs^{(j)}$. 

Similarly, the space of Hilbert–Schmidt operators\footnote{$\HSs (\Hs) $ is the space of bounded operators $A:
\Hs \to \Hs$ with finite Hilbert–Schmidt norm.} $\HSs (\Hs^{(j)}) $ can be decomposed into subspaces that transform as $\pointg{SU}(2)$ irreps,
\begin{equation}
\label{Eq.decomp.HS}
    \HSs (\Hs^{(j)}) = \bigoplus_{L=0 }^{2j} \BHs^{(L)} , 
\end{equation}
where each $L$-irrep $\BHs^{(L)}$ appears only once. Each subspace is spanned by multipolar operators, $\BHs^{(L)}=\mathrm{span}(\{ T_{LM} \}_{M=-L}^{L} )$~\cite{Fan:53,Var.Mos.Khe:88}. The complete set $\{T_{LM} :L=0,\ldots,2j; M=-L, \dots ,L\}$ forms an orthonormal basis of $\HSs (\Hs^{(j)})$ with respect to the Hilbert–Schmidt inner product. By definition, multipolar operators transform according to the spin-$L$ irrep under rotations  $\Rr \in \pointg{SO}(3)$, \ie,
\begin{equation}
\label{prop.sh}
D^{(j)}(\Rr)\, T_{L M}\, D^{(j) \dagger}(\Rr)  = \sum\limits_{M'=-L}^{L} D_{ M' M}^{(L)}(\Rr)\, T_{L M'} ,
\end{equation}
where $D_{ M' M}^{(L)}(\Rr)$ are the entries of the rotation matrix in the $L$-irrep, also called Wigner-$D$ matrix~\cite{Var.Mos.Khe:88}. They also fulfill $T_{L M}^{\dagger} = (-1)^{M} T_{L \,-M}$. In terms of angular momentum operators $J_a$ ($a=x,y,z$), $T_{LM}$'s are expressed as a sum of monomials of $J_a$ up to degree $L$~\cite{Var.Mos.Khe:88}, see Table~\ref{Table.1} for some examples.
\section{Point groups of operators}
\label{Sec.Point}
In this section,  we first review the Majorana representation of Hermitian operators presented in Ref.~\cite{Ser.Bra:20} and explain how it can be used to associate a point group to a Hermitian operator according to its symmetries. We then generalize our approach to arbitrary operators, not necessarily Hermitian.

\subsection{Majorana representation of Hermitian operators}
\label{subsec.Maj.Rep.Mixed}
The Majorana representation was originally introduced for pure spin states~\cite{Maj:32} (see Appendix~\ref{App.Majorana}). It has recently been generalised to Hermitian operators in~\cite{Ser.Bra:20}. The following presentation is rather different from that of Ref.~\cite{Ser.Bra:20} but more appropriate to this work.

We start with the expansion of a general Hamiltonian in the multipolar basis,
\begin{equation}
\begin{aligned}
H = \sum_{L=0}^{2j} \sum_{M=-L}^L h_{LM} T_{LM}
\end{aligned}
\end{equation}
with $h_{LM} = \Tr (T^{\dagger}_{LM}H)$ the multipolar components of $H$. It can be rewritten as 
\begin{equation}
\begin{aligned}
\label{rhoTLMexpansion}
 H = \sum_{L=0}^{2j} \mathbf{h}_L \boldsymbol{\cdot} \mathbf{T}_{L} = \sum_{L=0}^{2j} h_L \hat{h}_{L} \boldsymbol{\cdot} \mathbf{T}_{L}   , 
\end{aligned}
\end{equation}
where we gathered components with the same $L$ into vectors $\mathbf{h}_L = (h_{LL} , \dots , h_{L-L}) $ with Euclidean norm $h_L=\norm{\mathbf{h}_L}_2$ and corresponding unitary vector $\hat{h}_L=\mathbf{h}_L/h_L$. 
$\mathbf{T}_{L}= (T_{L L} , \dots , T_{L  -L})$ is a vector whose entries are multipolar operators, and the dot product in $\mathbf{h}_L\boldsymbol{\cdot} \mathbf{T}_{L}$ is the shorthand notation for $\sum_{M=-L}^L h_{L M}T_{L M}$. The properties of $T_{LM}$ imply that $h^{*}_{LM} = (-1)^M h_{L -M} $ and that each vector $\textbf{h}_L$ transforms as a $L$-spinor under $\pointg{SU}(2)$ rotations\footnote{More precisely, we have from Eqs.~\eqref{prop.sh} and \eqref{rhoTLMexpansion}
\begin{equation*}
    D^{(j)}(\Rr)\, H\, D^{(j) \dagger}(\Rr)  = \sum_{L=0}^{2j} \tilde{\mathbf{h}}_L \boldsymbol{\cdot} \mathbf{T}_{L} 
\end{equation*}
with $\tilde{\mathbf{h}}_L = ( \tilde{h}_{LL} , \dots ,  \tilde{h}_{L-L}) $ where 
\begin{equation*}
\tilde{h}_{LM}=\sum\limits_{M'=-L}^{L} D_{ M M'}^{(L)}(\Rr)\, h_{L M'}.
\end{equation*}
}. Reference~\cite{Ser.Bra:20} describes how to use these properties to uniquely characterize any Hermitian operator $H$ by
\begin{enumerate}
    \item The Majorana representation, or constellation, of each constituent spinor $\textbf{h}_L$, denoted $\co_L(H)$ or simply $\co_L$. It is a geometric object with antipodal symmetry made of $2L$ points (stars) on the sphere~\cite{Ser.Bra:20} (see also Appendix~\ref{App.Majorana}). When $h_L =0$, there is no associated constellation.
    \item An equivalence class $[\co_L]$ of star colorings (with two different colors available, for example black and red) of $\co_L$ such that each pair of antipodal stars is made up of stars of different colors. We say that two colorings belong to the same equivalence class if they differ by an even number of color exchanges in their antipodal pairs. Therefore, independently of the number and configuration of the stars, there can be only two equivalence classes for each constellation.
    \item The norms $h_L$ associated with each $\mathbf{h}_L$, which can be considered as the radii of the spheres of each constellation. Note that for $L=0$, $h_0 = \Tr (H)/\sqrt{2j+1}$.
\end{enumerate}
In summary, $\{h_0 , h_1 , [\co_1], \dots , h_{2j} , [\co_{2j}] \}$ is the set of parameters required to completely and bijectively characterize a Hermitian operator~\eqref{rhoTLMexpansion}, where $\co_{L}$ has $2L$ stars.
\subsection{Point groups of Hermitian operators via Shubnikov groups}
\label{Subsec.Shubnikov}
\begin{table}[t]
    \centering
    \begin{tabular}{c|c|l}
        $L$ 
        & $\group$ 
        &
        Invariant operator of $\BHs^{(L)}$
        \\
        \hline
        \\[-8pt]
       0 & $\mathrm{SO(3)}$ & $\, T_{00}~ \propto \mathds{1}$
        \\[6pt]
        1 & $\pointg{C}_{\infty}$ & $\, T_{10}~  \propto J_z$
        \\[6pt]
        2 & $ \pointg{D}_{\infty} $ & $\, T_{20}~ \propto 3J_z^2 -\mathbf{J}^2$
        \\[6pt]
        3 & $\left\{ \begin{array}{l}
             \pointg{C}_{\infty} \\
             \pointg{D}_3 \\ 
             \pointg{T} 
        \end{array} \right. $ & 
        $\begin{array}{l}
            T_{30}~\propto \left( 1-3 \mathbf{J}^2+5 J_z^2 \right)  J_z \\[2pt]
            T_{33} - T_{3-3}~ \propto J_+^3 + J_-^3 \\[2pt] 
            T_{32} + T_{3-2}~ \propto \{J_+^2 + J_-^2,J_z\}
        \end{array} $
       \\[18pt]
        4 & $ \left\{ \begin{array}{l}
             \pointg{D}_{\infty}  \\
             \pointg{O} 
        \end{array}  \right.$ 
        & 
        $\begin{array}{l}
            T_{40} \\
            T_{44} + T_{4-4}  + \sqrt{\frac{14}{5}} \,T_{40}  
        \end{array} $
        \\[12pt]
        5 & $\left\{ 
        \begin{array}{l}
             \pointg{C}_{\infty}  \\
             \pointg{D}_{5}
        \end{array}
        \right.
        $ & 
        $\begin{array}{l}
            T_{50} \\
            T_{55}-T_{5-5} 
        \end{array}$
        \\[12pt]
        6 &
        $\left\{ 
        \begin{array}{l}
             \pointg{D}_{\infty}  \\
             \pointg{O} \\
             \;\pointg{I}
        \end{array}
        \right. $
        & 
        $\begin{array}{l}
             T_{60} \\
              T_{64} + T_{6-4}  -\sqrt{\frac{2}{7}}\, T_{60}  \\
              T_{65} - T_{6-5} + \sqrt{\frac{11}{7}}\,T_{60} 
        \end{array} $
        \\[18pt]
        7 & $\left\{ 
        \begin{array}{l}
             \pointg{C}_{\infty}  \\
             \pointg{D}_{7}        \\
             \pointg{T}
        \end{array}
        \right.
        $ & 
        $\begin{array}{l}
            T_{70} \\
            T_{77}-T_{7-7} \\
            T_{76}-T_{7-6}  + \sqrt{\frac{13}{11}} \left( T_{72} - T_{7-2} \right)
        \end{array}$ 
    \end{tabular}
\caption{Two left-hand columns: Largest point groups $\group$ of different irreps $\mathscr{B}^{(L)}$ of dimensions $2L+1$ for angular momentum quantum numbers $L=1,\dots, 7$. Right-hand column: example of a Hermitian operator in $\BHs^{(L)}$ with $\group$ symmetry. For each $L$, the set of point groups shown is denoted $\SetG_{\mathrm{max}}\left(\BHs^{(L)} \right)$. For example, $\SetG_{\mathrm{max}}\left(\BHs^{(4)} \right)=\{\pointg{D}_{\infty},\pointg{O}\}$. Any operator in $\BHs^{(L)}$ necessarily has a point group equal to a subgroup of an element of $\SetG_{\mathrm{max}}\left(\BHs^{(L)} \right)$.}
    \label{Table.1}
\end{table}
\begin{table*}[t]
    \centering
    \begin{tabular}{@{\hskip 0.04in} c @{\hskip 0.04in} | | @{\hskip 0.04in} c @{\hskip 0.04in} |@{\hskip 0.04in} c @{\hskip 0.04in}|@{\hskip 0.04in} c @{\hskip 0.04in}|@{\hskip 0.04in} c @{\hskip 0.04in}|@{\hskip 0.04in} c @{\hskip 0.04in}|@{\hskip 0.04in} c @{\hskip 0.04in}}
        $j$ & 1/2 & 1 & 3/2 & 2 & 5/2 & $j\geqslant 3$
        \\
        \hline
        & & & & & 
        \\[-8pt]
        $\SetG_{\mathrm{max}}$   
        & $\big\{ \pointg{C}_{\infty} \big\}$
        & $\big\{ \pointg{D}_{\infty} \big\}$
        & $\big\{ \pointg{D}_{\infty} , \pointg{T} \big\}$
        & $\big\{ \pointg{D}_{\infty} , \pointg{O} \big\}$
        & $\big\{ \pointg{D}_{\infty} , \pointg{O} \big\}$
        & $\big\{ \pointg{D}_{\infty} , \pointg{O} , \pointg{I} \big\}$
    \end{tabular}
    \caption{Minimal sets of largest point groups $\SetG_{\mathrm{max}} \left(\HSs \left( \Hs^{(j)} \right) \right)$ for all possible values of the spin quantum number $j$. These sets are valid for both Hermitian and non-Hermitian operators.}
    \label{Table.2}
\end{table*}
The bijection between Hermitian operators and sets of two-color star constellations provided by Majorana's representation implies that the symmetry point group $\group_H$ of a Hermitian operator $H$\footnote{We should formally consider the symmetry subgroups of $\pointg{SU}(2)$, the double cover of $\pointg{SO}(3)$. However, the two transformations $\pm U \in \pointg{SU}(2)$ that are mapped to the same rotation have the same action on any operator as $(\pm U) H (\pm U^{\dagger}) = U H U^{\dagger}$. We can therefore restrict ourselves to subgroups of $\pointg{SO}(3)$.} is equal to the intersection of all the point groups $\group_{[\co_L]}$:
\begin{equation}
\label{Group.cal}
\group_H = \bigcap_{L=1}^{2j} \group_{[\co_{L}]} ,
\end{equation}
where $\group_{[\co_{L}]} = \pointg{SO}(3)$ if $h_{L}=0$. 
The possible constellations of Hermitian operators and their corresponding point groups can be studied for each subspace $\BHs^{(L)}$ separately (see Eq.~\eqref{Eq.decomp.HS}). 

Let us now explain a systematic method to obtain the point group of an arbitrary Hermitian operator and to identify the largest point groups that can appear in each subspace $\BHs^{(L)}$. We begin by introducing the Shubnikov groups~\cite{shubnikov1964} following the presentation given in \cite{Bra.Cra:10}. Consider a set of $N$ points on the sphere with an additional coordinate taking two possible values, for example, a color (black or red) as in our case. We now consider, in addition to the usual action of $\pointg{SO}(3)$ on the sphere, an abstract operation $I$ acting on the additional coordinate such that: i) $I$ commutes with all elements of $\pointg{SO}(3)$, $I  g=gI$ for $g\in \pointg{SO}(3)$,  and ii) $I^2=E$. All products of $\pointg{SO}(3)$ elements and $I$, $\langle \pointg{SO}(3) , I \rangle$, define a group $\pointg{M}$ which, by the properties of $I$, reduces to~\cite{shubnikov1964}
\begin{equation}
\pointg{M}  = \pointg{SO}(3) \cup I (\pointg{SO}(3)) ,
\end{equation}
with $I (\pointg{SO}(3)) = \{ Ig | g \in \pointg{SO}(3) \}$. In particular, we can choose $I$ as the operation that switches from one equivalence class $[\co_L]$ to the other, \ie,~that exchanges the colors of the stars in one antipodal pair. At the level of operators, the action of $I$ on $H \in \BHs^{(L)}$ would yield $-H$~\cite{Ser.Bra:20}. 
 Alekseĭ V. Shubnikov~\cite{shubnikov1964} has classified abstractly all the possible symmetry groups that a set of $N$ points with an additional two-valued coordinate can have, which are called Shubnikov groups or magnetic point groups. Each equivalence class $[\co_L]$ has its corresponding Shubnikov group and proper point group, which can be calculated systematically as follows:
\begin{enumerate}
\item First, we consider the Majorana's constellation $\co_L$ as a geometric object, without coloring, and determine its corresponding point group, denoted by $\group_{\co_L}$.
\item Then, we act with an element $g\in \group_{\co_L}$ on $[\co_L ]$. If the equivalence class has not changed, $g$ is a symmetry. If it has changed, $g$ is no longer a symmetry, but $Ig$ is. By repeating this procedure on each element of $\group_{\co_L}$, and adding $I$ whenever necessary to obtain a symmetry operation, we get a set of elements that leave $[\co_L]$ invariant and form the Shubnikov point group of $[\co_L]$ denoted by $\pointg{M}_{[\co_L]} \subseteq\pointg{M}$.
\item Finally, the point group $\group_{[\co_L]}$ is the subgroup of $\pointg{M}_{[\co_L]}$ lying in $\pointg{SO}(3)$, $\group_{[\co_L]} = \pointg{M}_{[\co_L]} \cap \group_{\co_L}$.
\end{enumerate}

At the end of the procedure, we find a group $\group_{[\co]} \subseteq\group_{\co}$ which is independent of the equivalence class, $\group_{[\co]} = \group_{[\co]'}$. Let us look at some examples. For the point group of a regular $2n$-gon on a great circle, $\group_{[\co]}= \pointg{D}_n$. For an octahedron, $\group_{[\co]}= \pointg{T}$. For a cube, $\group_{[\co]}= \pointg{O}$. For a dodecahedron or icosahedron, $\group_{[\co]}= \pointg{I}$. Lastly, for a pair of $n$ coincident points in antipodal directions, $\group_{[\co]}= \pointg{C}_{\infty}$ or $\pointg{D}_{\infty}$ for $n$ odd or even, respectively.

We list in Table~\ref{Table.1} examples of Hermitian operators in $\BHs^{(L)}$ having the largest point groups for $L=1,\dots , 7$. This Table also shows the smallest $L$ required for an operator to exhibit some point group symmetry. 
\subsection{Point groups of non-Hermitian operators}

\begin{figure}[t!]
\begin{center}
\includegraphics[width=.45\textwidth]{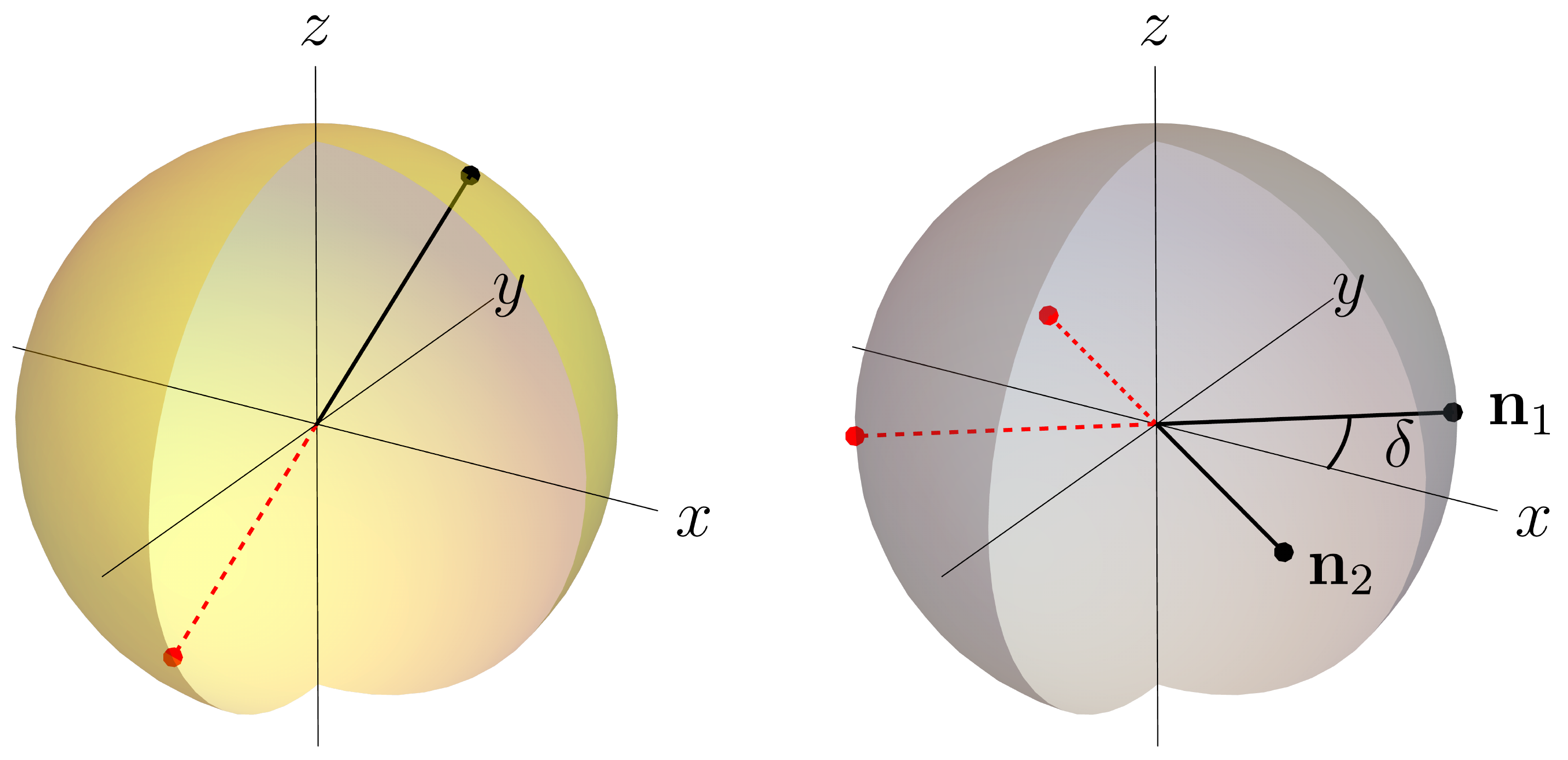} 
\end{center}
\caption{\label{Figure.1} Left: the constellation class $[\co_1 ]$ of a Hermitian operator $H \in\BHs^{(1)}$. Right: the constellation class $[\co_2 ]$ of a Hermitian operator $H \in \BHs^{(2)}$, oriented so that the constellation lies in the $xy$ plane.
}
\end{figure}
All the developments so far are valid only for Hermitian operators. However, they can be generalized to any operator $S$ using its decomposition into a Hermitian and an anti-Hermitian component
\begin{equation}
\label{Eq.Non.Hermitian}
    S = S_H + i S_A  ,
\end{equation}
where both operators $S_H = (S + S^{\dagger})/2$ and $S_A = (S - S^{\dagger})/2i$ are Hermitian. The decomposition is unique, and any unitary transformation applied to $S$ preserves it by linearity. Thus, the Majorana representation of a generic operator $S$ consists of the Majorana representation of the two parts $S_H$ and $S_A$, which is again a bijective characterisation of the operator. In particular, the point group of $S$, $\group_S = \group_{S_H}\cap \group_{S_A}$, can be calculated using the techniques mentioned previously. A direct consequence is that the only admissible point groups of an arbitrary operator are the same point groups as for Hermitian operators or some of their subgroups. Thus, the results presented in Table~\ref{Table.1} also hold for non-Hermitian operators.
\subsection{Set of largest point groups and examples}

\begin{table*}[t]
\begin{equation*}
\begin{array}{c ||c |c |c |c |c |c |c |c}
\group_{\co_2} &   E & \Rr (\mathbf{z}, \pi/2) & \Rr (\mathbf{z}, \pi) & \Rr (\mathbf{z}, 3\pi/2) & \Rr(\mathbf{x}, \pi) & \Rr(\mathbf{y}, \pi)  & \Rr(\mathbf{n}_1, \pi) & \Rr(\mathbf{n}_2, \pi)  
\\
\hline
\pointg{M}_{[\co_2]} & E & I \Rr (\mathbf{z}, \pi/2)  & \Rr (\mathbf{z}, \pi) & I \Rr (\mathbf{z}, 3\pi/2) & \Rr(\mathbf{x}, \pi) & \Rr(\mathbf{y}, \pi)  & I \Rr(\mathbf{n}_1, \pi) & I \Rr(\mathbf{n}_2, \pi) 
\end{array}
 , 
 \end{equation*}
    \caption{Elements of the groups associated with the constellation $\co_2$ shown in Fig.~\ref{Figure.1} (right sphere) with $\delta= \pi/4$. The elements of $\group_{\co_2}$ leave the constellation invariant if we ignore the color of the stars, whereas the elements of $\pointg{M}_{[\co_2]}$ leave the constellation invariant when we take the color of the stars into account. The point group of the equivalence class $[\co_2]$ is the intersection of the two sets, \ie,~$\group_{[\co_2]} = \{ E , \Rr (\mathbf{z}, \pi) ,  \Rr(\mathbf{x}, \pi) , \Rr(\mathbf{y}, \pi) \} = \pointg{D}_2$. }
    \label{Table.3}
\end{table*}
Let $\SetG(V)$ be the set of point groups that can appear as the symmetry group of any element in the operator space $V$, excluding multiples of the identity operator.
Additionally, let $\SetG_{\mathrm{max}}(V)$ be the minimal subset of $\SetG(V)$, such that each element of $\SetG (V)$ is a subgroup\footnote{Here, we use the usual convention that a set (group) is a subset (subgroup) of itself.} of an element of $\SetG_{\mathrm{max}}(V)$\footnote{The set $\SetG_{\mathrm{max}}(V)$ can be understood as a generalization, to sets, of the concept of maximal subgroups. A proper subgroup $\group'$ of a group $\group$ is called \emph{maximal} if there is no other subgroup of $\group$ that contains $\group'$ strictly.}.

The point groups listed in Table~\ref{Table.1} correspond to $\SetG_{\mathrm{max}}(\BHs^{(L)})$ for $L\leqslant 7$. Using Eqs.~\eqref{Eq.decomp.HS} and \eqref{Group.cal}, we can now obtain $\SetG_{\mathrm{max}}(\HSs(\Hs^{(j)}))$ for any spin value. These sets are listed in Table~\ref{Table.2}. Since any proper point group (excluding $\pointg{SO}(3)$) is a subgroup of at least one of the groups $\pointg{D}_{\infty}$, $\pointg{O}$, $\pointg{I}$, we have $\SetG_{\mathrm{max}}(\HSs(\Hs^{(j)})) = \SetG_{\mathrm{max}}(\HSs(\Hs^{(3)})) $ for $j\geqslant 3$.
Note that the results summarized in Tables~\ref{Table.1} and \ref{Table.2} are valid for both Hermitian and non-Hermitian operators. They will prove useful for the design of DD sequences.

For clarity, we show how to obtain the sets $\SetG$ and $\SetG_{\mathrm{max}}$ of $V=\BHs^{(L)}$ for $L=1$ and $2$. We consider first $L=1$ with a generic Hamiltonian of the form $H_{1} = \gamma\, \mathbf{n}\boldsymbol{\cdot} \mathbf{J} $, where we can assume $\gamma > 0$ without loss of generality. The decomposition \eqref{rhoTLMexpansion} of $H_1$ gives $h_0=0$ and $h_1 = \gamma \sqrt{j(j+1)}$, with only one constellation class $[\co_1]$ made of a pair of antipodal black and red stars (see Fig.~\ref{Figure.1}, left sphere). The black (or red) star points to $ \pm \mathbf{n}$, where the sign is associated with the coloring of the pair, or, in other words, with the equivalence class. It is now easy to see that $\SetG ( H_{1}) = \SetG_{\mathrm{max}} ( H_{1}) = \{\pointg{C}_{\infty} \}$.

We now turn to $L=2$ with a generic Hamiltonian of the form $H_{2} = \mathbf{h}_2 \bcdot \mathbf{T}_2$. The operators of $\BHs^{(2)}$ have constellation classes $[\co_2]$ made up of two pairs of antipodal black and red stars (see Fig.~\ref{Figure.1}, right sphere). We can identify the two black stars by the unit vectors $\{ \mathbf{n}_1 , \, \mathbf{n}_2 \}$. The equivalence class $[\co_2]$ has two elements corresponding to constellations with black stars pointing to $\{ \mathbf{n}_1 , \, \mathbf{n}_2 \}$ and $\{ -\mathbf{n}_1 , \, -\mathbf{n}_2 \}$. All possible constellations can be parametrized by an angle $\delta \in [0,\pi/2]$. 
As an example of the procedure explained in Subsection~\ref{Subsec.Shubnikov}, we consider the case where $[\co_2]$ forms a square on the equator ($\delta =\pi/4 $). We list in Table~\ref{Table.3} the point group of a square, which contains 8 elements. We can now determine the group $\pointg{M}_{[\co_2]}$ and shown in the second row of Table~\ref{Table.3}. From this we conclude that $\group_{[\co_2]} = \pointg{D}_2$. We can obtain all possible point groups for the Hermitian operators $H \in \BHs^{(2)}$ by inspecting all possible constellations $[\co_2]$, which yields
\begin{equation}
\label{PG.2}
\begin{aligned}
\group_{[\co_2]} = & \left\{  
\begin{array}{l l}
\pointg{D}_{\infty}  & \quad \text{ for } \delta = 0 \text{ and } \pi/2
\\[0.2cm]
\pointg{D}_2  & \quad \text{ for } \delta \in  ( 0 , \pi/2 )
\end{array}
\right. .
\end{aligned}
\end{equation} 
Now, the possible point groups of non-Hermitian operators $S =S_H + i S_A \in \BHs^{(2)}$ are defined by the intersection of the two points groups $\group_{S_H} \cap \group_{S_A}$. The axes of symmetry of the groups may or may not coincide. In particular, if $\group_{S_H} = \group_{S_A}= \pointg{D}_2$ and if only one axis of symmetry of each of these groups coincides, then $\group_{S_H} \cap \group_{S_A} = \pointg{C}_2$. On the other hand, if they share no axis of symmetry, then $\group_{S_H} \cap \group_{S_A} = E$. Thus, the set of point groups for generic operators (Hermitian and non-Hermitian) is $\SetG(\BHs^{(2)}) = \{ \pointg{E},\pointg{C}_2 , \pointg{D}_2 ,\pointg{D}_{\infty} \}$, and $\SetG_{\mathrm{max}}(\BHs^{(2)}) = \{ \pointg{D}_{\infty} \} $ since $\pointg{E},\pointg{C}_2$, and $\pointg{D}_2$ are subgroups of $\pointg{D}_{\infty}$.
\section{Decoupling groups for single spins}\label{Sec. single spin}

In this section, we apply the general framework presented in Sections~\ref{Sec.In.Sym} and \ref{Sec.Gen.Frame}, using the point groups of operators introduced in the previous section, to find the decoupling groups of a single spin-$j$ based on its interaction subspace. The main result of this section is summarized in Proposition~\ref{Result.1} and Table~\ref{Table.4} and is applied to several systems.
\subsection{General theory}
Consider a single spin-$j$ system, with Hilbert space $\Hs_S=\Hs^{(j)}$, interacting with its environment via the interaction Hamiltonian \eqref{Schmidtoperator}, leading to the interaction subspace $\mathcal{I}_S =\mathrm{span}(\qty{S_{\alpha}})$. The subspace $\mathcal{I}_S$, which may not be closed under SU$(2)$ operations, is contained in a direct sum of irreps 
\begin{equation}
    \mathcal{I}_S \subseteq V = \bigoplus_{L=0}^{L_{\mathrm{max}}} \BHs^{(L)} \subseteq \HSs(\mathcal{H}_S) 
\end{equation}
where $L_{\mathrm{max}}\leq 2j$ is the smallest $L$ value for which the first inclusion is valid, and where the irrep $\BHs^{(0)}$ ($0$-irrep) contains only operators proportional to the identity $\mathds{1}_S$. 
\par 
As explained in Sec.~\ref{Sec.Gen.Frame}, a rotational symmetry that does not appear in any of the irreps $\BHs^{(L)}$ ($1\leq L \leq L_{\mathrm{max}}$) will be a decoupling group. Using Tables~\ref{Table.1} and ~\ref{Table.2}, we can first find the set of largest rotational symmetries $\mathcal{F}_{\mathrm{max}}(V)$ by considering $j_{\mathrm{eff}} = L_{\max}/2$. We can then find the smallest decoupling group by selecting the smallest rotational symmetry group which is not a subgroup of an element of $\mathcal{F}_{\mathrm{max}}(V)$. We summarize the results in the following proposition. 

\begin{table}[t]
    \centering
    \begin{tabular}{c||c|c|c|c}
       $L_{\mathrm{max}}$  & $\mathcal{F}_{\mathrm{max}}(V)$  &$\group$ & $\abs{\group}$ & Operators $S_{\alpha}$ \\[2pt]\hline \hline \rule{0pt}{1\normalbaselineskip}
       1 \rule{0pt}{1\normalbaselineskip} & $\big\{ \pointg{C}_{\infty} \big\}$ & $\pointg{D}_2$ & 4 & $\mathbf{J}\boldsymbol{\cdot} \mathbf{n}$
       \\[2pt]
       2  & $\big\{ \pointg{D}_{\infty} \big\}$ & $\pointg{T}$ & 12 & $J_z^2 ,  J_x^2- J_y^2 $
       \\[2pt]
       3 &$\big\{ \pointg{D}_{\infty}, \pointg{T} \big\}$ & $\pointg{O}$ & 24
       & $J_{z}^3, J_x J_z J_x $
       \\[2pt]
       4 or 5 & $\big\{ \pointg{D}_{\infty}, \pointg{O} \big\}$ & $\pointg{I}$ & 60 & $J_z^4,J_z^5$ \\
       $\geq$ 6 & $\big\{ \pointg{D}_{\infty}, \pointg{O},\pointg{I} \big\}$ & none & $-$ & $-$
    \end{tabular}
    \caption{The sets of largest point groups for any interaction subspace $\mathcal{I}_S \subseteq \bigoplus_{L=0}^{L_{\mathrm{max}}} \BHs^{(L)}$ as a function of $L_{\mathrm{max}}$, along with the corresponding smallest decoupling group $\group$ and its cardinality. For $L_{\mathrm{max}} \geq 6$, there is no decoupling group composed solely of rotations. The point groups listed are universal decoupling groups for single spin systems with spin quantum number $j\leq L_{\mathrm{max}}/2$. The last column shows examples of the system operators $S_\alpha$ in the interaction Hamiltonian~\eqref{Schmidtoperator} that are decoupled.}
    \label{Table.4}
\end{table}
\begin{Proposition}
\label{Result.1}
Consider a single quantum system of spin $j$ interacting with its environment, whose interaction subspace $\mathcal{I}_S \subseteq \bigoplus_{L=0}^{L_{\mathrm{max}}} \BHs^{(L)}$ for some integer $L_{\mathrm{max}}\geq 1$. Any point group that is not a subgroup of an element of $\SetG_{\mathrm{max}}\qty(\bigoplus_{L=1}^{L_{\mathrm{max}}} \BHs^{(L)})$ is a decoupling group of  $\mathcal{I}_S$. For $L_{\mathrm{max}}=2j$, this is moreover a universal decoupling group.
\end{Proposition}

\par Note that Proposition \ref{Result.1} refers to a subgroup (not a proper subgroup), which may be the group itself. The smallest decoupling groups for the first values of $L_{\mathrm{max}}$ are listed in Table~\ref{Table.4}, derived from the results contained in Tables~\ref{Table.1} and ~\ref{Table.2}. Four interesting groups emerge, namely the 2-Dihedral ($\pointg{D}_2$), Tetrahedral ($\pointg{T}$), Octahedral ($\pointg{O}$) and Icosahedral ($\pointg{I}$) point groups. Each of the last three corresponds to the symmetry group of one of the Platonic solids.
\\
\indent An interesting observation is that when $L_{\mathrm{max}}\geq 6$, the set of largest point groups is $\big\{ \pointg{D}_{\infty}, \pointg{O},\pointg{I} \big\}$. Any finite proper point group is a subgroup of an element of this set such that the smallest rotational symmetry inaccessible at this point is the entire rotation group $\mathrm{SO(3)}$, which would result in a DD sequence of infinite number of pulses. The following corollaries follow.
\begin{Corollary}\label{Corollary}
    There is no universal decoupling group based solely on rotations when $j\geq 3$.
\end{Corollary}
\begin{Corollary}\label{Corollary2}
    If the interaction subspace is such that $\SetG_{\mathrm{max}} = \big\{ \pointg{D}_{\infty}, \pointg{O},\pointg{I} \big\}$, then there is no decoupling group based solely on rotations.
\end{Corollary}
\begin{table*}[t]
    \centering
    \begin{tabular}{c|c|c|c|c|c}
    \hline 
      \multirow{2}{*}{Decoupling group}& Platonic & \multicolumn{2}{c|}{Generators} &\multirow{2}{*}{\# pulses} & \multirow{2}{*}{Sequence} \\\cline{3-4} 
         & sequence & $a$ & $b$ && \\ \hline 
         \rule{0pt}{1\normalbaselineskip} 2-Dihedral ($\pointg{D}_2$) \rule{0pt}{1\normalbaselineskip} & EDD & $\qty(\qty(1,0,0), \pi)$& $\qty(\qty(0,1,0), \pi)$& 8 & $abab^2aba$\\[2pt]
         Tetrahedral ($\pointg{T}$) & TEDD & $\qty(\qty(0,0,1),\frac{2\pi}{3})$ & $\qty(\qty(\frac{\sqrt{2}}{3},\sqrt{\frac{2}{3}},\frac{1}{3}),\frac{2\pi}{3})$& 24 & $aba^2bab^3a^2bab^3a^2bab^2a^2$ \\[4pt]
         Octahedral ($\pointg{O}$) & OEDD  & $\qty(\qty(0,0,1),\frac{\pi}{2})$ & $\qty(\frac{1}{\sqrt{3}}\qty(1,1,1), \frac{2\pi}{3})$ & 48& see Appendix \ref{Ap.CayleyDiagram}, Eq.~\eqref{platonicseq_OEDD}\\[2pt]
         Icosahedral ($\pointg{I}$) & IEDD & $\left( \frac{(0,-1,\phi)}{\sqrt{\phi+2}} ,\frac{2\pi}{5} \right)$ & $\left( \frac{(1-\phi , 0, \phi)}{\sqrt{3}} ,\frac{2\pi}{3} \right)$ & 120& see Appendix \ref{Ap.CayleyDiagram}, Eq.~\eqref{platonicseq_IEDD}\\[2pt]
         \hline 
    \end{tabular}
    \caption{Summary of the Platonic DD sequences constructed from the point groups $\pointg{T}$, $\pointg{O}$ and $\pointg{I}$. The generators are specified in the axis-angle notation and correspond to the two types of pulses required to implement each sequence. The two shortest sequences are given in their condensed notation and we refer to Appendix~\ref{Ap.CayleyDiagram} for the two longest sequences. In the IEDD sequence generators, $\phi=\frac{\sqrt{5}+1}{2}$ is the golden ratio.}
    \label{Table.5}
\end{table*}
\begin{figure}[t!]
    \centering
    \includegraphics[width=\linewidth]{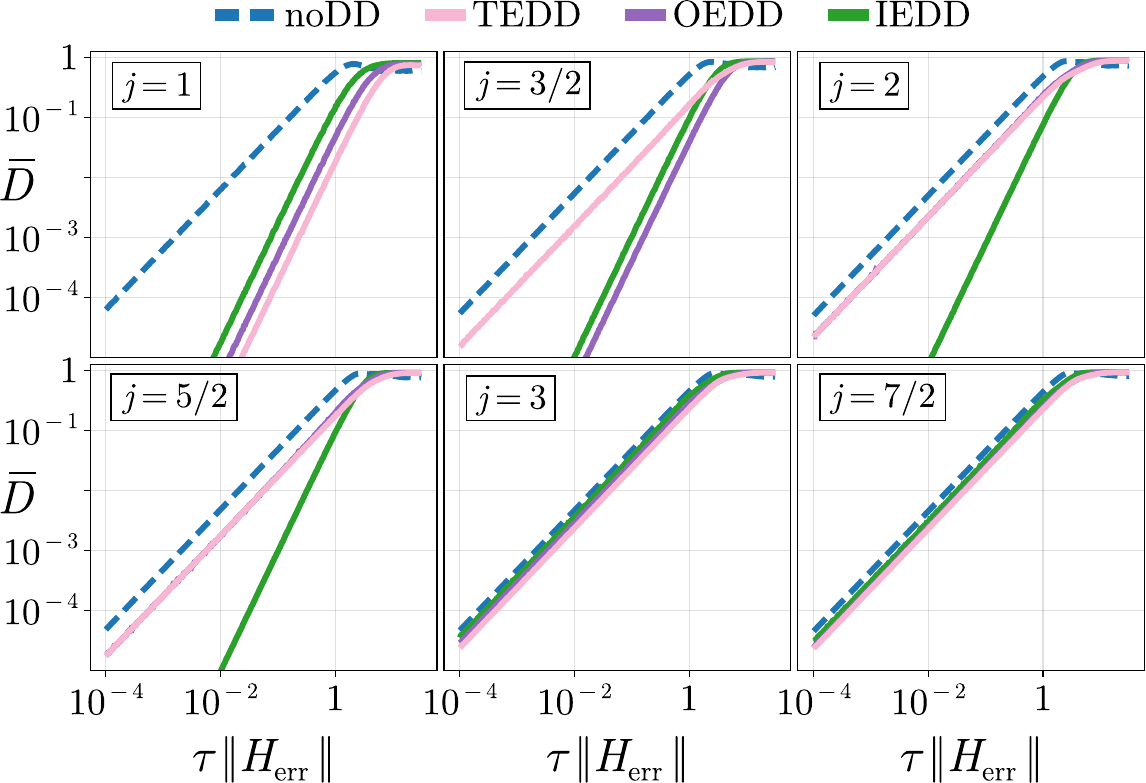}
    \caption{Average distance $\overline{D}$ between the identity propagator and the noisy propagator for a single spin-$j$ system (where $j$ ranges from 1 to $7/2$), with and without DD sequence. $\overline{D}$ is plotted as a function of $\tau \norm{H_{\mathrm{err}}}$ for the TEDD (purple), OEDD (red) and IEDD (green) sequences and for the free evolution (noDD, blue). $H_{\mathrm{err}}$ is a Hermitian error Hamiltonian and $\tau$ is the total time of the free propagation (and also the duration of the TEDD sequence, see main text).}
    \label{Figure.2}
\end{figure}
\subsection{Applications}
\par We are now in a position to construct a sequence of pulses for each decoupling group of Table~\ref{Table.4} by choosing an Eulerian path on the corresponding Cayley graph. In particular, each group appearing in Table~\ref{Table.4}, and thus its corresponding Cayley graph, has two generators $\qty(a,b)$, such that the sequences will be composed of only two different types of pulses. The exact form of the resulting sequences and details of their construction can be found in Appendix \ref{Ap.CayleyDiagram}, but the results are summarized in Table~\ref{Table.5}. For the most elementary system, the qubit (spin-$1/2$), the smallest universal decoupling group is $\pointg{D}_2$. The corresponding Eulerian sequence has already been studied in the literature and has been called \textit{Eulerian Dynamical Decoupling} (EDD)~\cite{Viola_2003}. The novelty here lies in the three other exceptional groups appearing in Table~\ref{Table.4}. We call their Eulerian sequences \textit{Platonic DD sequences} because their respective symmetries are the same as those of Platonic solids. We denote them by XEDD, with $\mathrm{X}\in\{\mathrm{T}, \mathrm{O}, \mathrm{I}\}$.

\par A direct application of our framework is the construction of universal decoupling sequences for a single spin $j$ (using Proposition~\ref{Result.1} with $L_{\mathrm{max}}=2j$). For spins with $j<3$, it is possible to find a universal decoupling group in Table~\ref{Table.4} since it suffices to choose any of the decoupling groups with $L_{\mathrm{max}}\geqslant 2j$. For example, the octahedral group is a universal decoupling group for spins $1/2$, $1$ and $3/2$. 

\par To demonstrate and validate the decoupling properties of the Platonic sequences, we use the DD performance quantifier introduced in Ref.~\cite{Lidar_2013} and defined as 
\begin{equation}
    D(\mathds{1}_S,U_S) = \sqrt{1-\frac{\abs{\Tr\qty[U_S]}}{2j+1}}.
\end{equation}
This quantifier $D$ was derived in Ref.~\cite{Grace_2010} and is a distance measure between the identity operator $\mathds{1}_S$ (the desired free evolution operator) and a noisy propagator $U_S$. The smaller $D$ is, the more freely the system evolves without perturbation, and the better the DD sequence. Here, we consider a noisy propagator $U_S=\exp(-i H_{\mathrm{err}} \tau)$ generated by a random error Hamiltonian $H_{\mathrm{err}}$ drawn from the Gaussian ensemble of random $(2j+1)\times(2j+1)$ Hermitian matrices. 
Figure~\ref{Figure.2} shows the average distance $\overline{D}(\mathds{1}_S,U_S)$ as a function of $\tau \norm{H_{\mathrm{err}}}$ in log-log scale for the free evolution (\ie, without the DD sequence) and with the application of each of the Platonic sequences. The average was computed over 5000 random Hamiltonians. The time $\tau$ is the duration of both the total free evolution (the $\mathrm{NoDD}$ protocol) and the shortest sequence (TEDD in this case). To enable a fair comparison of performance between the different sequences, the time interval between successive pulses was chosen to be identical for each sequence.

When there is no first-order decoupling, the average distance $\overline{D}$ should scale as $\overline{D} \propto \tau \norm{H_{\mathrm{err}}}$, according to the Magnus expansion. This is indeed the behavior we observe, for any spin, in the log-log plot in Fig.~\ref{Figure.2} in the absence of DD sequence (dashed blue curves with a slope of $1$) and for certain DD sequences. In this plot, a slope that is twice as steep indicates first-order decoupling, \ie, $\overline{D} \propto \qty(\tau \norm{H_{\mathrm{err}}})^2$~\cite{Lidar_2013}. 
It can also be seen that when several sequences achieve decoupling, those with a smaller decoupling group result in a smaller distance and therefore better performance. This can be understood by the fact that the time required to implement a DD sequence is proportional to the number of elements in the decoupling group.

\par
\begin{figure}[t!]
    \centering
    \includegraphics[width=0.4\textwidth]{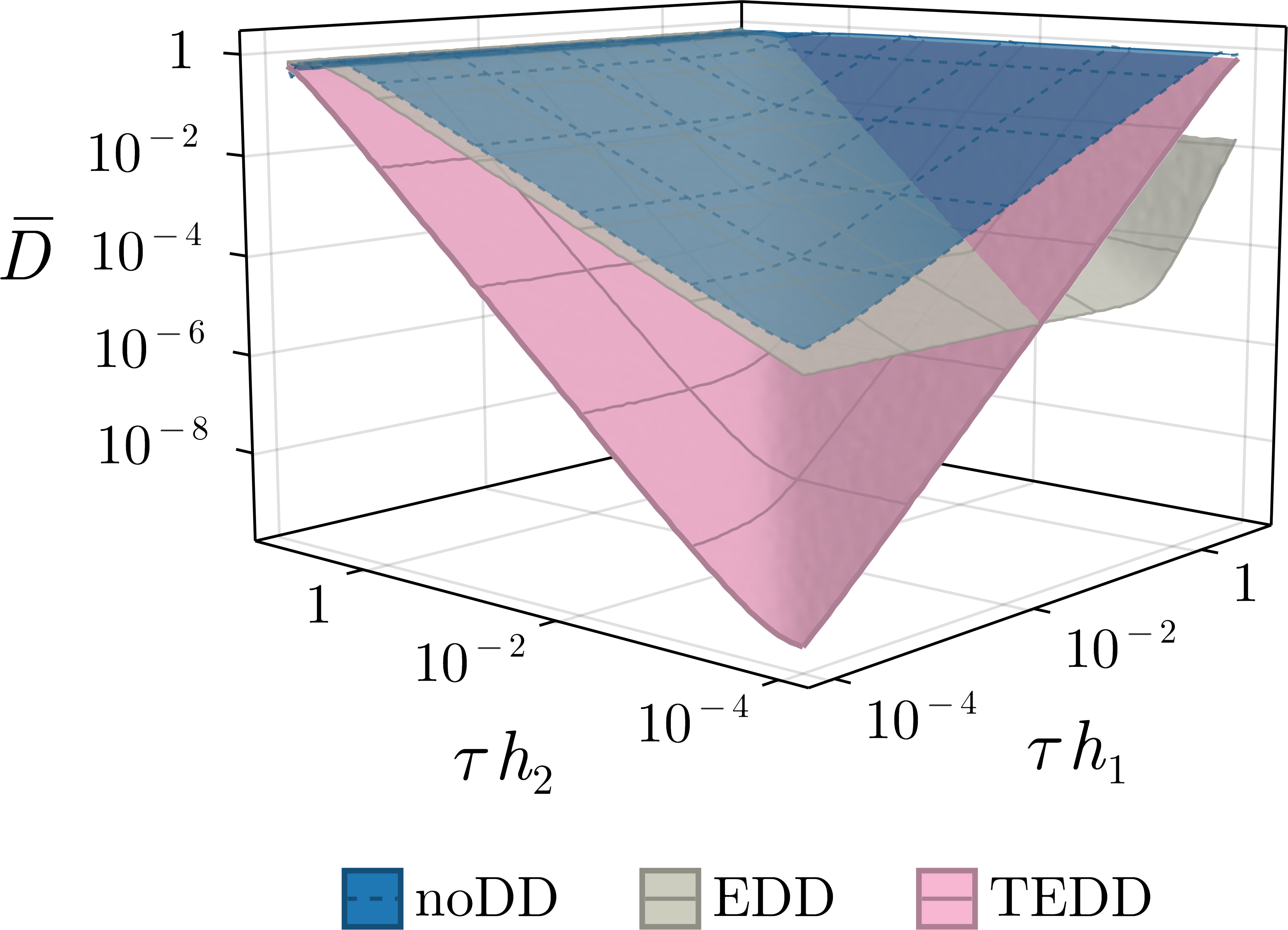}
    \caption{Average distance $\overline{D}$ between the identity propagator and the noisy propagator in the $(\tau h_1, \tau h_2)$ parameter space for a single spin-$1$ system, for the free evolution (blue) and the EDD (gray) and TEDD (pink) sequences. The averages were performed over 1000 randomly generated Hamiltonians. 
    }
    \label{Figure.3}
\end{figure}

\begin{figure}[t!]
    \centering
    \includegraphics[width=0.4\textwidth]{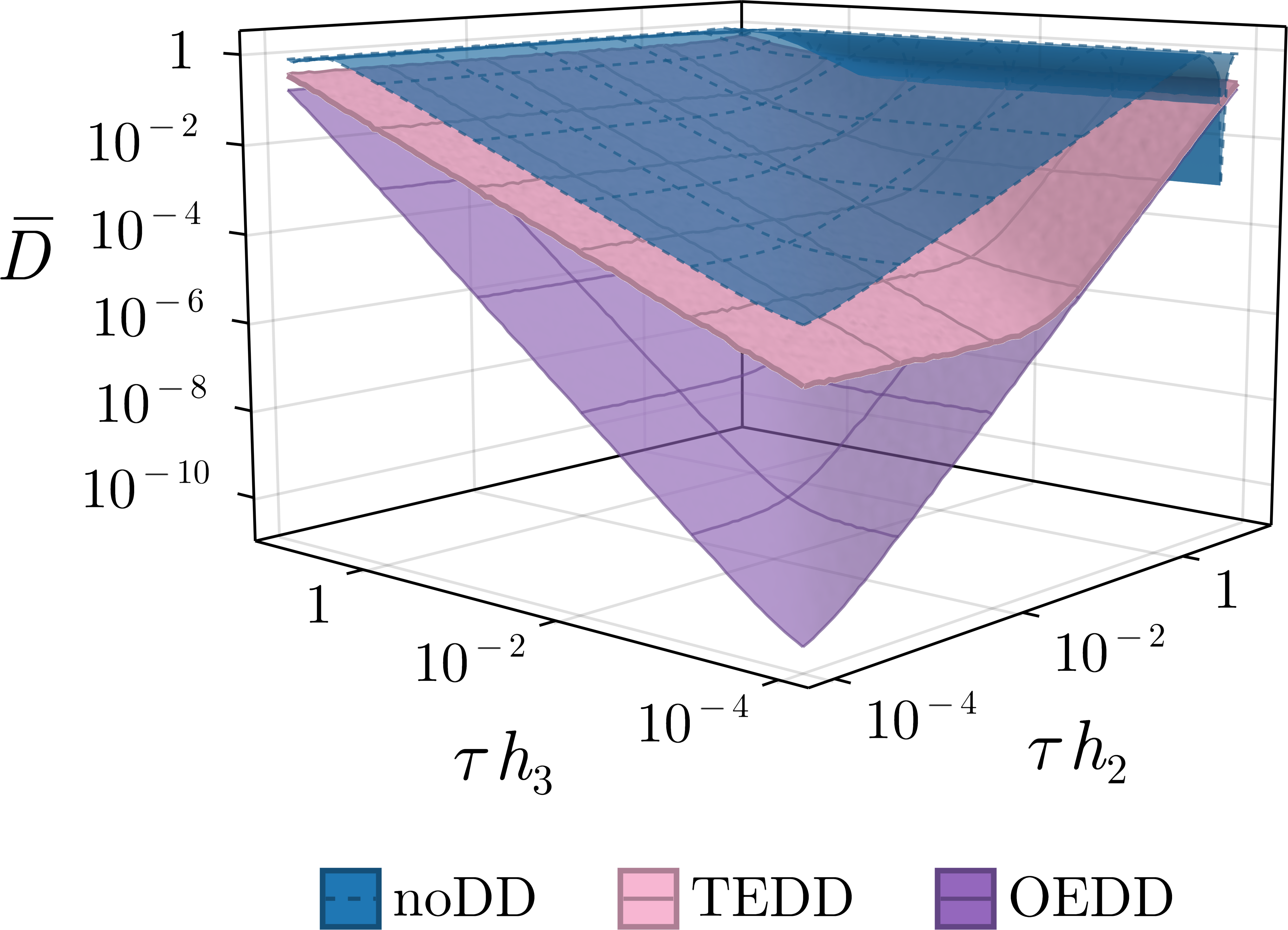}
    \caption{Same as in Fig.~\ref{Figure.3} but in the $(\tau h_2, \tau h_3)$ parameter space of a single spin-$3/2$ system, for the free evolution (blue) and the TEDD (pink) and OEDD (purple) sequences. The averages were performed over 1000 randomly generated Hamiltonians. 
    }
    \label{Figure.4}
\end{figure}

 Although platonic sequences do not achieve full decoupling for $j\geq 3$, they can nevertheless be of great practical interest for dynamical decoupling of large spins~\cite{omanakuttan_phd, Omanakuttan_2024}. Indeed, the TEDD sequence is capable of decoupling any Hamiltonian with $\mathcal{I}_S \subseteq \oplus_{L=1}^2 \BHs^{(L)} $, regardless the spin $j$ of the system. In this case, $\mathcal{I}_S$ consists of linear combinations of operators that are linear or quadratic with respect to the angular momentum operators. The EDD sequence, in contrast, only handles terms linear in the spin operators. As an example, we compare the performance of the EDD and TEDD sequences for a spin-$1$ on random error Hamiltonians of the form $H^{(1)} = \sum_{L=0}^2 \mathbf{h}_L\boldsymbol{\cdot} \mathbf{T}_{L}$. The average distance $\overline{D}$ is shown in the $(\tau h_1,\tau h_2)$ parameter space in Fig.~\ref{Figure.3}; when $h_1/h_2 \ll 1$ (resp. $h_1/h_2 \gg 1$), the distance scales as $\overline{D}\propto \qty(\tau h_2)^r$ (resp. $\overline{D} \propto \qty(\tau h_1)^r$) with $r=1$ when there is no decoupling and $r=2$ when first-order decoupling is achieved. As expected, we observe that the TEDD sequence outperforms the EDD sequence in cases where the quadratic components $h_2$ are not negligible compared to the linear components ($r=2$ for TEDD and $r=1$ for EDD in the regime $h_1/h_2 \ll 1$). 
A similar advantage of the OEDD sequence over the TEDD sequence is observed in Fig.~\ref{Figure.4} for random spin-$3/2$ Hamiltonians of the form $H^{(3/2)} = \sum_{L=2}^3 \mathbf{h}_L\boldsymbol{\cdot} \mathbf{T}_{L}$, where we show the average distance in the parameter space $(\tau h_2,\tau h_3)$. Here we set $h_0$ and $h_1$ equal to zero for convenience. It should be noted that similar results can be obtained for any spin, since the decoupling properties of the sequences depend only on $L_{\mathrm{max}}$ and not on $j$. 

\subsection{Comparison with existing schemes}
We first point out that the universal Platonic sequence designed for a spin-$j$ can also serve as a universal sequence for an arbitrary qudit with a number of levels $d = 2j+1$, provided that the two $\mathrm{SU(2)}$ pulses $a$ and $b$ can be implemented efficiently. Moreover, the generator $\mathbf{J}\boldsymbol{\cdot}\vec{n}$ of rotation pulses is tridiagonal in the energy eigenbasis, so it suffices to have control over transitions between adjacent energy levels\footnote{For example, the two generating pulses of $\mathrm{T}$ for a qutrit system $(d=3)$ are defined by the spin-1 generators $\frac{2\pi}{3}\mathbf{J}\boldsymbol{\cdot}\vec{n}$ with rotation axes given Table~\ref{Table.5}. Their respective $3\times 3$ matrices are equal to 
\begin{equation}
\begin{aligned}    
    \mathbf{J}\bm{\cdot}(0,0,1){}={}& \left(
\begin{array}{ccc}
 1 & 0 & 0 \\
 0 & 0 & 0 \\
 0 & 0 & -1 \\
\end{array}
\right) ,
\\[12pt]
   \mathbf{J}\bm{\cdot} \left( \frac{\sqrt{2}}{3},\sqrt{\frac{2}{3}},\frac{1}{3} \right){}={}&
\frac{1}{3} 
\left(
\begin{array}{ccc}
 1 & 1-i \sqrt{3} & 0 \\
 1+i \sqrt{3} & 0 & 1-i \sqrt{3} \\
 0 & 1+i \sqrt{3} & -1 \\
\end{array}
\right)
,
\end{aligned}
\end{equation}
Their entries can be modified by choosing different axes of rotation.}.

However, Platonic sequences may not be the most efficient DD sequences for an arbitrary qudit. For instance, the universal sequences in Ref.~\cite{tripathi_2024} are more efficient---although not being based solely on global rotations---as the number of pulses scales only with the square of the qudit dimension $d$.
For example, this sequence for a qudit with $d=6$ levels requires 36 pulses, which is significantly fewer pulses than the corresponding Platonic icosahedral sequence $\mathrm{IEDD}$ (120 pulses). On the other hand, when dephasing is the dominant decoherence error in a qudit, the basis cycling procedure introduced in Ref.~\cite{iiyama_2024} decouples the system with a pulse sequence whose size increases linearly with $d$. However, this is not a universal DD sequence.
\\
\indent Platonic sequences appear to be advantageous over the protocols of Refs.~\cite{tripathi_2024,iiyama_2024} in two cases. The first one is when the $d$-dimensional elementary quantum system can be easily controlled by global $\mathrm{SU(2)}$ pulses. For instance, such pulses are native in qudits defined in the hyperfine manifold of alkali atoms and are simply achieved by applying a magnetic field~\cite{Deutsch_2010, omanakuttan_phd, Omanakuttan_2024, Omanakuttan_2021}. In order to implement the protocols of Refs.~\cite{tripathi_2024,iiyama_2024} in this system, each transition between adjacent energy levels would need to be accessed independently, e.g.\ via optical transitions with an intermediate state from another hyperfine manifold~\cite{Ringbauer_2022}. The second case is when the interaction subspace of a large qudit only includes irreps of small dimension. This is, again, the case in the hyperfine manifold of alkali atoms where the dominant sources of errors involve operators linear and quadratic with respect to the angular momentum operators~\cite{Omanakuttan_2024}. In the 10-dimensional qudit presented in Refs.~\cite{Omanakuttan_2021, Omanakuttan_2024}, the universal sequence~\cite{tripathi_2024} would consists of 100 pulses while the dominant errors should be suppressed using the shorter $\mathrm{TEDD}$ sequence (24 pulses). This might also be the case for some decoherence model in a spinor or multi-component Bose-Einstein condensate that behaves as a spin-like system~\cite{PhysRevA.84.013606,Edri_2021}. 
\section{Decoupling spin-spin interactions with global rotations}\label{Sec. multi spin}
\label{Sec.Dec.group.composite}
We will now extend the framework of Sec.~\ref{Sec. single spin} to multispin systems where we seek to decouple the environment and/or some spin-spin interactions by using pulses composed only of global $\pointg{SU}(2)$ transformations, \ie, the same $\pointg{SU}(2)$ operation is applied to each spin. Our main theoretical results are summarized in Proposition~\ref{Result.2} and in Table~\ref{Table.6}, then applied to several systems. As explained below, some spin interactions cannot be decoupled using only global rotations; these are identified and listed in Appendix \ref{Ap.condition} for the first nontrivial systems.

\subsection{Interaction subspace of multispin systems}
\label{Subsec.interaction}
\begin{figure*}[t]
    \centering
    \includegraphics[width=\linewidth]{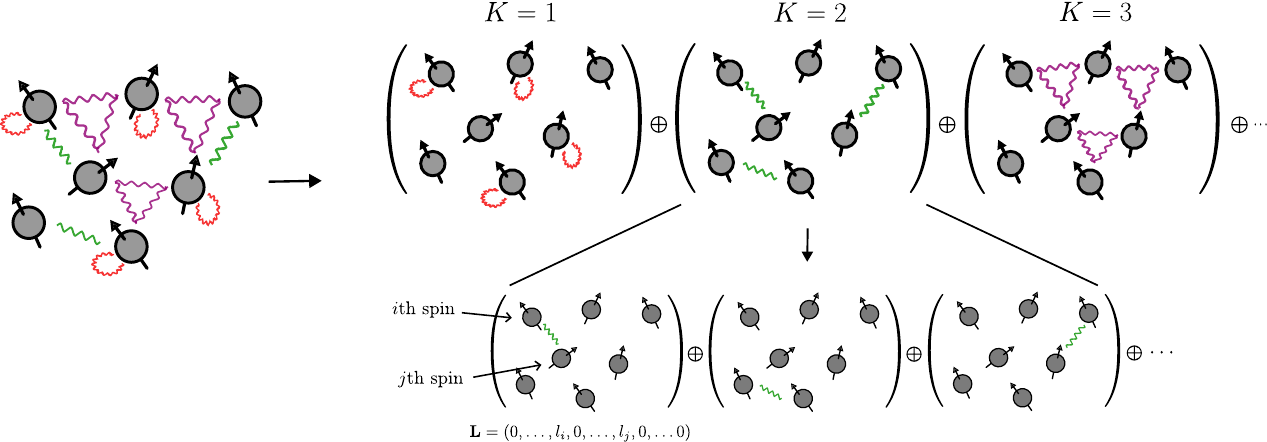}
    \caption{Visual representation of the decomposition of the interaction subspace of a multispin system into smaller subspaces, facilitating the study of the $\mathrm{SU(2)}$ symmetries. The number of non-zero elements in $\mathbf{L}$, denoted $|\mathbf{L}|$, corresponds to the order $K$ of the many-body expansion \eqref{eq:int_sub_multi}--\eqref{IKbody}.}
    \label{fig:fig9}
\end{figure*}
Consider an ensemble of $N$ interacting spins (system $S$), of spin quantum numbers $(j_1 , \dots ,j_N)$, coupled to an environment (bath $B$). Depending on the interactions we want to decouple, the Hamiltonian $H$ will include interactions between the system and the bath, interactions between the spins, or both. The most general interaction subspace contains all operators acting on the global Hilbert space associated with the spin ensemble. It can be written as tensor products of individual operator subspaces
\begin{equation}
\label{Eq.Int.Multispin}
\begin{aligned}    
\mathcal{I}_S =  
\bigotimes_{k=1}^N  \HSs (\Hs^{(j_k)}) 
= & \bigotimes_{k=1}^N \left(  \bigoplus_{L_k=0}^{2j_k} \BHs_{j_k}^{(L_k)} \right)
\\
  =  & \bigoplus_{L_1=0}^{2j_1} \dots \bigoplus_{L_N=0}^{2j_N} \qty(\bigotimes_{k=1}^N \BHs^{(L_k)}_{j_k})
\\
   = & \bigoplus_{\vec{L}} \qty(\bigotimes_{k=1}^N \BHs^{(L_k)}_{j_k}),
\end{aligned}
\end{equation}

with $\vec{L} \equiv \qty(L_1,\dots ,L_N)$ a vector-index. The possible interaction operators of $\mathcal{I}_S$ can be grouped into $K$-body terms (with $K=0,\dots , N$\footnote{When $K \geq 2$, we call the $K$-body terms also $K$-body interactions.}) according to the number of zero components in each index $\vec{L}$ of the direct sum in Eq.~\eqref{Eq.Int.Multispin}. Fig.~\ref{fig:fig9} depicts a graphical representation of this decomposition. 
Note that 1-body terms correspond to 1-local Hamiltonians and 0-body terms to operators proportional to the global identity. Thus,
\begin{equation}
    \mathcal{I}_S = \bigoplus_{K=0}^{N} \mathcal{I}^{K-\mathrm{body}}
    \label{eq:int_sub_multi}
\end{equation}
with
\begin{equation}\label{IKbody}
    \mathcal{I}^{K-\mathrm{body}} =     \bigoplus_{|\vec{L}|=K} \qty(\bigotimes_{k=1}^N \BHs^{(L_k)}_{j_k}) 
\end{equation}
and $|\mathbf{L}|$ equal to the number of nonzero components of $\mathbf{L}$. For each $\vec{L}$, the corresponding subspace decomposes entirely into irreps $\tilde{\BHs}_{\mathbf{L}}^{(\tilde{L},\alpha)}\subset \HSs(\mathcal{H}_S)$ as follows
\begin{equation}
    \bigotimes_{k=1}^N \BHs^{(L_k)}_{j_k} = \bigoplus_{(\tilde{L},\alpha)} \tilde{\BHs}_{\mathbf{L}}^{(\tilde{L},\alpha)},
    \label{eq:furtherirreps}
\end{equation}
where $\tilde{L}\in \qty{\tilde{L}_{\mathrm{min}},\tilde{L}_{\mathrm{min}}+1,\dots , \tilde{L}_{\mathrm{max}} }$, with
$\tilde{L}_{\mathrm{max}}=\sum_{k=1}^N L_k$ and $\tilde{L}_{\mathrm{min}}$ the minimum angular momentum that can be obtained according to the angular momenta coupling rules~\cite{OpticallyPolarizedAtoms,Var.Mos.Khe:88}. Different subspaces with the same value of $\tilde{L}$ may appear, hence the use of the superscript $\alpha$ to distinguish them. The irrep decomposition for each $\vec{L}$ with $K$-body terms is independent and none of them contains the global identity, except for $\vec{L}=\vec{0}$ ($K=0$). For better understanding, we begin by analyzing a few examples of this tensor product decomposition before we continue with the general theory of decoupling groups for multispin systems.

\subsubsection{Arbitrary interactions in a pair of spins}
Consider a system of two interacting spins $j_1$ and $j_2$. The 0-body and 1-body terms written in the notations of Eqs.~\eqref{eq:int_sub_multi}--\eqref{eq:furtherirreps} read
\paragraph{\underline{$0$-body term}}
\begin{equation}
\label{Eq.0.body}
    \begin{aligned}    
        &\mathbf{L}=(0,0) \; \to \; \mathcal{I}^{0-\mathrm{body}}=\tilde{\BHs}^{(0)}_{\mathbf{L}} = \BHs^{(0)}_{j_1} \otimes \BHs^{(0)}_{j_2} 
    \end{aligned}
\end{equation}
\paragraph{\underline{$1$-body term}}
\begin{equation}
\label{Eq.1.body}
    \begin{aligned}
         & \mathbf{L}=(l_1,0) \;\to \;   \tilde{\BHs}^{(l_{1})}_{\mathbf{L}} =  \BHs^{(l_1)}_{j_1} \otimes \BHs^{(0)}_{j_2} ,
        \\[5pt]
         & \mathbf{L}=(0,l_2) \;\to \;
\tilde{\BHs}^{(l_{2})}_{\mathbf{L}} = \BHs^{(0)}_{j_1} \otimes \BHs^{(l_2)}_{j_2} ,\\[5pt]
    & \mathcal{I}^{1-\mathrm{body}}=(\BHs^{(l_1)}_{j_1} \otimes \BHs^{(0)}_{j_2})\oplus (\BHs^{(0)}_{j_1} \otimes \BHs^{(l_2)}_{j_2} )
    \end{aligned}
\end{equation}
with $l_1,l_2 \neq 0$.
On the other hand, the two-body interactions are labeled by $\mathbf{L}=(l_1 , l_2)$. Its respective irrep decomposition reads
\paragraph{\underline{$2$-body term}}
\begin{equation}
\label{Eq.2.body}
    \begin{aligned}    
        \mathbf{L}=(l_1,l_2) \; \to \;\mathcal{I}^{2-\mathrm{body}}=\bigoplus_{\tilde{L}=\left| l_1-l_2 \right|}^{l_1+l_2} \tilde{\BHs}^{(\tilde{L})}_{\mathbf{L}}  
    \end{aligned}
\end{equation}
where the explicit expression of $\tilde{\BHs}^{(\tilde{L})}_{\mathbf{L}}$ in terms of the operators belonging to $\BHs^{(l_1)}_{j_1}$ and $\BHs^{(l_2)}_{j_2}$ are given in Appendix~\ref{Ap.condition}. It is also essential to note that the \rhs~of the last equation may have 0-irreps for $l_1 = l_2$ which do not correspond to the global identity operator. Instead, they correspond to other SU$(2)$ invariant subspaces whose operators cannot be averaged to zero by global $\pointg{SU}(2)$ operations. In particular, they are insensitive to the DD sequences studied in this work.
We list a set of conditions for the interaction Hamiltonian not to have such an isotropic component in Appendix~\ref{Ap.condition}. We call these conditions \emph{anisotropy conditions}. The simplest example of an SU$(2)$ invariant operator other than the identity arises when $l_1=l_2=1$ and is given by the isotropic interaction term\paragraph{Isotropic interaction}
\begin{equation}
\label{Eq.Two.body.isotropic}
    \mathbf{J}^1 \bm{\cdot} \mathbf{J}^2 \;\in\; \mathcal{I}^{2-\mathrm{body}} ,
\end{equation}
where $\mathbf{J}^1$ (resp.\ $\mathbf{J}^2$) is the angular momentum operator acting on the first (resp.\ second) spin.
\subsubsection{Tripartite spin-1/2 system}
\label{subsub.tripartite}
Here, the system has $(j_1,j_2,j_3)= (\tfrac{1}{2},\tfrac{1}{2},\tfrac{1}{2})$ spins, and the irreps in their individual operator space can have $L_k=0,1$ for any $k$. The $K$-body terms for $K<3$ are essentially the same as those listed in Eqs.~\eqref{Eq.0.body}--\eqref{Eq.2.body} with an additional tensor product with a 0-irrep subspace. For example, the 1-body term associated with $\mathbf{L}=(0,0,l_3)$ is equal to $\BHs_{j_1}^{(0)} \otimes \BHs_{j_2}^{(0)} \otimes \BHs_{j_3}^{(l_3)}$. On the other hand, the $K=3$-body term $\mathbf{L}=(1,1,1)$ is obtained by applying the triangular rule for addition of angular momentum (see Fig.~\ref{fig:fig9}) to give
\paragraph{\underline{$3$-body term}}
\begin{equation}
    \begin{aligned}
        &\mathcal{I}^{3-\mathrm{body}}= \BHs_{j_1}^{(1)} \otimes \BHs_{j_2}^{(1)}  \otimes \BHs_{j_3}^{(1)} = \left( \bigoplus_{l'=0}^2 \BHs^{(l')}
       \right) \otimes \BHs_{j_3}^{(1)} 
      \\
      & =
\underbrace{\tilde{\BHs}_{\mathbf{L}}^{(1)}}_{l'=0} \oplus \underbrace{\tilde{\BHs}_{\mathbf{L}}^{(0)}  \oplus \tilde{\BHs}_{\mathbf{L}}^{(1)} \oplus
      \tilde{\BHs}_{\mathbf{L}}^{(2)}}_{l'=1} \oplus 
    \underbrace{\tilde{\BHs}_{\mathbf{L}}^{(1)}  \oplus \tilde{\BHs}_{\mathbf{L}}^{(2)} \oplus \tilde{\BHs}_{\mathbf{L}}^{(3)}}_{l'=2}
    \\
    & =
    \tilde{\BHs}_{\mathbf{L}}^{(0)}\oplus
    \left( \bigoplus_{\alpha=1}^3  \tilde{\BHs}_{\mathbf{L}}^{(1,\alpha)}
    \right) \left( \bigoplus_{\alpha=1}^2
    \tilde{\BHs}_{\mathbf{L}}^{(2,\alpha)}
    \right) \oplus 
    \tilde{\BHs}_{\mathbf{L}}^{(3)}
    \end{aligned}
\end{equation}
where the isotropic interaction operator that spans the irrep $\tilde{\BHs}_{\mathbf{L}}^{(0)}$ is given by (see Appendix~\ref{Ap.condition})
\begin{equation}
\label{Eq.Three.body.isotropic} 
 \mathbf{J}^1 \bm{\cdot} \left( \mathbf{J}^2 \times \mathbf{J}^3 \right) .
\end{equation}
\subsection{Decoupling groups for multispin systems}

We now have the tools needed to study the decoupling groups of an ensemble of $N$ spins $(j_1, \dots, j_N)$ interacting with each other and with a bath. For a generic system Hamiltonian $H$, we can first compute its Schmidt decomposition with respect to the bipartition of the spin ensemble and the bath~\eqref{Schmidtoperator}. The corresponding system operators will be the elements that span the interaction subspace $\mathcal{I}_S$, which can be grouped in $K$-body terms as in Eq.~\eqref{eq:int_sub_multi}. The SU$(2)$-irrep decomposition of each subspace is calculated by applying tensor products and irrep decompositions recursively, as done in the examples of Subsection~\ref{Subsec.interaction}. The interaction subspace associated to the $K$-body terms is contained in a direct sum of irreps
\begin{equation}
\label{Kbody.case}
    \mathcal{I}^{K-\mathrm{body}} \subseteq V^{K-\mathrm{body}}=\bigoplus_{|\vec{L}|=K}\bigoplus_{(\tilde{L},\tilde{\alpha})}\BHs^{(\tilde{L},\tilde{\alpha})} ,
\end{equation}
where now the $L_k$ components of $\mathbf{L}$ run from zero to $L_{\max}^k$ with $L_{\max}^k \leq 2j_k$ the minimum value of $L_{k} $ to guarantee the inclusion $\mathcal{I}^{K-\mathrm{body}} \subseteq V^{K-\mathrm{body}}$.
Crucially, the largest irreps involved in the decomposition has dimension $\tilde{L}_{\max}$ equal to the sum of the $K$ largest $L_{\max}^k$'s. This leads to the observation that the set of point groups of $V$ cannot exceed that of the single spin system of effective spin $j_{\mathrm{eff}} = \tilde{L}_{\max}/2$, or more precisely 
\begin{equation}
    \SetG\qty(V^{K-\mathrm{body}} ) \subseteq\SetG\qty(\HSs\qty(\Hs^{(j_{\mathrm{eff}})})).
\end{equation}
In particular, a symmetry inaccessible for the space of operators of the effective spin $j_{\mathrm{eff}}$ will be inaccessible for $V^{K-\mathrm{body}}$, and thus will be a decoupling group for $\mathcal{I}^{K-\mathrm{body}}$, according to Proposition~\ref{Result.0}. All this remains valid as long as $\mathcal{I}^{K-\mathrm{body}}$ fulfills the anisotropy conditions.
\\
\indent
The general results described above, based on Eq.~\eqref{Kbody.case}, are formulated as follows:
\begin{Proposition}
\label{Result.2}
Consider an ensemble of $N$ spins of quantum numbers $(j_1,\dots ,j_N)$ and a Hamiltonian of $K$-body terms such that its interaction subspace $\mathcal{I}^{K-\mathrm{body}}$ contains no $0$-irreps (isotropic components) except the global identity. Any group that is not a subgroup of $\SetG_{\mathrm{max}}\qty(\HSs\qty(\mathcal{H}^{(j_{\mathrm{eff}}(K))}))$, where $j_{\mathrm{eff}}(K)$ is half the sum of the $K$ largest $L^k_{\mathrm{max}}$'s, is then a decoupling group for $\mathcal{I}^{K-\mathrm{body}}$. 
\end{Proposition}
\begin{Corollary}
\label{Coro.2} 
    Under the same conditions as in Proposition~\ref{Result.2}, any point group that is not a subgroup of an element of $\SetG_{\mathrm{max}}\qty(\HSs\qty( \Hs^{j_{\mathrm{eff}}}))$, where $j_{\mathrm{eff}}=\max_{k=1}^Kj_{\mathrm{eff}}(k)$, is a decoupling group for $\mathcal{I}_S$.
\end{Corollary}

The smallest decoupling groups for $K$-body terms where the interaction subspace of each constituent has the same $L^k_{\mathrm{max}}= L_{\mathrm{max}}$ are listed in Table \ref{Table.6a} for different values of $K$ and $L_{\mathrm{max}}$. The smallest decoupling groups for two-body interactions are listed in Table \ref{Table.6b} for different pairs $\qty(L^1_{\mathrm{max}}, L^2_{\mathrm{max}})$.
A case often encountered in applications is that of systems with multilinear interactions (see Subsection~\ref{Sub.Appl.nultispin}). $K$-body multilinear interactions are defined as interactions between $K$ spins which can be expressed as a sum of $K$ tensor products between the angular momentum operators of each spin, \ie, $L_{\max}^k=1$ for any spin $j_k$ in Eq.~\eqref{Kbody.case}. For instance, the most general 2-body multilinear interaction between two spins $j_1$ and $j_2$ is $H = \sum_{\alpha,\beta} h_{\alpha \mu}^{ij} J^1_{\alpha}\otimes J^2_{\alpha}$.  In this case, the anisotropy conditions for the $K$-body terms of multilinear interactions can be fully characterized with the help of the isotropic tensors~\cite{Kearsley_1975}. We do this in Appendix~\ref{Ap.condition} for $K\leq 4$, but the results can be easily extended to $K\leq 8$ by using the results of Ref.~\cite{Kearsley_1975}.
The corollary~\ref{Coro.2} is simplified for Hamiltonians with only multilinear terms:
\begin{Corollary}
Consider an ensemble of $N$ spins with quantum numbers $(j_1,\dots ,j_N)$ and a Hamiltonian with at most $K$-body multilinear terms, such that its interaction subspace contains as only $0$-irrep the global identity. Then, any group that is not a subgroup of an element of $\SetG_{\mathrm{max}}\qty(\HSs\qty(\mathcal{H}^{(j_{\mathrm{eff}})}))$, where $j_{\mathrm{eff}} = K/2$, is a decoupling group for $\mathcal{I}_S$. 
\end{Corollary}

\subsection{Applications}
\label{Sub.Appl.nultispin}
\begin{figure}[t!]
    \centering
    \includegraphics[width=0.4\textwidth]{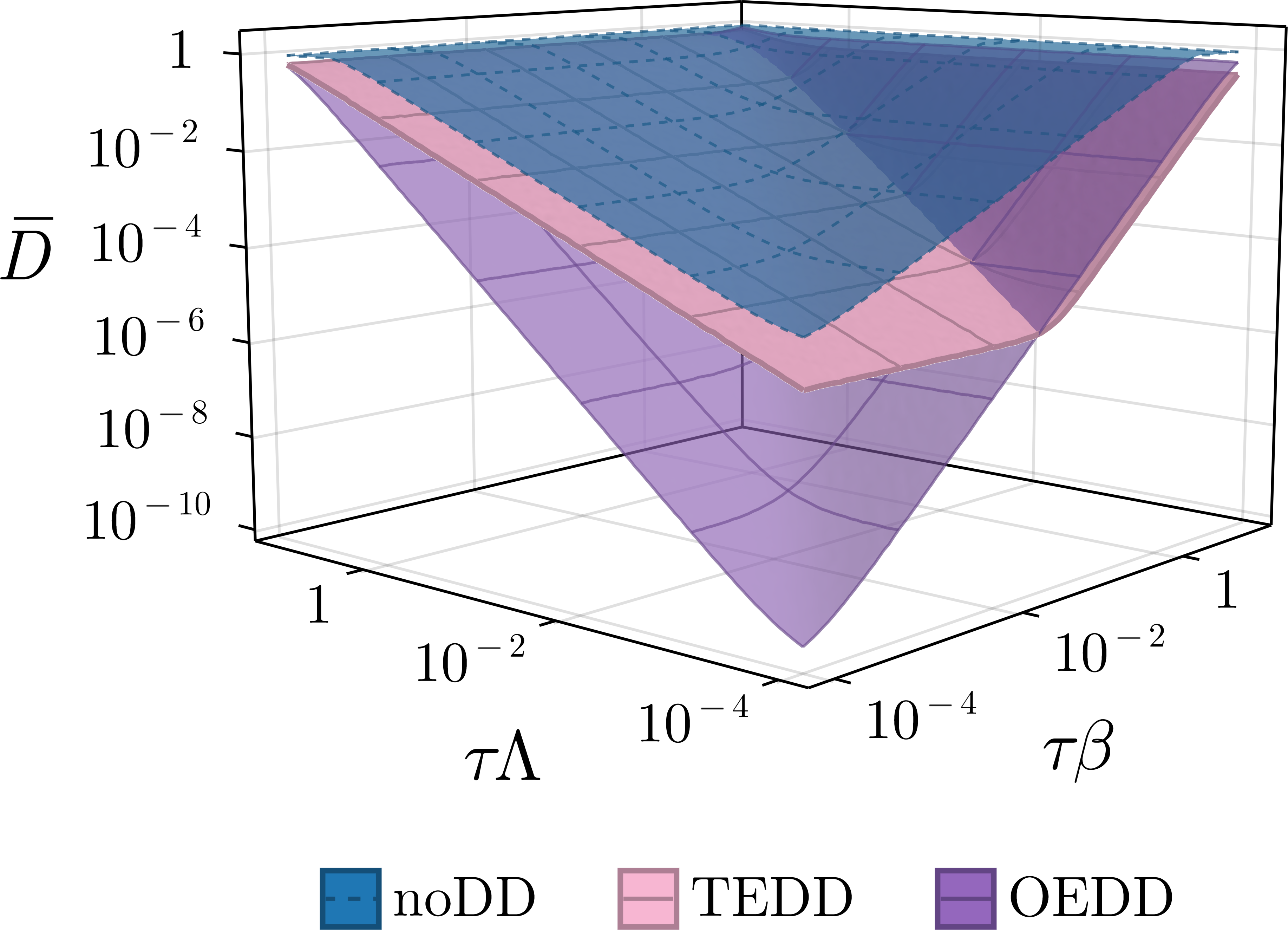}
    \caption{Average distance $\overline{D}$ between the identity propagator and the noisy propagator in the $\qty(\tau\beta,\tau\Lambda)$ parameter space for an ensemble of four interacting spin-$1/2$ with two and three-body interactions, for the free evolution (blue) and the TEDD (pink) and OEDD (purple) sequences. Averages were performed on 1000 randomly generated Hamiltonian satisfying the anisotropy conditions ~\eqref{eq:any.condition.two} and \eqref{eq:any.condition.three}.}
    \label{Figure.5}
\end{figure}

\begin{table}[t]
    \centering
    \begin{subtable}{0.5\textwidth}\centering
    \caption{\phantom{$K=2$}\hspace*{5cm}}
            \begin{tabular}[t]{c||cccccc}
         \multirow{2}{*}{$L_{\mathrm{max}}$} & \multicolumn{5}{c}{$\qquad \; K$}
         \\[2pt]
          & 1 &  2 & 3 & 4 & 5 &$\geq 6$
         \\ \cline{2-7}
        \\[-10pt] \cline{2-7} 
        \rule{0pt}{1\normalbaselineskip} 1 \rule{0pt}{1\normalbaselineskip} & $\pointg{D}_2$ & $\pointg{T}$ & $\pointg{O}$ & $\pointg{I}$ & $\pointg{I}$ & $-$ \\
        2  & $\pointg{T}$& $\pointg{I}$ & $-$ & $-$ & $-$ &$-$\\
        $ 3$ & $\pointg{O}$ & $-$ & $-$ & $-$ & $-$&$-$
    \end{tabular}
    \label{Table.6a}
    \end{subtable}
    \newline
    \vspace*{10pt}
    \newline
    \begin{subtable}{0.5\textwidth}\centering
    \caption{$K=2$\hspace*{5cm}}
    \begin{tabular}[t]{c||ccccc}
        \multirow{2}{*}{$L_{\mathrm{max}}^1$} & \multicolumn{5}{c}{$L_{\mathrm{max}}^2$} 
        \\[3pt]
        & $1$ & $2$ & $3$ & $4$ &$\geq 5$
        \\ \cline{2-6}
        \\[-10pt] \cline{2-6} 
        \rule{0pt}{1\normalbaselineskip}$1$\rule{0pt}{1\normalbaselineskip} & T & O & I & I &$-$\\
        $2$  &  & I & I & $-$ &$-$\\
        $3$  & & & $-$ & $-$&$-$\\
        $4$  & & & & $-$&$-$
    \end{tabular}
    \label{Table.6b}
    \end{subtable}
    \caption{(a) Decoupling groups for an ensemble of spins of quantum numbers $(j_1,\dots, j_N)$ with $K$-body interactions satisfying $\mathcal{I}^{K\text{---body}} \subseteq \bigoplus_{|\vec{L}|=K}\bigoplus_{(\tilde{L},\tilde{\alpha})}\BHs^{(\tilde{L},\tilde{\alpha})}$ where each $L_k$ runs from zero to $L_{\max}$. The first column coincides with the decoupling groups of a single spin presented in Table~\ref{Table.4}. (b) Decoupling groups for two-body interactions and for different values of $L_{\mathrm{max}}$,  $L^1_{\mathrm{max}}\leq L^2_{\mathrm{max}}$. In both cases, we assume that the anisotropy conditions are fulfilled, so that the Hamiltonian does not contain isotropic components. }
    \label{Table.6}
\end{table}
\begin{table}
\centering

    \begin{tabular}[t]{c||c c}
         \multirow{2}{*}{$K$} & System  & Decoupling 
         \\
         & interactions & sequence
        \\ \cline{1-3}
        \\[-10pt] \cline{1-3} 
        \rule{0pt}{1\normalbaselineskip} 1 \rule{0pt}{1\normalbaselineskip}  
         & On-site disorder~\cite{10.1093} 
         & EDD
        \\[7pt]
         \multirow{2}{*}{2}  & Antisymmetric exchange, & \multirow{2}{*}{TEDD}
         \\ & 
         Dipole-Dipole~\cite{levitt2008spin,10.1093} 
        \\[7pt]
         \multirow{2}{*}{3} & Nearest-neighbour 3-body
         & \multirow{2}{*}{OEDD}\\ & interaction~\cite{buchler2007three}
    \end{tabular}    
\caption{Examples of typical $K$-body interactions in spin systems and the respective platonic DD sequences to decouple them.} 
    \label{Table.6c}
\end{table}
We present in Fig.~\ref{Table.6c} a list of examples of $K$-body interactions that may appear in multispin systems and can be decoupled by Platonic sequences.
Proposition~\ref{Result.2} allows us to obtain decoupling groups for certain multispin systems with $K$-body terms, which we present in Table \ref{Table.6a}. For example, we find that the group $\pointg{T}$ (resp.\ $\pointg{O}$) is a decoupling group for any spin ensemble with only two-body (resp.\ two- and three-body) multilinear interaction terms, provided that the Hamiltonian satisfies the anisotropy conditions. For two-body and three-body interactions, this means ensuring that the Hamiltonian has no isotropic components given by  \eqref{Eq.Two.body.isotropic} and \eqref{Eq.Three.body.isotropic}, respectively. Thus, the anisotropy conditions for the most general Hamiltonian with multilinear two- and three-body interactions, written as
\begin{equation}
\label{Eq.Ham.23}
\begin{aligned}
    H ={}&  \overbrace{\sum_{i<j} \sum_{\alpha,\mu} h^{ij}_{\alpha\mu} J_{\alpha}^i\otimes J_{\mu}^j}^{\equiv~H^{2-\mathrm{body}}} \\ &+  \underbrace{\sum_{i<j<k}\sum_{\alpha,\mu,\lambda }h^{ijk}_{\alpha\mu\lambda} J_{\alpha}^i\otimes J_{\mu}^j \otimes J_{\lambda}^k}_{\equiv~H^{3-\mathrm{body}}},
    \end{aligned}
\end{equation}
are
\paragraph{Two-body anisotropy conditions:}
\begin{equation}
\label{eq:any.condition.two}
\sum_{\alpha,\beta}h^{ij}_{\alpha\beta}\,\delta_{\alpha\beta} = 0 , \quad \forall\:i\neq j 
\end{equation}
\paragraph{Three-body anisotropy conditions:}
\begin{equation}\label{eq:any.condition.three}
\sum_{\alpha,\beta,\lambda} h^{ijk}_{\alpha\beta\lambda}\,\epsilon_{\alpha\beta \lambda} = 0,  \quad \forall\:i\neq j \neq k \neq i .
\end{equation} 
To illustrate the decoupling properties of Platonic sequences on multispin systems, we again compute the average distance $\overline{D}$, this time for an ensemble of four spin-$1/2$ with interaction Hamiltonian of the form~\eqref{Eq.Ham.23}, for the free evolution and with the application of the OEDD and TEDD sequences.
Figure~\ref{Figure.5} shows the distance in the $(\tau \beta, \tau\Lambda)$ parameter space, where $\beta$ and $\Lambda$ are defined as 
\begin{equation}
\label{Eq.beta.lambda}
        \beta = \norm{H^{2-\mathrm{body}}},\quad 
        \Lambda = \norm{H^{3-\mathrm{body}}}.
\end{equation}
The tensors $h^{ij}$ and $h^{ijk}$ appearing in the Hamiltonian~\eqref{Eq.Ham.23} are randomly generated, while ensuring that they satisfy the anisotropy conditions \eqref{eq:any.condition.two} and \eqref{eq:any.condition.three}. The decoupling properties of the $\pointg{T}$ and $\pointg{O}$ groups for multiqubit interactions can be read in Fig.~\ref{Figure.5} by the slope of $\overline{D}$ as a function of $\tau\beta$ for $\tau\Lambda\ll 1$, and as a function of $\tau\Lambda$ for $\tau\beta\ll 1$. The TEDD and OEDD sequences both provide first-order decoupling for dominating two-body multilinear interactions ($\Lambda\ll \beta$), while the OEDD sequence is the only one to also provide first-order decoupling for dominating three-body multilinear interactions ($\Lambda\gg \beta$). Numerical calculations (data not shown) confirm that first-order decoupling is not achieved when the conditions \eqref{eq:any.condition.three} are not satisfied. Table~\ref{Table.6a} also shows that no decoupling group (consisting only of global rotations) exists for interactions involving simultaneously 6 or more bodies in an ensemble of interacting qubits. 
\par
\begin{figure}[t!]
    \centering
    \includegraphics[width=0.4\textwidth]{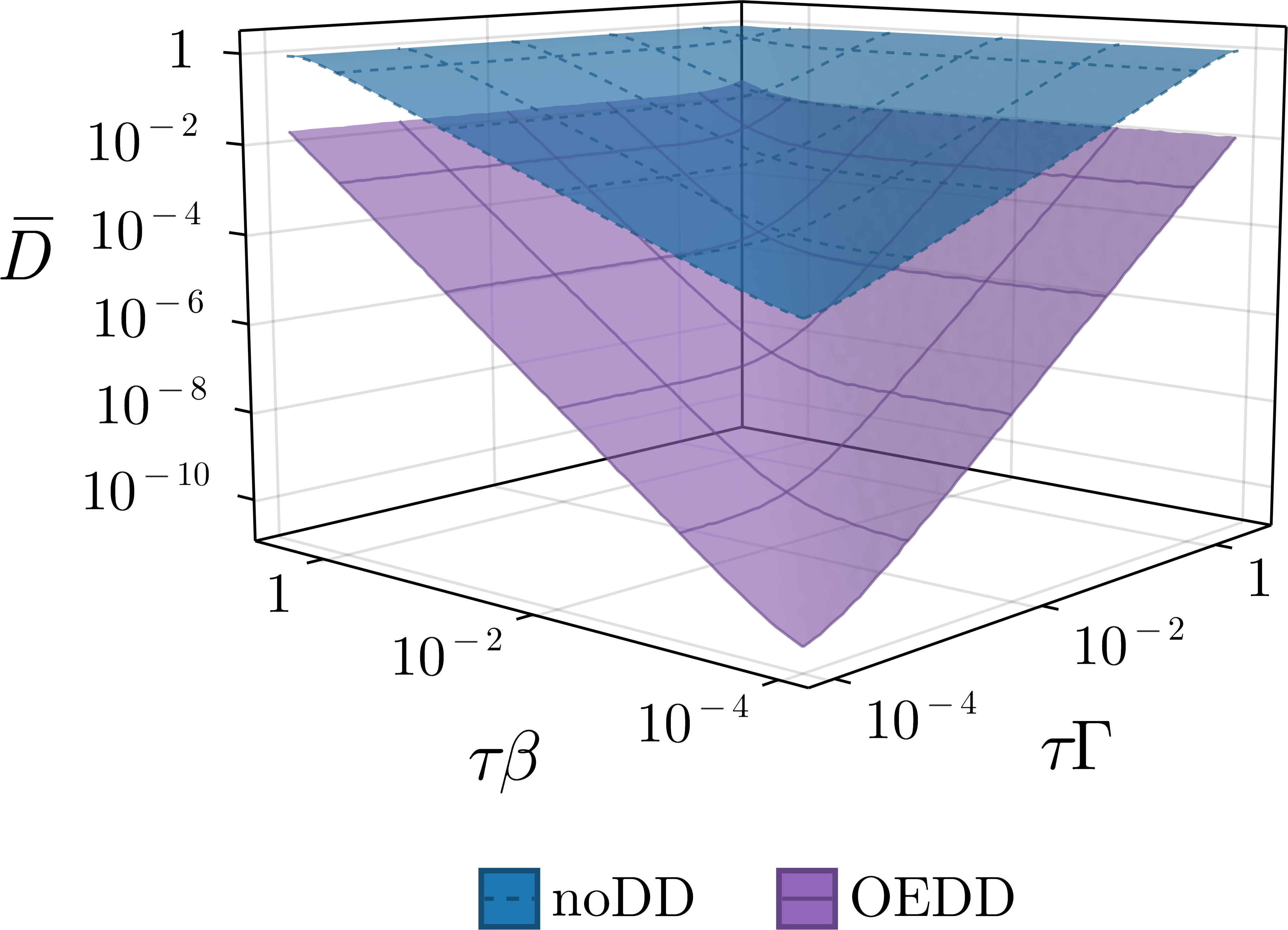}
    \caption{Same as in Fig.~\ref{Figure.5} in the ($\tau\Gamma$, $\tau\beta$) parameter space for the free evolution (blue) and the OEDD sequence (purple) for a qubit-qutrit composite spin system.}
    \label{Figure.6}
\end{figure}

 On the other hand, Table~\ref{Table.6b} shows the decoupling groups for two-body interactions of a composite system. Without loss of generality, we can  assume a bipartite system. As an example, we consider a qubit-qutrit system, which is equivalent to a two spin system with $j=1/2$ and $1$. Its most general Hamiltonian can be expanded as
\begin{equation}\label{Hqubitqutrit}
    H = \sum_{L_1=0}^1\sum_{L_2=0}^2 \sum_{M_1=-L_1}^{L_1}\sum_{M_2=-L_2}^{L_2}w_{M_1M_2}^{L_1L_2}T_{L_1M_1}\otimes T_{L_2M_2}
\end{equation}
where the tensors $w^{L_1L_2}$ should satisfy the hermiticity conditions
\begin{equation}
    w^{L_1L_2}_{-M_1-M_2} = (-1)^{M_1+M_2} \qty(w^{L_1L_2}_{M_1M_2})^* .
\end{equation}
The SU$(2)$ isotropic component of \eqref{Hqubitqutrit} is necessarily of the form (see Appendix~\ref{Ap.condition}) 
\begin{equation}
\begin{aligned}    
    H_\mathrm{isotropic}^{2-\mathrm{body}} & \propto \sum_{M=-1}^1 (-1)^M T_{1M}\otimes T_{1-M} 
    \\
    & \propto \mathbf{J}^{\mathrm{qubit}}\boldsymbol{\cdot} \mathbf{J}^{\mathrm{qutrit}}
\end{aligned}
\end{equation}
The anisotropy condition on the Hamiltonian will therefore be equivalent to the condition~\eqref{eq:any.condition.two}. Defining $\Gamma$ (resp.\ $\beta$) as the supremum operator norm of the one-body (resp.\ two-body) Hamiltonian, we compare in Fig.~\ref{Figure.6} the OEDD sequence with the free evolution in the $(\tau\Gamma,\tau\beta)$ parameter space, where $\tau$ is the duration of the free evolution and the pulse sequence. Once again, we observe the first-order decoupling obtained with the OEDD sequence through the slope of $\overline{D}$ as a function of $\tau\Gamma$ and $\tau\beta$.
\subsection{Comparison with existing schemes }\label{Sec.Comp.Multi}
\subsubsection{Disorder and dipole-dipole interactions}
We have shown that the TEDD sequence has decoupling properties for multispin systems with $K$-body multilinear terms for $K=1,2$. This case encompasses the well-known scenario of multispin systems with on-site disorder and dipole-dipole interactions~\cite{Lukin_2020, Cappellaro_2022, cory_1969}, with a Hamiltonian given in its most general form by~\cite{levitt2008spin}  
\begin{equation}
\label{Eq.Ham.gen}
    H = H_{\mathrm{dis}} + H_{\mathrm{dd}}
    \end{equation}
with
\begin{flalign}
   & H_{\mathrm{dis}} =   
   \sum_i \delta_{i}\, \mathbf{m}_i \bm{\cdot} \mathbf{J}^{i} ,
   \label{Eq.disorder.gen}
   \\
     & H_{\mathrm{dd}} =
      \sum_{i<j} \Delta_{ij} \left( 3 \left[ \mathbf{e}_{ij} \bm{\cdot} \mathbf{J}^i \right] \left[ \mathbf{e}_{ij} \bm{\cdot} \mathbf{J}^j \right] - \mathbf{J}^i \bm{\cdot} \mathbf{J}^j \right) ,
   \label{Eq.dipole.gen}
\end{flalign}
where $\mathbf{m}_i$ is the unit vector giving the disorder direction for the $i$th spin and $\mathbf{e}_{ij}$ is the unit vector parallel to the line passing through the spins $i$ and $j$. The Hamiltonian~\eqref{Eq.Ham.gen} satisfies the anisotropy conditions\footnote{The only multilinear isotropic two-body terms are of the form $I_2 \propto \mathbf{J}^i \bm{\cdot} \mathbf{J}^j$ (see appendix~\ref{Ap.condition}). By direct calculation, we obtain that $\Tr (H_{\mathrm{dd}} \mathbf{J}^i \bm{\cdot} \mathbf{J}^j)=0$ for any pair of spins $i$ and $j$.} and includes the case where the rotating wave approximation does not hold, which occurs in certain physical scenarios such as zero-field NMR~\cite{theis2011parahydrogen}. The TEDD sequence guarantees first order decoupling with the bath for any ensemble of $N$ spins with Hamiltonian~\eqref{Eq.Ham.gen}. Other interactions such as spin-exchange interactions~\cite{Lukin_2020} or $J$-couplings~\cite{levitt2008spin} are decoupled by the TEDD sequence, with the exception of their isotropic parts.
\\
\indent Several sequences have been constructed for the above Hamiltonian under the rotating wave approximation~\cite{levitt2008spin}, \ie, by considering all the unit vectors pointing in the same direction, let us say $\mathbf{m}_i = \mathbf{e}_{ij} = \mathbf{z}$. In this case, the Hamiltonian \eqref{Eq.Ham.gen} becomes
\begin{equation}
    H = \sum_{\phantom{j} i\phantom{j}}\delta_i J_z^i + \sum_{i<j}\Delta_{ij}\qty(3J_z^iJ_z^j - \vec{J}^i\boldsymbol{\cdot} \vec{J}^j)
    \label{eq:Dis.Dip.} .
\end{equation}
The sequences built to cancel this Hamiltonian also typically consist of global rotations.
The smallest sequence that cancels the disorder term is the two-pulse CPMG sequence~\cite{Viola_1998}. On the other hand, the smallest sequence that mitigates the dipole-dipole interaction term is the 6-pulse WAHUHA sequence that was constructed in the context of NMR spectroscopy~\cite{Waugh_1968}. Other slightly longer sequences, such as REV-8~\cite{10.1063/1.1679423} and Echo+WAHUHA~\cite{Lukin_2020}, with 8 and 6 pulses respectively, are capable of decoupling the two terms. More recently, longer sequences have been designed to improve robustness against various imperfections, such as finite-duration errors or certain control errors such as flip-angle (systematic over- or under-rotation)  or axis-misspecification errors (systematic error in the axis of rotation)~\cite{Lidar_2023_survey}. For example, the 48-pulses Cory-48 sequence\footnote{The Cory-48 sequence has actually 72 pulse intervals as it includes 24 identity pulses.}~\cite{cory_1969,Lukin_2020} offers great robustness to dipolar interactions during pulses of finite duration. Another example is the 60-pulse DROID-60 sequence~\cite{Lukin_2020}, which is highly robust to disorder during the finite-duration pulses, while offering some robustness to flip-angle errors and dipolar interaction during the finite-duration pulses. Other sequences (\textit{e.g.} 24-pulses yxx24 and 48-pulses yxx48) presented in Ref.~\cite{Cappellaro_2022} were designed using a deep-learning algorithm and shown to provide some robustness to dipolar interactions and disorder during pulses of finite duration, as well as to flip angle errors. In addition, the DROID60 and yxx24 sequences cancel some higher-order terms in the Magnus expansion and therefore outperform the TEDD sequence for the specific Hamiltonian~\eqref{eq:Dis.Dip.} and when only flip-angle errors and finite-duration pulses are considered. We present a comparison of the TEDD, DROID-60 and yxx48 sequences in Fig.~\ref{fig:comparison} (upper left plot) of Appendix~\ref{Ap.Comp.} in several regimes of the Hamiltonian parameters~\eqref{eq:Dis.Dip.}.
\par 
The TEDD sequence, by its Eulerian design, is robust to all finite-duration errors and to other systematic control errors, including flip-angle and axis-misspecification, but also more exotic systematic errors (see Sec.~\ref{Sec.robustness}). This implies that, with the addition of non-negligible axis-misspecification errors or disorder around the $x$ and $y$ axis, the $\mathrm{TEDD}$ sequence achieves or exceeds the performance of its competitors in a certain regime of disorder and dipole-dipole parameters. Consequently, the TEDD sequence can lead to potential applications in systems where the high-order errors arising from the Magnus expansion of the model Hamiltonian~\eqref{eq:Dis.Dip.} are less important than the uncorrected, first-order  errors. We compare the TEDD and other sequences under this scenario in Appendix~\ref{Ap.Comp.}.

\par 
A straightforward method to increase the decoupling order and robustness of a sequence while keeping the unparalleled versatility is through concatenation~\cite{Khodjasteh_2005, concatenatedPlatonic} and reflection symmetry~\cite{Lukin_2020}. In particular, we can construct a new 48-pulse time-antisymmetric sequence by applying a $\mathrm{TEDD}$ sequence followed by its inverse\footnote{Meaning that the sequence is run in the opposite direction, where each rotation is replaced by its inverse rotation. This ensures that the Hamiltonian in the toggling frame is time-symmetric, in which case all even-order terms of the Magnus expansion are equal to zero.}, which we denote $\qty(\mathrm{TEDD})(\mathrm{TEDD})^{\dagger}$, or simply $\mathrm{TT}^{\dagger}$. This new sequence achieves second-order decoupling over its entire correctable subspace, increasing its robustness properties as well. In the finite pulse regime, $\mathrm{TT}^{\dagger}$ outperforms both DROID-60 and yxx24 for the Hamiltonian~\eqref{eq:Dis.Dip.} if the amplitude of the unwanted Hamiltonian is sufficiently small (see Appendix~\ref{Ap.Comp.}). The addition of small perturbations such as axis-misspecification errors further extends the parameter regime for which $\mathrm{TT}^{\dagger}$ performs better than state-of-the-art sequences.

\subsubsection{Cross-Kerr Hamiltonian}
\par There are, however, multiqudit systems for which the Platonic sequences cannot achieve decoupling from the bath, partly due to the anisotropy conditions. For example, in transmon qudits, a common interaction between a pair of qudits is described by the Cross-Kerr Hamiltonian~\cite{tripathi_2024,Goss_2022}, 
\begin{equation}
    H_{\mathrm{CK}} =  \sum_{i, j = 1}^{d-1} \alpha_{ij}\ket{ij}\bra{ij}
\end{equation}
where $\alpha_{ij} = \omega_{ij} + \omega_{00} - \omega_{i0}-\omega_{0j}$ and $\omega_{ij}$ is the transition frequency between the $i$th and $j$th energy levels. This interaction generalizes the Ising interaction between coupled qubits $H_{\mathrm{Ising}} = \sum_{i,j}\Delta_{ij} \sigma_z^{i}\sigma_z^{j}$ which does not satisfy the anisotropy conditions. For the decoupling of two-qudit interactions, we refer to Table~\ref{Table.6b} where we find that this is only possible (using Platonic sequences) for qubits and qutrits using the Tetrahedron and Icosahedral sequence, respectively. Furthermore, applying the Platonic sequence to this Hamiltonian leads to a residual term that survives the symmetrization, given by
\begin{equation}
\begin{aligned}
H_{\mathrm{CK}}^{\mathrm{sym}} \propto {}&{}
\sum_{L=0}^{d-1} \Tr\qty[I_LH_{\mathrm{CK}}]I_L
\\
\propto {}&{} \frac{1}{3}\qty(\alpha_{11} + \alpha_{12} + \alpha_{21} + \alpha_{22})I_0 + \frac{\alpha_{22}}{2\sqrt{3}}I_1 \\
    {}&{} + \frac{1}{6\sqrt{5}}\qty(4\alpha_{11} -2(\alpha_{12} + \alpha_{21}) + \alpha_{22}) I_2
    \end{aligned}
\end{equation}
where the isotropic operators $I_L$ are defined in Appendix~\ref{Ap.condition} (Eq.~\eqref{Eq.Isotropic.twospins}). In particular, $I_0$ is proportional to global identity, and $I_1 \propto \vec{J}^1 \boldsymbol{\cdot} \vec{J}^2$. Consequently, Platonic sequences, as well as any sequence composed solely of global rotation pulses, cannot decouple $H_{\mathrm{CK}}^{\mathrm{sym}}$. Other strategies must then be used to mitigate Cross-Kerr interactions, such as the DD sequences presented in Ref.~\cite{tripathi_2024}, which requires individual addressing of each qudit. Similarly, the Ising interaction between a pair of qubits is not totally suppressed by the $\mathrm{TEDD}$ sequence, and the remaining term will be proportional to $\boldsymbol{\sigma}^1 \boldsymbol{\cdot}  \boldsymbol{\sigma}^2$.

\section{Robustness to control errors and finite pulse duration errors}\label{Sec.robustness}
So far, we have considered infinitely short and strong DD pulses, without imperfections due to control errors. However, in a realistic setting, the pulses take some time to implement (a time during which decoherence occurs) and may not be perfect; consequently, the resulting decoupling may also not be perfect. We therefore carry out a robustness analysis of the Platonic sequences in this section.

\par Consider a dynamical decoupling sequence designed from a certain group $\group$ (of correctable subspace $\mathcal{C}_{\group}$) with certain generators $\qty{P_{\lambda}}$, where each pulse $P\in\qty{P_{\lambda}}$ has a certain finite duration $\tau_P\in\qty{\tau_{P_{\lambda}}}$. Assume, without loss of generality, that there is no free evolution between the pulses \ie, once a pulse is completed, the next one starts\footnote{Since no assumptions are made about the shape of the pulses, we can always add a free evolution time by simply turning off the control Hamiltonian for a certain period of time.}. During the implementation of each pulse, three Hamiltonians contribute to the dynamics, namely (i) the ideal pulse Hamiltonian $H_P(t)\in\HSs(\mathcal{H}_S)$, (ii) the pulse error Hamiltonian $H_P^{\text{err}}(t)\in \HSs(\mathcal{H}_S)$, and (iii) the decoherence Hamiltonian $H_{SB}=\sum_{\alpha}S_{\alpha}\otimes B_{\alpha}\in \HSs(\mathcal{H}_{SB})$, from which the interaction subspace $\mathcal{I}_S$ is usually defined. The pulse error and decoherence Hamiltonians will cause the pulse to slightly deviate from the intended unitary; this deviation can be quantified by an EPO $\Phi_P$ by moving to the toggling frame with respect to $H_P(t)$. The EPO is defined by the following equation
\begin{equation}
    \begin{aligned}
        e^{-i\Phi_P} \equiv \mathcal{T} \exp\biggl\{&-i\int_0^{\tau_P}\!\biggl[\sum_{\alpha}P^{\dagger}(t)S_{\alpha}P(t)\otimes B_{\alpha}  \\&+ P^{\dagger}(t)H_P^{\text{err}}(t)P(t)\otimes \mathds{1}_B  \biggr]dt\biggr\},
    \end{aligned}\label{eq:pulseerror}
\end{equation}
where $P(t)$ is the propagator associated with $H_P(t)$, with $P(\tau_P)=P$, and $\mathcal{T}$ is the time-ordering operator. The exact form of the EPO can be formally calculated by performing a Magnus expansion; when both the decoherence and the pulse errors are small enough, it is well approximated by its first-order contribution, \ie, 
\begin{multline}
    \Phi_P \approx \Phi_P^{[1]} =  \sum_{\alpha}\mathrm{F}_{(P,\tau_P)}\qty[S_{\alpha}]\otimes B_{\alpha}\\ +\mathrm{F}_{(P,\tau_P)}\qty[H_P^{\text{err}}]\otimes \mathds{1}_B ,
    \label{eq:pulseerror1}
\end{multline}
where we define the function $\mathrm{F}_{(P,\tau_P)}[\cdot]: \HSs(\mathcal{H}_S)\to \HSs(\mathcal{H}_S)$ as 
\begin{equation}\label{eq:operationrobust}
    \mathrm{F}_{(P,\tau_P)}[S] = \int_0^{\tau_P}P^{\dagger}(t)S(t)P(t)\,dt. 
\end{equation}
A key result of Ref.~\cite{Viola_2003} is that if we choose an Eulerian sequence and if the errors $H_P^{\text{err}}(t)$ are systematic, \ie,~the same error Hamiltonian $H_P^{\text{err}}(t)$ occurs each time $H_P(t)$ is turned on, the first-order EPO of the total sequence is given by 
\begin{multline}
        \Phi_{\text{\tiny EDD}}^{[1]} = \sum_{\lambda}\biggl\{\sum_{\alpha}\Pi_{\group}\qty[\mathrm{F}_{(P_{\lambda},\tau_{P_{\lambda}})} \qty[S_{\alpha}]]\otimes B_{\alpha}\\
        + \Pi_{\group}\qty[\mathrm{F}_{(P_{\lambda},\tau_{P_{\lambda}})} \qty[H^{\text{err}}_{P_{\lambda}}]]\otimes \mathds{1}_B\biggr\},
\label{eq:EDDEP}
\end{multline}
and $\Phi_{\text{\tiny EDD}}\approx\Phi_{\text{\tiny EDD}}^{[1]}$ if decoherence is small enough. We can now define the subspace $\mathcal{I}_{P_{\lambda}}$ as the subspace spanned by the pulse error Hamiltonian at different times, 
\begin{equation}
    \mathcal{I}_{P_{\lambda}} \equiv \mathrm{span}\qty(H_{P_{\lambda}}^{\text{err}}(t)\, \forall t)
\end{equation}
and define for the pulse $\lambda$ an extended interaction subspace $\mathcal{I}_S\oplus\mathcal{I}_{P_{\lambda}}$ which includes both decoherence and pulse errors. By linearity of \eqref{eq:operationrobust}, $\mathrm{F}_{(P_{\lambda},\tau_{\lambda})}\qty[\mathcal{I}_S\oplus\mathcal{I}_{P_{\lambda}}]$ is also a subspace. Consequently, first-order decoupling is still achieved if, for all generators $P_{\lambda}$, we have 
\begin{equation}
\mathrm{F}_{(P_{\lambda},\tau_{\lambda})}\qty[\mathcal{I}_S\oplus\mathcal{I}_{P_{\lambda}}] \subseteq \mathcal{C}_{\group}.
    \label{eq:robustnessanalysis}
\end{equation}
\par Let us now consider the specific case of the Platonic sequences which are, by construction, Eulerian, and where the generators are simply global rotations. A decoupling Platonic sequence must be chosen according to Proposition~\ref{Result.2} such that
\begin{equation}
    \mathcal{I}_S \subseteq V = \bigoplus_{(L,\alpha)} \BHs^{(L,\alpha)} \subseteq \mathcal{C}_{\group}.
\end{equation}
First of all, as global rotations do not couple different irreps, it follows that 
\begin{equation}
    \mathcal{I}_S\oplus \mathcal{I}_{P_{\lambda}} \subseteq \mathcal{C}_{\group} \Rightarrow   \mathrm{F}_{(P_{\lambda},\tau_{\lambda})}\qty(\mathcal{I}_S\oplus\mathcal{I}_{P_{\lambda}}) \subseteq \mathcal{C}_{\group} \quad \forall \lambda.
\end{equation}
Therefore, Platonic sequences are intrinsically robust to finite pulse duration errors. 
\par It is also clear that Platonic sequences are robust to systematic errors satisfying $\mathcal{I}_{P_{\lambda}} \subseteq V$ $\forall \lambda$. In particular, they are all robust to \textit{flip-angle errors}, defined as over- or under-rotation, and \textit{axis specification errors}, where a target rotation about an axis $\mathbf{n}$ is misimplemented about an axis $\mathbf{n}' = \sqrt{1-\epsilon}\,\mathbf{n} + \sqrt{\epsilon} \,\mathbf{n}_{\perp}$~\cite{Lidar_2023_survey}, because both types of errors are linear with respect to the spin operators and thus belong to the $L=1$-irrep, which is in the correctable subspace of the three Platonic groups. This observation makes the sequences robust to disorder induced by the pulses, where the amplitude of the control field implementing the pulses is not perfectly homogeneous across the spin ensemble, leading to all spins being rotated at slightly different rates. Each Platonic sequence is also robust to systematic pulse errors which are quadratic with respect to the spin operators, as the $L=2$-irrep also belongs to the correctable subspace of each group (except the $\pointg{D}_2$ group). Moreover, if the Platonic sequence is the universal decoupling strategy for a quantum system, then robustness to arbitrary and systematic pulse errors is guaranteed. 
\section{Dynamically corrected gates}\label{Sec.DCG}
While dynamical decoupling is commonly used to extend the lifetime of an idle qudit, a prescription for designing pulse sequences which mitigate decoherence while performing a non-trivial operation was introduced in Ref.~\cite{Khodjasteh_2009}. In this section, we discuss the potential application of the framework presented in Secs.~\ref{Sec. single spin} and \ref{Sec. multi spin} for the design of such pulse sequences. 

\subsection{General theory}
Consider a spin ensemble that interacts with an environment through an interaction Hamiltonian $H_{SB} = \sum_{\alpha}S_{\alpha}\otimes B_{\alpha}$ such that the interaction subspace satisfies 
\begin{equation}
    \mathcal{I}_S \subseteq V = \bigoplus_{(L,\alpha)} \BHs^{(L,\alpha)},
\end{equation}
and suppose we want to implement a control protocol described by the propagator 

\begin{equation}
    U_S(t) = \mathcal{T} \exp{-i\int_0^t H_u(t')dt'}, \quad U_S(\tau_u) \equiv U_S,
\end{equation}
which implements the unitary $U_S$ in a finite time $\tau_u$ on the system of interest. During the implementation of the protocol, decoherence occurs and the actual propagator deviates slightly from the ideal unitary due to finite duration errors. This deviation can again be quantified through an EPO $\Phi_u$, as defined in Eq.~\eqref{eq:pulseerror}, which is well approximated (if decoherence is small enough) by~\cite{Khodjasteh_2009} 
\begin{equation}
    \Phi_u \approx \Phi_u^{[1]} \equiv \sum_{\alpha} \mathrm{F}_{(U,\tau_u)}\qty[S_{\alpha}] \otimes B_{\alpha}
\end{equation}
with the operation $\mathrm{F}_{(U,\tau_u)}\qty[\cdot] $ as defined in Eq.~\eqref{eq:operationrobust}.
\par To reduce this deviation, we can use the formalism presented in Refs.~\cite{Khodjasteh_2009,Khodjasteh_2009_PRA,Khodjasteh_2010} and construct a so-called \textit{dynamically corrected gate} (DCG). To this end, we first need to construct a \textit{balanced pair} ($\mathds{1}_{U}(t)$, $U^*(t)$), defined as a pair of pulses acting on the system and satisfying the three conditions below,
\begin{equation}
    \begin{aligned}
& \mathrm{(i)}~\sum_{\alpha}\mathrm{F}_{(\mathds{1}_{U},\tau_{\mathds{1}})}\qty[S_{\alpha}]\otimes B_{\alpha} = \sum_{\alpha} \mathrm{F}_{(U^*,\tau_{u^*})}\qty[S_{\alpha}]\otimes B_{\alpha},\\
& \mathrm{(ii)}~\mathds{1}_U(\tau_{\mathds{1}}) = \mathds{1}_S,\\[4pt]
& \mathrm{(iii)}~U^*(\tau_{u^*}) = U_S.
    \end{aligned}
\end{equation}
In other words, we need to find two pulses with identical EPOs, such that one implements the identity while the other implements the target unitary $U_S$. Once such a pair is constructed\footnote{A simple prescription to construct a balanced pair is presented in Refs.~\cite{Khodjasteh_2010,Khodjasteh_2009_PRA}.}, a DCG can be designed using the following four-step procedure:
\begin{enumerate}
    \item Pick a decoupling group $\group$ and draw its Cayley graph
    \item On each vertex, add the identity edge $\mathds{1}_U$
    \item Find an Eulerian path ending in an identity edge
    \item Swap this last edge with the corresponding $U^*$
\end{enumerate}
The resulting sequence then implements the target unitary $U_S$ while the first-order EPO of the DCG reads 
\begin{equation}
    \Phi_{\mathrm{DCG}}^{[1]} = \Phi_{\mathrm{EDD}}^{[1]} + \sum_{\alpha} \Pi_{\group}\qty(\mathrm{F}_{(U^*,\tau_{u^*})}\qty[S_{\alpha}] )\otimes B_{\alpha}
\end{equation}
where $\Phi^{[1]}_{\mathrm{EDD}}$ is simply the first-order error phase of the usual Eulerian path, as defined in Eq.~\eqref{eq:EDDEP}, and satisfies $\Phi^{[1]}_{\mathrm{EDD}}\propto \mathds{1}_S\otimes B$ if the interaction subspace belongs to the correctable subspace of $\group$, \ie, $\mathcal{I}_S\subseteq \mathcal{C}_{\group}$. The remaining condition for the DCG to provide first-order decoupling is then \begin{equation}
    \mathrm{F}_{(U^*,\tau_{u^*})}\qty[\mathcal{I}_S] \subseteq \mathcal{C}_{\group}.
\end{equation}
When the propagator $U^*(t)$ is solely composed of global rotations, we have that
\begin{equation}
    \mathcal{I}_S \subseteq V = \bigoplus_{(L,\alpha)} \BHs^{(L,\alpha)} \BHs^{(L)}_{j_k} \Rightarrow \mathrm{F}_{(U^*,\tau_{u^*})}\qty[\mathcal{I}_S] \subseteq V.
\end{equation}
Consequently, the same Platonic sequence that was used to mitigate decoherence can also be used to construct a DCG. However, when $U^*(t)$ is not a global rotation at all times, this is no longer true as $\mathrm{F}_{(U^*,\tau_u^*)}[\cdot]$ may couple different irreps. In this case, a universal decoupling group, if it exists, must be used to construct the DCG. The results mentioned above can be formulated as follows:
\begin{Proposition}
\label{Result.3}
Consider an ensemble of $N$ interacting spins with quantum numbers $(j_1,\dots,j_N)$ such that its interaction subspace satisfies
\begin{equation}
    \mathcal{I}_S \subseteq V = \bigoplus_{(L,\alpha)} \BHs^{(L,\alpha)}
\end{equation}
 and contains no $0$-irreps (isotropic components) except the global identity. Then, any group not contained as a subgroup of $\SetG_{\mathrm{max}}\qty(\HSs\qty(\mathcal{H}^{(j_{\mathrm{eff}})}))$, where $j_{\mathrm{eff}}=\sum_{k=1}^Nj_k$, can be used to construct a DCG. If the intended gate consists solely of global rotations, any group which is a decoupling group for $\mathcal{I}_S$ can be used to construct a DCG.
\end{Proposition}
We should point out, however, that when the intended gate is not a global rotation, the subspace $\mathrm{F}_{(U^*,\tau_{u^*})}\qty[\mathcal{I}_S]$ may overlap with rotation-invariant subspaces even though $\mathcal{I}_S$ does not. When constructing a DCG for a multispin system, we should then ensure that the balanced pair is designed in such a way this does not happen. 
\subsection{Applications}
For a single spin-$j$ system with an interaction subspace $\mathcal{I}_S\subseteq \bigoplus_{L=1}^{L_{\mathrm{max}}}\BHs^{(L)}$, we refer again to Table ~\ref{Table.4} to choose the appropriate decoupling group. In particular, we find that no Platonic group can be used to construct a DCG for an arbitrary quantum gate for $j \geq 3$ because there is no universal decoupling group composed only of rotations in this case. This limits the use of Platonic DCG to protect operations outside of $\pointg{SU}(2)$ in high-dimensional qudits~\cite{omanakuttan_phd,Omanakuttan_2024,Ringbauer_2022}.  
\par In the case of a set of identical spin-$j$ with up to $K$-body interactions, we now refer to Proposition~\ref{Result.2} and its corollaries to find the appropriate decoupling group. From Proposition~\ref{Result.3} and looking at Table~\ref{Table.6a}, we find that no Platonic group can be used to construct a DCG for an arbitrary quantum gate in an ensemble of qubits (resp. qutrits) if the ensemble contains more than five (resp. two) subsystems.
\par However, the $\pointg{T}$ group is sufficient to construct a DCG which performs an entangling gate between a pair of qubits. As mentioned above, one should however make sure that the balanced pair's EPO still respects the relevant anisotropy conditions. For example, for a pair of interacting qubits with the coupling Hamiltonian $H\in\HSs\qty(\mathcal{H}^{(1/2)})\otimes \HSs\qty(\mathcal{H}^{(1/2)})$ and on which we wish to perform an entangling gate, the propagator $U^*(t)$ must satisfy the following condition 
\begin{multline}
        \Tr \qty[\int_0^{\tau^*} \qty(U^*(t))^{\dagger}H U^*(t) \,\mathbf{J}^1\boldsymbol{\cdot}\mathbf{J}^2 dt] = 0\\
        \Leftrightarrow\quad\Tr \qty[ H \int_0^{\tau^*}U^*(t)\, \mathbf{J}^1\boldsymbol{\cdot}\mathbf{J}^2\,\qty(U^*(t))^{\dagger}dt] = 0.
\end{multline}
A sufficient condition, provided that $H$ itself satisfies the anisotropy conditions, is that $U^*(t)$ commutes with $\mathbf{J}^1\boldsymbol{\cdot}\mathbf{J}^2$ at all times. This is the case, for example, for a propagator of the form $U^*(t) = \exp{-i\chi(t)J_{\mu}^1\otimes J^2_{\mu}}$ for any axis $\mu$.

\par Finally, for a pair of spins with different quantum numbers $\qty{j_1,j_2}$, we refer to Table \ref{Table.6b}. In particular, the point group $\pointg{O}$ can be used to construct an arbitrary DCG for a qubit-qutrit pair. For example, one could use this formalism to perform an entangling gate protected from finite-duration errors that transfers quantum information between the electron qubit and the nitrogen nucleus qutrit in an NV center, where the nuclear spin can be used as a quantum memory and the electronic spin for quantum computation~\cite{Yang_2016, Kalb_2018}.
\section{Conclusions}
\label{Sec.Conclusions}
In this work, we have presented and studied three novel dynamical decoupling sequences (TEDD, OEDD and IEDD), called Platonic sequences, which are inspired by the three exceptional point groups describing the symmetries of the tetrahedron, octahedron and icosahedron. The information required for their construction can be found in Table~\ref{Table.5}. They are generated from only two specific global $\pointg{SU}(2)$ rotations, so there is no need for individual subsystem control.
Platonic sequences are distinguished not only by their decoupling capabilities but also by their structural simplicity and elegance, stemming from their construction as Eulerian cycles on Cayley graphs of exceptional point groups (see Fig.~\ref{Figure.8}). This construction underpins their inherent robustness to systematic pulse errors, finite duration effects, and other perturbations, which further enhances their practical applicability, making them highly relevant to a wide range of quantum systems. They exhibit remarkable decoupling properties for single- and multispin systems, as we summarized below.

For single-qudit systems, Proposition~\ref{Result.1} and Table~\ref{Table.4} summarize our results, highlighting decoupling groups that cancel out different types of system-bath interaction that can perturb the system. A key result is the identification of at least one Platonic DD sequence that is universal for individual qudits with a number of levels $d < 7$, or equivalently, for individual spin-$j$ with spin quantum number $j<3$. Although Platonic sequences are longer than other known sequences for single qudits~\cite{tripathi_2024,iiyama_2024}, we have showed that they can be advantageous in certain cases, such as when the $\mathrm{SU(2)}$ pulses are native to the quantum platform~\cite{omanakuttan_phd, Omanakuttan_2021} or when the interaction subspace involves only low-dimensional irreps~\cite{Omanakuttan_2024}. Furthermore, the number of pulses of each Platonic sequence can be reduced by a factor of two by selecting a Hamiltonian path\footnote{A Hamiltonian path is a cyclic path that visits each vertex exactly once.} on the Cayley graph, instead of an Eulerian, resulting in a sequence shorter but less robust.\\
\indent
Similarly, Platonic sequences decouple several types of interaction in multispin systems when the Hamiltonian does not contain isotropic components (see Proposition~\ref{Result.2} and Table \ref{Table.6}). Because they use global pulses, they are completely independent of the number of spins in the system, unlike the sequences constructed in Refs.~\cite{Stollsteimer_2001,Wocjan_2002,Rötteler_Wocjan_2013} (resp.\ Ref.~\cite{Wocjan_2006}) whose lengths grow linearly (resp.\ quadratically) with the number of subsystems. In particular, we found Platonic sequences that decouple up to five-body, non-isotropic, multilinear interactions in an ensemble of $N$ spins. Isotropic terms cannot be canceled by sequences based solely on global rotations. Exploiting improper rotations (e.g.\ reflections and inversion) to mitigate pseudoscalar isotropic term such as the three-body terms~\eqref{Eq.Three.body.isotropic} could be an avenue for future work.
\par The simplest Platonic sequence is the $\mathrm{TEDD}$ sequence, which has remarkable decoupling properties and requires a reasonable number of pulses to be implemented. This makes it a versatile sequence with many potential applications in quantum information processing. Firstly, the TEDD sequence can decouple linear and quadratic interactions in spin operators, leading to potential applications in quantum computation using large spins~\cite{omanakuttan_phd, Omanakuttan_2021, Omanakuttan_2024}, where these errors are dominant. It is also a universal decoupling sequence for a single qutrit, which appears naturally in NV centers~\cite{Yang_2016,Kalb_2018, Liu_2022}.
In addition, it decouples any linear one-body term and bilinear two-body interaction that is not isotropic~\eqref{Eq.Ham.gen}, such as disorder, dipole-dipole interactions and anisotropic spin exchange, among others~\cite{Zakharov_2008,Lukin_2020,Lukin_2023}. These interaction Hamiltonians appear in several physical scenarios and quantum technologies, such as NMR spectroscopy~\cite{Waugh_1968,10.1063/1.1679423, BHATTACHARYYA_2020,cory_1969}, quantum sensing~\cite{Zhou_2023}, and quantum computation using circular Rydberg atoms~\cite{Cohen_2021}. As discussed in Sec.~\ref{Sec.Comp.Multi}, the robustness and versatility of $\mathrm{TEDD}$ surpass those of state-of-the-art sequences~\cite{cory_1969,Lukin_2020,Cappellaro_2022,Lukin_2023} relevant in many-body spin systems. Furthermore, a simple optimization procedure can be carried out to improve the decoupling and robustness properties of the Platonic sequence to compete with advanced DD sequences such as DROID-60~\cite{Lukin_2020} and yxx24~\cite{Cappellaro_2022}. For example, the time-antisymmetric variant of $\mathrm{TEDD}$ constructed by reflection symmetry, which we denote by $\mathrm{TT}^{\dagger}$, already outperforms them in certain parameter regime.

\par To arrive at our findings, we had to generalize the Majorana representation of Hermitian operators to non-Hermitian operators, which allowed us to study the possible point groups of bounded operators acting on a finite-dimensional Hilbert space. This may be useful beyond the main focus of this work, for example in the study of quantum correlations, where extremal quantum states~\cite{Mar.Gir.Bra.Bra.Bas:10,Chr.etal:21} and extremal quantum gates~\cite{PhysRevA.105.012601} for spin systems have a high degree of rotational symmetry.

\par Overall, the results presented in this study contribute to the expansion of the frontiers of dynamical decoupling and Hamiltonian engineering by providing novel sequences with both theoretical and practical advantages. Platonic sequences offer a promising avenue for future research and applications, e.g., in quantum computing, particularly in environments where robustness to errors is crucial. Their appeal also lies in the fact that they are compatible with more advanced dynamical decoupling strategies, such as dynamically corrected gates~\cite{Khodjasteh_2009,Khodjasteh_2009_PRA}, as discussed in Sec.~\ref{Sec.DCG}, but also concatenated dynamical decoupling~\cite{Khodjasteh_2005,concatenatedPlatonic}.
\section*{Acknowledgments}
We would like to thank the referees, whose constructive comments helped to improve the presentation of our work. ESE acknowledges support from the postdoctoral fellowship of the IPD-STEMA program of the University of Liège (Belgium). JM acknowledges the FWO and the F.R.S.-FNRS for their funding as part of the Excellence of Science program (EOS project 40007526). CR is a Research Fellow of the F.R.S.-FNRS. Computational resources were provided by the Consortium des Equipements de Calcul Intensif (CECI), funded by the Fonds de la Recherche Scientifique de Belgique (F.R.S.-FNRS) under Grant No. 2.5020.11. The numerical calculations and figures in this manuscript were produced using the Julia programming language, in particular the Makie package~\cite{Makie}.
\bibliographystyle{quantum}
\bibliography{References}
\appendix 
\section{Exceptional point groups and their generators}
\label{App.Three.exceptional}
Consider a group $\group$ and a subset of elements $\Gamma = \qty{a,b,\dots}\subseteq \group$. $\Gamma$ is called a \textit{generating set}, and its elements are called \textit{generators}, if each element of $\group$ can be uniquely expressed as a product of elements of $\Gamma$. Furthermore, the group generated by $\Gamma$ admits a \textit{presentation} in terms of its generators and a set of \textit{defining relations}~\cite{Bollobás1998,rotman2012introduction}. A defining relation is a sequence of generators that implements the identity. For example, we write $\group = \bra{a,b}\ket{a^2,b^2, ab}$ the group generated by the generators $\qty{a,b}$ satisfying $a^2=b^2=ab = E$ where $E$ is the identity element. Each proper exceptional group associated with the Platonic solids has two generators, and a presentation of these is given by~\cite{rotman2012introduction}
\begin{equation}
\label{EQ.Point.groups.defining.relations}
\begin{aligned}
\pointg{T} = & \bra{a,b}\ket{a^3,b^3,(ab)^2} ,
\\
\pointg{O} = & \bra{a,b}\ket{a^4, b^3, (ab)^2} ,
\\
\pointg{I} = & \bra{a,b}\ket{a^5,b^3,(ab)^2} .    
\end{aligned}
\end{equation}
Other presentations of the groups are $\bra{a,b}\ket{a^2,b^3,(ab)^k}$ where $k=3,4,5$ for $\pointg{T}$, $\pointg{O}$ and $\pointg{I}$, respectively~\cite{rotman2012introduction}.
\subsection{Tetrahedral group T}
The tetrahedral point group consists of 12 transformations $\pointg{T}= \{ E , 8\,\rot_3 , 3\,\rot_2 \}$ where we use the notation defined in Section~\ref{Sec.Mathematical.tools}. The $C_3$ rotations are performed around an axis passing through the vertices (or the barycentre of the faces) of the tetrahedron, and the $C_2$ rotations are performed around axes passing through the midpoints of two complementary edges (those without common vertices). We write below the twelve explicit rotations sorted in each class in terms of the axis-angle notation $\left( \mathbf{n} , \theta \right)$. Setting
\begin{equation*}
\begin{aligned}    
& \mathbf{n}_0=(0, \, 0 , \, 1), && \mathbf{n}_1^{\pm}=\left( \frac{\sqrt{2}}{3}, \pm\sqrt{\frac{2}{3}}, \frac{1}{3} \right) ,
\\ & \mathbf{n}_2=\left( \frac{2 \sqrt{2}}{3},0,-\frac{1}{3} \right),
&& \mathbf{n}_3=\left( \sqrt{\frac{2}{3}},0,\frac{1}{\sqrt{3}}\right), 
\\ & \mathbf{n}_4^{\pm}=\left( -\frac{1}{\sqrt{6}}, \pm\frac{1}{\sqrt{2}},\frac{1}{\sqrt{3}} \right), &&
\end{aligned}
\end{equation*}
we have
\begin{equation*}
\end{equation*}
\begin{equation*}
\begin{aligned}    
 E={}& \left( \mathbf{n}_0 , \, 0 \right)  ,
 \\
8\,C_3 ={}&  \Bigg\{
 \left( \pm \mathbf{n}_0 ,  \frac{2 \pi }{3} \right) , 
 \left( \pm \mathbf{n}_1^{\pm} ,  \frac{2 \pi }{3} \right)   ,
 \left( \pm \mathbf{n}_2,  \frac{2 \pi }{3} \right)   \Bigg\},
\\
3\,C_2 ={}& \Big\{
  \left( \mathbf{n}_3, \pi \right) , 
  \left( \mathbf{n}_4^{\pm} , \pi \right)  \Big\}.
\end{aligned}
\end{equation*}
The presentation~\eqref{EQ.Point.groups.defining.relations} of $\pointg{T}$ can be obtained with the generators 
\begin{equation}
    a = \left( \mathbf{n}_0 , \frac{2 \pi }{3} \right) ,  \quad b = \left( \mathbf{n}_1^+ , \, \frac{2 \pi }{3} \right) .
\end{equation}
The whole group is thus given by $E$ and
\begin{equation}
\begin{aligned}   
    8\,C_3 =  &\left\{ a , a^2 \right\} \bigcup \left\{ a^{-j} b^k a^j \right\}_{\substack{k=1,2 \\ j=0,1,2} } 
    \\
    3\,C_2 = & \left\{
    a^{-j} b a^{j+1}
    \right\}_{j=0,1,2} 
\end{aligned}
\quad .
\end{equation}
\subsection{Octahedral group O}
The octahedral point group (equivalent to the point group of a cube) has 24 elements. It is made up as follows $\pointg{O}=\{ E , 8\,C_3 , 6\,C_2 , 6\,C_4 , 3\,C_2 (= C_4^2) \}$. The octahedron can be oriented such that its vertices are in the Cartesian axes
\begin{equation}
    ( \pm 1 ,0,0) , \quad 
    (0, \pm 1 ,0) , \quad 
    (0,0, \pm 1) . 
\end{equation}
The generators of its point group can be taken as
\begin{equation}
    a = \left( \mathbf{n}_0 , \frac{ \pi }{2} \right) , \quad b = \left( \frac{1}{\sqrt{3}} \left( 1,1,1 \right) ,  \frac{2 \pi }{3} \right) .
\end{equation}
The rotations $6\,C_4$ and $3\,C_2$ are the $2\pi n/4$ rotations about the axes of symmetry passing through the vertices of the octahedron. They are spanned by the operations
\begin{equation}
    \left\{ b^{-j} a^{k+1} b^{j} \right\}_{j,k=0,1,2} .
\end{equation}
The $8\,C_3$ rotations are the $2\pi n/3$ rotations about an axis passing through the faces of the octahedron, generated by
\begin{equation}
    \left\{ a^{-j} b^{k} a^{j} \right\}_{\substack{k=1,2 \\ j= 0,1,2,3}} .
\end{equation}
Finally, the $6\,C_2$ ($2\pi/2 = \pi$) rotations are about an axis passing through the edges of the octahedron, and can be generated by 
\begin{equation}
    \left\{ b^{-j} a^{k+1}  b a^{-k} b^j \right\}_{\substack{j=0,1,2 \\ k= 0,1}} .
\end{equation}
\subsection{Icosahedral group I}
The last exceptional proper point group is $\pointg{I} = \{ E , 12\,\rot_5 , 12\,\rot_5^2 , 20\,\rot_3 , 15\,\rot_2\}$ with 60 elements and associated to the proper symmetries of the icosahedron or the dodecahedron. A possible orientation of the icosahedron corresponds to the following $12$ non-normalised vertices~\cite{coxeter1973regular}
\begin{equation}
    ( \pm 1 , \pm \phi , 0) , \;\;
    ( 0 , \pm 1 , \pm \phi ) , \;\;
    ( \pm \phi , 0 ,  \pm 1  ) ,
\end{equation}
where $\phi=\frac{\sqrt{5}+1}{2}$ is the golden ratio. 
One type of generators for the presentation~\eqref{EQ.Point.groups.defining.relations} of $\pointg{I}$ is given by
\begin{equation}
a = \left( \frac{(0,-1,\phi)}{\sqrt{\phi+2}} ,\frac{2\pi}{5} \right) ,\;
b = \left( \frac{(1-\phi , 0, \phi)}{\sqrt{3}} ,\frac{2\pi}{3} \right).
\end{equation} 
The rotations $12\,C_5$ and $12\,C_5^2$ are the symmetries associated to $2\pi n/5$ rotations about an axis passing through the vertices of the icosahedron
with elements
\begin{equation}
    \left\{ a^k \right\}_{k=1}^4 \bigcup \left\{ 
a^k b a^j b^{-1} a^{-k}
    \right\}_{\substack{j=1,2,3,4 \\ k= 0,1,2,3, 4}} .
\end{equation}
The $20\,C_3$ rotations are the rotations about an axis passing through the barycentre of the faces, and they are generated by
\begin{equation}
   \left\{ a^j A^{-(l-1)} b^k A^{(l-1)} a^{-j}
   \right\}_{\substack{k,l=1,2 \\ j=0,1,2,3,4 }}
\end{equation}
with $A= b a b^{-1}$. Lastly, the symmetries associated to rotations about an axis passing through the midpoint of the edges of the icosahedron 
are generated by
\begin{equation}
    \{ 
    a^{k+1} b a^{-k} ,
    a^k X a^{-k} ,
    a^k Y a^{-k}
    \}_{k=0,1,2,3,4} ,
\end{equation}
with 
\begin{equation}
X= b^{-1} a b^2 , \,  Y= (a^{-1} b a)^{-1} X (a^{-1} b a) .
\end{equation}
\begin{figure*}[h!]
    \centering
    \includegraphics[width=0.8\textwidth]{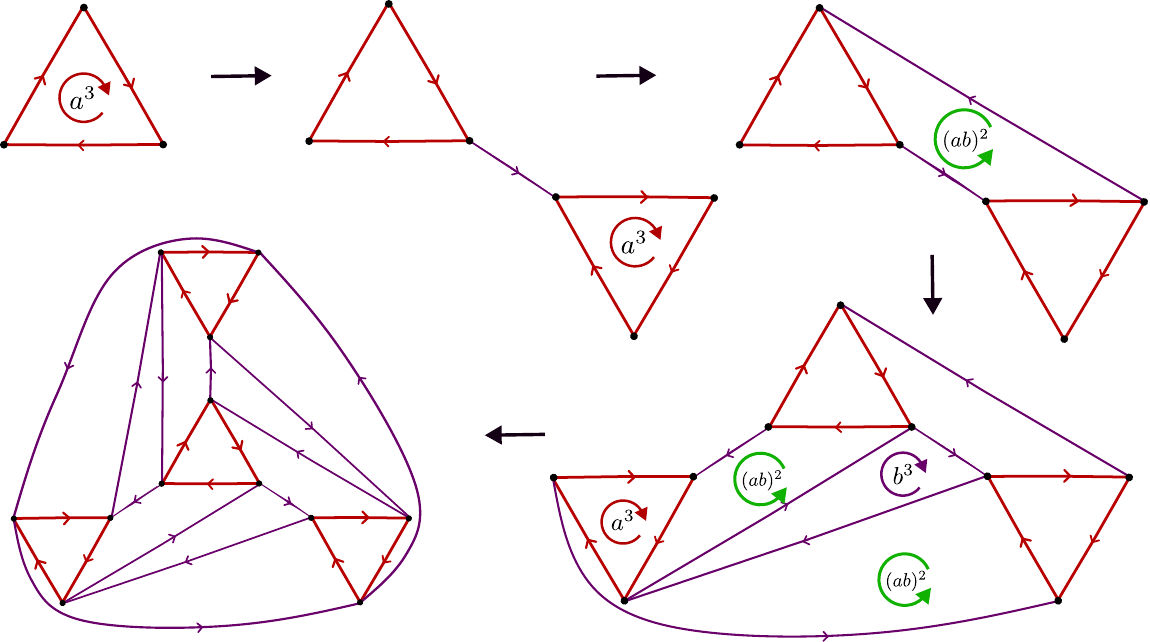}
    \vspace{10pt}
    \caption{Construction process of the Cayley graph of $\group=\bra{a,b}\ket{a^3,b^3,(ab)^2}$. Each loop corresponds to a defining relation and is added one by one until every vertex has the right number of outgoing and incoming edges. Edges of color red (resp.\ purple) refer to the generator $a$ (resp. $b$).}
    \label{Figure.7}
\end{figure*}
\begin{figure*}[h!]
    \centering
    \includegraphics[width=0.25\textwidth]{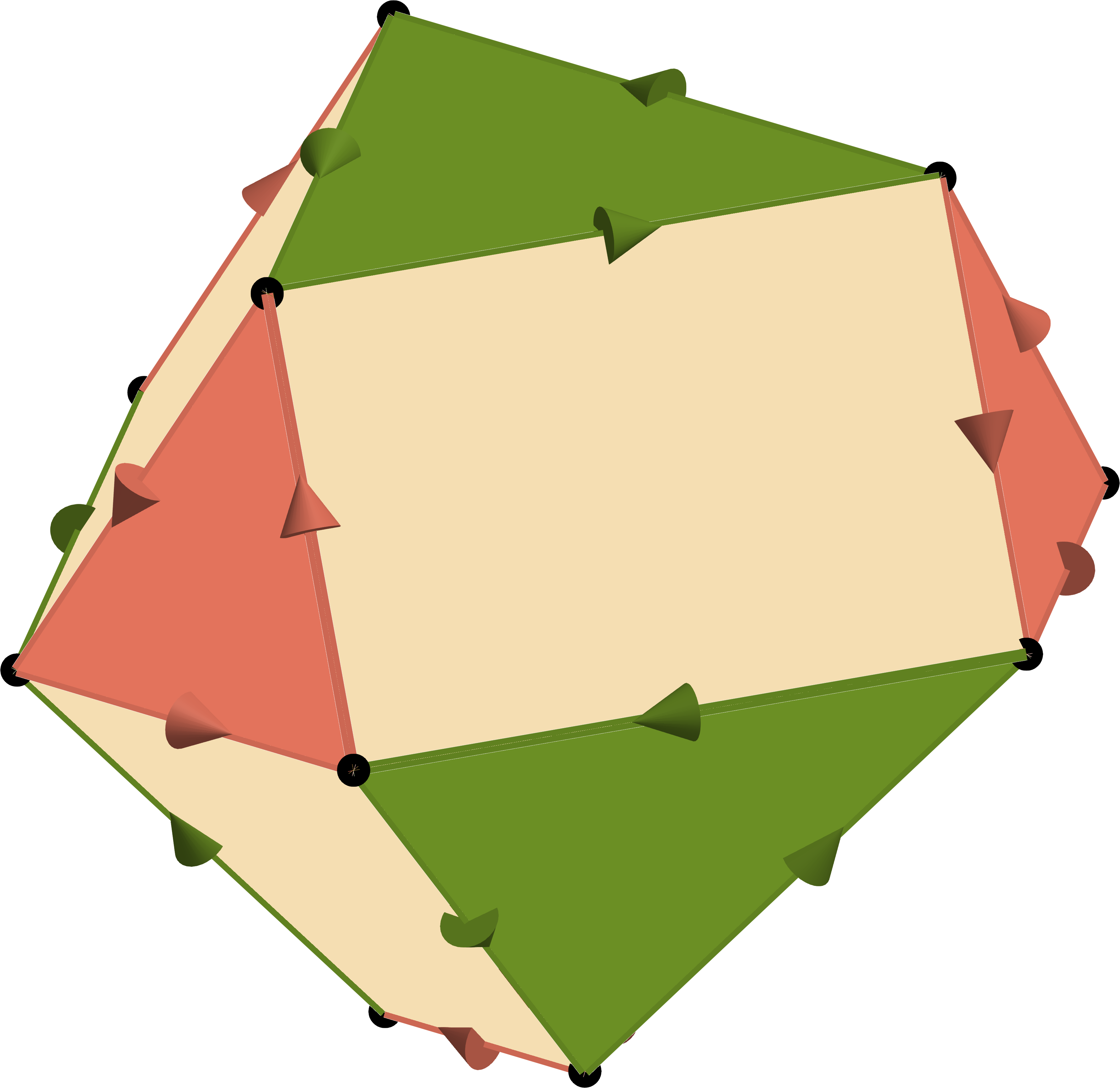}
    \hskip 30pt
    \includegraphics[width=0.25\textwidth]{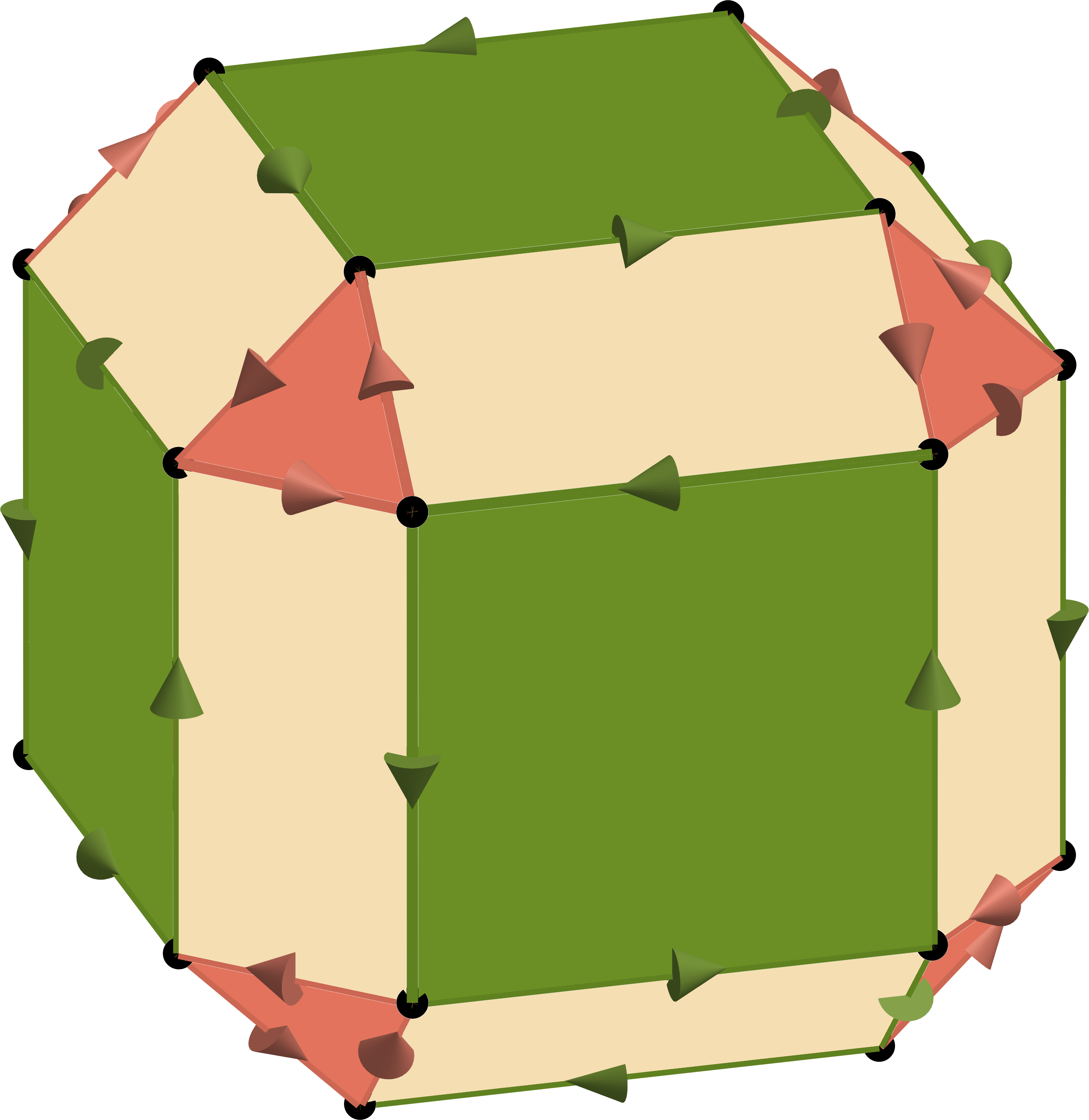}
    \hskip 30pt
    \includegraphics[width=0.25\textwidth]{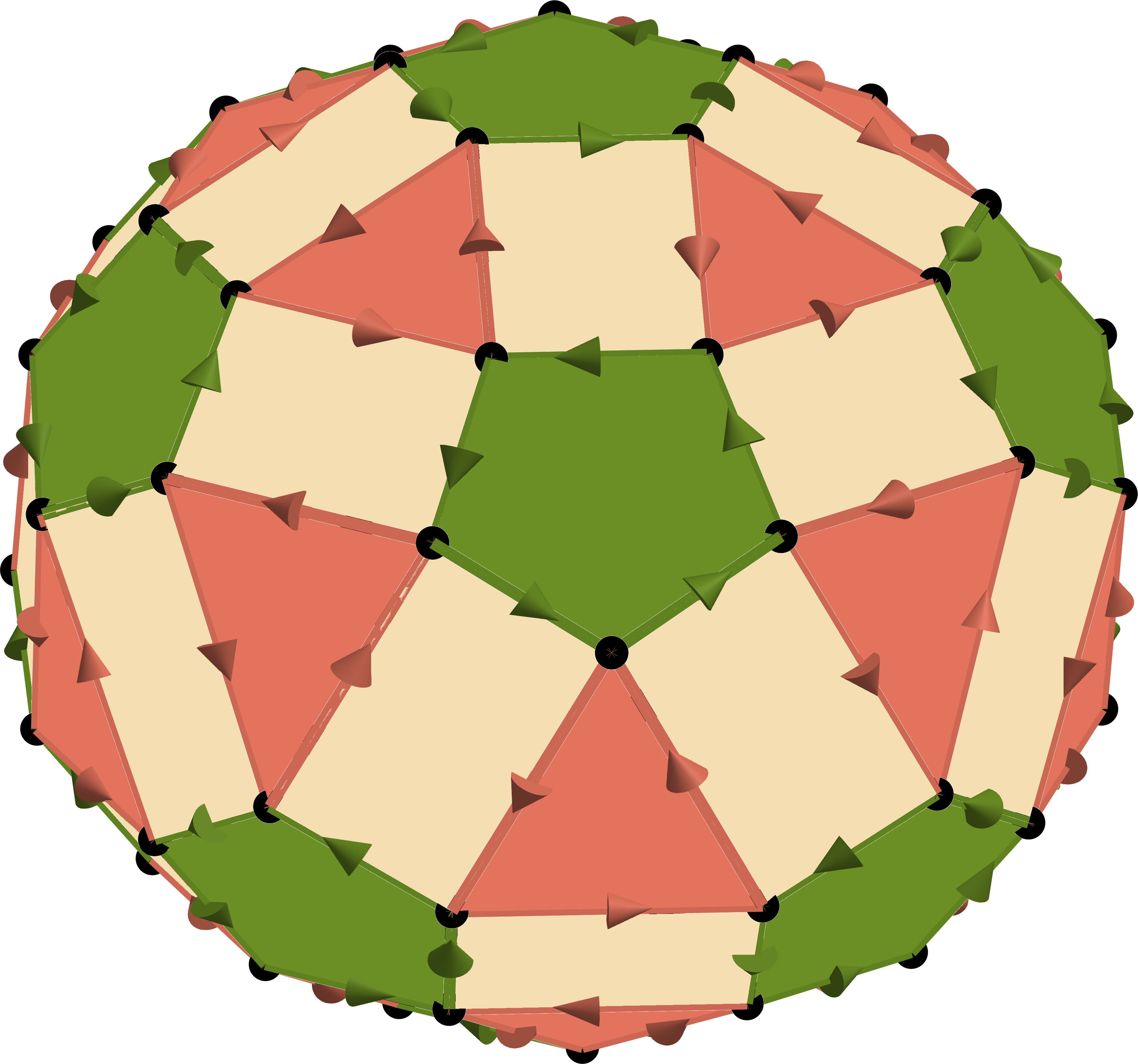}
    \vspace{15pt}
    \caption{Three-dimensional representation of the Cayley graph of the groups (left) $\pointg{T}=\bra{a,b}\ket{a^3,b^3,(ab)^2}$, (middle) $\pointg{O} = \bra{a,b}\ket{a^4, b^3, (ab)^2}$ and (right) $\pointg{I}=\bra{a,b}\ket{a^5,b^3,(ab)^2}$. The generator $a$ (resp.\ $b$) is represented by a green (resp.\ red) arrow. Each colored surface corresponds to a loop generated by one of the defining relations.}
    \label{Figure.8}
\end{figure*}

\section{Cayley graphs and Eulerian cycles}
\label{Ap.CayleyDiagram}
\par We call the \textit{Cayley graph} (or diagram) of $\group$ with respect to $\Gamma$, $G(\group,\Gamma)$, the graph constructed by assigning a vertex to each element of $\group$ and linking pairs of vertices by directed, colored edges, where each color represents a generator. An edge departing from $x$ and heading to $y$ has the color of the generator $g$ iff $y = gx$. On such a graph, each vertex has $\abs{\Gamma}$ ingoing and departing edges, where $\abs{\Gamma}$ is the cardinality of $\Gamma$ (the number of generators). This implies that there exists an Eulerian cycle\footnote{An Eulerian cycle is a closed loop that uses each edge of the graph exactly once.} on such graph~\cite{Bollobás1998}. Furthermore, in some cases a Hamiltonian cycle\footnote{A Hamiltonian cycle is a closed loop that visits each node of the graph exactly once.} could also exist~\cite{Pak_2009}. In the Cayley graph representation, the defining relations are identified by closed loops. 
\par A simple prescription for constructing a Cayley graph of the group $\group$ based on its generating set and defining relations is presented in Ref.~\cite{Bollobás1998}. The basic idea is to start with a single vertex and expand the graph by noting that at each vertex, two properties must be satisfied : (i) there must be exactly $\abs{\Gamma}$ outgoing and incoming edges, one of each color, and (ii) every defining relation must be satisfied. So we add the vertices one by one, and each time a vertex is added, we add the edges by closing the loops corresponding to the different defining relations. The procedure is illustrated in Fig.~\ref{Figure.7} for $\pointg{T} = \bra{a,b}\ket{a^3,b^3,(ab)^2}$. Each presentation~\eqref{EQ.Point.groups.defining.relations} of the exceptional point groups has an elegant three-dimensional Cayley graph representation (see Fig.~\ref{Figure.8}) with Eulerian cycles that can easily be found using Hierholzer's algorithm~\cite{Hierholzer_1873}); such cycles for the groups T, O and I are, for example, given by  
\begin{equation}\label{platonicseq_TEDD}
    \begin{aligned}
    \mathrm{TEDD} \equiv\,& abaababbbaababbbaababbaa\\
     =\,& aba^2bab^3a^2bab^3a^2bab^2b^2,
\end{aligned}
\end{equation}
\begin{equation}\label{platonicseq_OEDD}
    \begin{aligned}
    \mathrm{OEDD} \equiv \,&abaaabbbabaabbbaababbaaa\\
    & ababbbabaabbaaaababbbabb\\
    =\,& aba^3b^3aba^2b^3a^2bab^2a^4bab^3\\&aba^2b^2a^4bab^3ab^2,
\end{aligned}
\end{equation}
\begin{equation}\label{platonicseq_IEDD}
    \begin{aligned}
    \mathrm{IEDD} \equiv \,& baaabbaabaaaaabbaaab\\
    & abbbabaabbaabbabbabb\\
    & abbbaaaababbbaaababb\\
    & baaababbbaababbaabba\\
    & abbaabbbabbbaababbba\\
    & ababbbaababbbabaaaaa\\
    =\,& ba^3b^2a^2ba^5b^2a^3bab^3aba^2\\& 
    b^2a^2b^2ab^2ab^2ab^3a^4bab^3a^3\\& 
    bab^3a^3bab^3a^2bab^2a^2b^2a^2\\& 
    b^2a^2b^3ab^3a^2bab^3a^2bab^3a^2\\&
    bab^3aba^5.
\end{aligned}
\end{equation}
where the generators for each group are summarized below (see also Table~\ref{Table.5})
\begin{align}
    \pointg{T} & \;\to\; \begin{cases}
        \; a =\qty(\qty(0,0,1),\frac{2\pi}{3})\\[3pt]
        \; b = \qty(\qty(\frac{\sqrt{2}}{3},\sqrt{\frac{2}{3}},\frac{1}{3}),\frac{2\pi}{3})
    \end{cases}\\[5pt]
        \pointg{O} &\; \to\; \begin{cases}
        \; a = \qty(\qty(0,0,1),\frac{2\pi}{4})\\[3pt]
        \; b = \qty(\frac{1}{\sqrt{3}}\qty(1,1,1),\frac{2\pi}{3})
    \end{cases}\\[5pt]
        \pointg{I} & \;\to\; \begin{cases}
        \; a = \qty(\frac{1}{\sqrt{\phi+2}}\qty(0,-1,\phi),\frac{2\pi}{5})\\[3pt]
        \; b = \qty(\frac{1}{\sqrt{3}}\qty(1-\phi,0,\phi),\frac{2\pi}{3})
    \end{cases}
\end{align}
with $\phi = \frac{\sqrt{5}+1}{2}$. We specify in Eqs.~\eqref{platonicseq_TEDD}--\eqref{platonicseq_IEDD} the full sequence for each group as well as a (slightly) more compact and intelligible formulation.
  \section{Majorana representation of pure spin states}
  \label{App.Majorana}
The Majorana or \emph{stellar} representation for pure spin-$j$ states~\cite{Maj:32,Chr.Guz.Ser:18} maps each element $\ket{\psi} \in \Hs^{(j)}$ of the Hilbert space $\Hs^{(j)}$ of dimension $2j+1$ to $N = 2j$ points on the sphere $S^2$. This representation contains all the information about the state after removing its normalization and the global phase factor. Majorana~\cite{Maj:32} introduced this representation via a polynomial constructed from the expansion of the state in the $J_z$ eigenbasis, $\ket{\psi} = \sum_{m=-j}^j \la_m \ket{j,m}$ and given by
\begin{equation}
    p_{\psi}(z) = \sum_{m=-j}^j (-1)^{j-m} \sqrt{\binom{2j}{j-m}} \la_m \, z^{j+m} .
\end{equation}
The complex roots of the polynomial are complemented by as many roots at the infinity as are needed to be $2j$ in number. This set of roots $\{ \zeta_k \}_{k=1}^{2j}$ is then mapped to a collection of points on the sphere via stereographic projection from the South Pole, poetically referred to as a \emph{constellation} $\co_{\psi}$ of $\ket{\psi}$. Specifically, each root $\zeta = \tan (\theta/2) e^{i \phi}$ is projected onto a point (\emph{star}) on the sphere with polar and azimuthal angles $(\theta,\phi)$.

In contrast to the Majorana representation for Hermitian operators, the point group of a pure state $\ket{\psi}$ can be obtained simply by analyzing this standard Majorana constellation where there is no distinction (coloring) among the stars. As an example, the Majorana representation of the spin-2 pure state
\begin{equation}
    \ket{\psi} = \frac{1}{\sqrt{3}} \left( 
    \ket{2,2} + \sqrt{2} \ket{2,-1}
    \right) .
\end{equation}
has constellation equal to the tetrahedron. Consequently, the point group of $\ket{\psi}$ is equal to the tetrahedral group $\pointg{T}$. 
\section{Correspondence between constellation colouring and a Hermitian operator}
Following Ref.~\cite{Ser.Bra:20}, we present here how to associate an equivalence class to a Hermitian operator $H = \mathbf{h}_L \boldsymbol{\cdot} \mathbf{T}_L\in \BHs^{(L)}$ with constellation $\co_L$. More precisely, we want to associate an equivalence class $[\co_L]$ of colorings of $\co_L$ with the unit vector $\hat{h}_L=\mathbf{h}_L/h_L$. First, we denote by $\pm \mathbf{n}_k$ the stars of $\co_L$. Each coloring $\mathbf{c}$ of $\co_L$ can be now defined by an $L$-tuple of these points as follows
\begin{equation}
\mathbf{c} = \{  \gamma_1 \mathbf{n}_1 , \gamma_2 \mathbf{n}_2 ,  \dots  , \gamma_{L} \mathbf{n}_{L} \}  ,
\end{equation}
with $ \gamma_k= \pm 1 $. The two equivalence classes, denoted by $[\co_L^{\pm}]$, are now defined as\footnote{The choice of signs for $\pm \mathbf{n}_k$ is not uniquely defined and is associated with the labelling of the equivalence classes. This is analogous to a gauge freedom to specify the classes $[\co_L]$. Once $\pm \mathbf{n}_k$ has been chosen, the classes $[\co_L^{\pm}]$ are uniquely defined.}
\begin{equation}
\label{class.p}
[\co_L^{\pm}] \equiv  \Bigg\{ \mathbf{c} \, \Big| \prod_{k=1}^{L} \gamma_k = \pm 1  \Bigg\} .
\end{equation}
On the other hand, each tuple $\mathbf{c}^{\pm} \in [\co_L^{\pm}]$ defines a unique spin-$L$ pure state as~\cite{Ser.Bra:20}
\begin{equation}
\label{subc.sta}
h_{LM}^{\pm} = 
N_{\phi} \bra{j,j,L,M }
(\ket{\phi}\otimes A\ket{\phi}) ,
\end{equation}
with $\ket{\phi}=\sum_{m} \la_{m}\ket{j, m}$ an $(L/2)$-spinor with Majorana constellation defined by $\mathbf{c}^{\pm}$, $\ket{\phi^A}$ the corresponding antipodal state
\begin{equation}
\label{antipodal.def}
\ket{\phi^A} \equiv \sum_{m} (-1)^{j+m} \la^*_{-m}\ket{j, m},
\end{equation}
$N_{\phi}$ a positive factor that guarantees the normalization of $\hat{h}_{L}^{\pm} = ( h_{LL}^{\pm} , \dots, h_{L-L}^{\pm} )$, and 
$\ket{j,j,L,M}$ the common eigenvectors of the total angular momentum operators $\mathbf{J}^2$ and $J_z$ where $J_{a}= J_{a}^{1}+ J_{a}^{2}$ with $a=x,y,z$. It turns out that $\hat{h}_L^{\pm} $ depends only on the equivalence class of $\mathbf{c}^{\pm} $~\cite{Ser.Bra:20}. Thus, we have uniquely specified $\hat{h}_L^{\pm}$ to each equivalence class $[\co_L^{\pm}]$, the two differing only by a sign
$\hat{h}_L^{+} = - \hat{h}_L^{-}$~\cite{Ser.Bra:20}.  
With the vectors $\hat{h}_L^{\pm}$, we can now uniquely associate a class $[\co_L^{\pm}]$ to any operator $H\in \BHs^{(L)}$ with uncolored constellation $\co_L$. The corresponding class $[\co_L]$ of an operator $H$, with associated vector $\hat{h}_L$, is $[\co_L^{+}]$ (resp. $[\co_L^{-}]$) if $\hat{h}_L = \hat{h}_L^{+}$ (resp. $\hat{h}_L = \hat{h}_L^{-}$).
\begin{table*}[t]
    \centering
    \begin{tabular}{c|c|c|c}
       \multirow{2}{*}{Rank}  & Linearly independent & Most general & Anisotropy  \\
        & isotropic tensors & isotropic Hamiltonian & condition \\\hline \hline && \\
        2 &  $\delta_{ij}$ & $\lambda\, \mathbf{J}^{1}\boldsymbol{\cdot} \mathbf{J}^{2}$ & $\Tr\qty[h] = 0$ \\ &&&\\
        3 & $\epsilon_{ijk}$ & $\lambda \,\mathbf{J}^{1}\boldsymbol{\cdot} \qty(\mathbf{J}^{2}\cross \mathbf{J}^{3})$ & $\sum_{ijk}\epsilon_{ijk}h_{ijk} = 0$ \\ &&& \\
        \multirow{3}{*}{4} & $\delta_{ij}\delta_{kt}$ & $\phantom{+~}\lambda \,\qty(\mathbf{J}^{1}\boldsymbol{\cdot} \mathbf{J}^{2})\qty(\mathbf{J}^{3}\boldsymbol{\cdot} \mathbf{J}^{4})$ & $\sum_{ij} h_{iijj} = 0$, \\ 
       & $\delta_{it}\delta_{jk}$ & $+~\eta \,\qty(\mathbf{J}^{1}\boldsymbol{\cdot} \mathbf{J}^{4})\qty(\mathbf{J}^{3}\boldsymbol{\cdot} \mathbf{J}^{2})$ & $\sum_{ij} h_{ijij} = 0$,\\
       & $\delta_{ik}\delta_{jt}$ & $+~\gamma \,\qty(\mathbf{J}^{1}\boldsymbol{\cdot} \mathbf{J}^{3})\qty(\mathbf{J}^{2}\boldsymbol{\cdot} \mathbf{J}^{4})$ & and $\sum_{ij} h_{ijji} = 0$
    \end{tabular}
    \caption{List of linearly independent isotropic tensors (in Cartesian coordinates) of dimension three and rank 2, 3 and 4, and the corresponding most general multilinear isotropic Hamiltonian and anisotropy conditions on the interaction tensor \eqref{htensor}.}
    \label{Table.7}
\end{table*}
\section{Rotation-invariant component of a Hamiltonian}
\label{Ap.condition}
Applying global rotations on an ensemble of interacting subsystems will leave a part of the Hamiltonian invariant called the rotation-invariant or isotropic component of the Hamiltonian. In order to identify them, we first determine how the total Hamiltonian transforms under rotation and find the conditions for the Hamiltonian to be invariant under these transformations. This approach is similar to that of Ref.~\cite{Lukin_2017}, where the authors identified SU$(d)$-invariant components in the interaction Hamiltonian of a pair of qudits.
\par Here we describe how to obtain the rotation-invariant part of a generic Hamiltonian for the special cases of only multilinear interaction terms between the constituents of a $N$-spin system and for any two-body interactions.
\subsection{Arbitrary two-body interactions}
Consider now a two-body interaction Hamiltonian between two spins of quantum numbers $j_1 \leq j_2$. The most general Hamiltonian belongs to (see Sec.~\ref{Sec.Dec.group.composite})
\begin{equation}
\begin{aligned}    
    H \in  \bigoplus_{\vec{L}}  \BHs^{(L_1)}_{j_1} \otimes \BHs^{(L_2)}_{j_2}  = \bigoplus_{\vec{L}} \bigoplus_{\tilde{L}_{\mathbf{L}} }  \tilde{\BHs}^{\tilde{L}_{\mathbf{L}}} ,
\end{aligned}
\end{equation}
with $\vec{L}=(L_1,L_2)$ and where $L_k$ runs from 0 to $2j_k$ and $\tilde{L}_{\mathbf{L}}$ from $|L_1- L_2|$ to $L_1+L_2$. As we explained in Sec.~\ref{Sec.Dec.group.composite}, the anisotropy conditions are related to the 0-irrep subspaces which only appear for $L_1 = L_2$. There are therefore $2j_1+1$ different rotation-invariant subspaces in $\BHs^{(L)}_{j_1} \otimes \BHs^{(L)}_{j_2}$ for $L=0, \dots ,2j_1$. Their exact expressions can be calculated using the theory of addition of angular momentum~\cite{Var.Mos.Khe:88}; the spin-$0$ state formed from two spin-$L$ states is given by
\begin{multline}
\label{Eq.New.Op}
\sum_{M_1 , M_2 =-L}^L C_{L M_1 L M_2}^{00} \ket{L , M_1} \otimes \ket{L,M_2} 
\\ 
= \frac{(-1)^L}{\sqrt{2L+1}} \sum_{M =-L}^L (-1)^{-M} \ket{L , M} \otimes \ket{L,-M} ,
\end{multline}
where $C_{L M_1 LM_2}^{00} = (-1)^{L-M_1} \delta_{M_1,- M_2}/\sqrt{2L+1}$ is a Clebsch-Gordan coefficient. We now replace the states $\ket{L,M}$ with multipolar tensor operators $T_{L M}$. The resulting Hermitian operators, denoted as
\begin{equation}
\label{Eq.Isotropic.twospins}
    I_L \equiv \sum_{M =-L}^L \frac{(-1)^{-M}}{\sqrt{2L+1}}  T_{L  M} \otimes T_{L-M},
\end{equation}
are $\pointg{SU}(2)$ invariant because they will transform as the corresponding spin-$0$ state. All $I_L$ are linearly independent and $I_0$ is proportional to the identity operator. Thus, a Hamiltonian free of rotation invariant components except the identity must fulfill $2j_1$ anisotropy conditions given by 
\begin{equation}
   \Tr ( I_L H) =0 , \quad L=1, \dots , 2j_1 .
\end{equation}
For the case of a multispin system with more than two parties, $H$ must satisfy the same conditions for any pair of constituents.
\subsection{Multilinear interactions between $N$ spins}
We consider an ensemble of $N$ spins of quantum numbers $\qty{j_k}_{k=1}^N$ with an interaction Hamiltonian $H$ with fixed $K$-body terms multilinear in the spin operators. By simplicity, let us start by considering the case $K=N$. Since we are only interested in multilinear interactions, the Hamiltonian belongs to the following interaction subspace
\begin{equation}
   H \in \mathcal{I}_S = \bigotimes_{k=1}^N \mathcal{I}^k_S \quad\text{with}\quad \mathcal{I}_S^k \subseteq \BHs_{j_k}^{(1)}
\end{equation}
where $\BHs_{j_k}^{(1)} = \mathrm{span} (\{ J_{\alpha_k}\}_{\alpha_k=x,y,z})$, with $J_{\alpha_k} $ are the corresponding $(2j_k+1) \times (2j_k+1)$ matrices representing the angular momentum operators. The interaction Hamiltonian can be decomposed in the orthonormal operator basis $\qty{\frac{1}{\sqrt{\Gamma}}\bigotimes_{k=1}^{N}J_{\alpha_k} }$ where $\Gamma=\prod_{k=1}^N \frac{j_k(j_k+1)(2j_k+1)}{3}$ is a normalization factor\footnote{The basis is orthonormal with respect to the trace norm.}; we define the interaction tensor $h$ as the rank-$N$ tensor $h_{\alpha_1 , \dots ,\alpha_N}$ resulting from the folllowing decomposition 
\begin{equation}
        H = \frac{1}{\sqrt{N}} \sum_{\boldsymbol{\alpha}} h_{\boldsymbol{\alpha}} \bigotimes_{k=1}^{N}J_{\alpha_k}, 
\end{equation}
or
\begin{equation}\label{htensor}
    h_{\boldsymbol{\alpha}} = \frac{1}{\sqrt{\Gamma}} \Tr\qty[H \bigotimes_{k=1}^{N}J_{\alpha_k} ] ,
\end{equation}
where $\boldsymbol{\alpha}= (\alpha_1 , \dots , \alpha_N)$. Suppose that we apply an identical rotation $\Rr$ on each subspace via its corresponding irrep through the Wigner-D matrices $D^{(j_k)}(\Rr)=D^{(j_k)}$, and set $G=\bigotimes_{k=1}^N D^{(j_k)}$. The interaction tensor transforms as follows 
\begin{equation}\begin{aligned}
        \tilde{h}_{\boldsymbol{\alpha}} & = \frac{1}{\sqrt{\Gamma}} \Tr\qty[G^{\dagger}H G  \bigotimes_{k=1}^{N}J_{\alpha_k} ]\\ 
        & = \frac{1}{\Gamma}\sum_{\boldsymbol{\beta}} h_{\boldsymbol{\beta}} \Tr \qty[\qty( \bigotimes_{k=1}^{N} D^{(j_k) \dagger}J_{\beta_k } D^{(j_k)} )  \bigotimes_{k=1}^{N}J_{\alpha_k} ] \\
        & = \frac{1}{\Gamma}\sum_{\boldsymbol{\beta}} h_{\boldsymbol{\beta}} \prod_{k=1}^N\Tr \qty[D^{(j_k) \dagger}J_{\beta_k } D^{(j_k)}  J_{\alpha_k} ].
\end{aligned}
\end{equation}
Since a SU$(2)$ operation $D$ acts on the spin operator $\vec{J}$ with its corresponding physical three-dimensional rotation matrix $\Ro(\Rr) =\Ro \in \pointg{SO}(3)$,
\begin{equation}
    D^{(j_k) \dagger} J_{\alpha_k} D^{(j_k)} = \sum_{\beta_k  =1}^3 \Ro_{\beta_k \alpha_k  } J_{\beta_k} ,
\end{equation}
 we can write 
\begin{equation}
    \tilde{h}_{\boldsymbol{\alpha}} = \frac{1}{\Gamma}\sum_{\boldsymbol{\beta}} h_{\boldsymbol{\beta}} \prod_{k=1}^N \qty(\sum_{\boldsymbol{\lambda}} \Ro_{\lambda_k\beta_k} \Tr \qty[J_{\lambda_k} J_{\alpha_k}]).
\end{equation}
Using the relation 
\begin{equation}
    \Tr \qty[J_{\lambda_k} J_{\alpha_k}] = \frac{j_k(j_k+1)(2j_k+1)}{3}\delta_{\lambda_k\alpha_k}, 
\end{equation}
we find 
\begin{equation}
    \tilde{h}_{\boldsymbol{\alpha}} =  \sum_{\boldsymbol{\beta}} h_{\boldsymbol{\beta}} \qty(\prod_{k=1}^N \Ro_{\alpha_k\beta_k}).
\end{equation}
A tensor that transforms under rotations according to the previous equation is called a Cartesian tensor of rank $N$ and dimension three~\cite{Appleby_Duffy_Ogden_1987}. Thus, the rotation-invariant component of an interaction Hamiltonian is equivalent to the component of its Cartesian tensor that is invariant under rotations, \ie, under any change of coordinates. Rotationally invariant tensors are called \textit{isotropic tensors} and are formed by a sum of products of Kronecker deltas and Levi-Civita symbols~\cite{Jeffreys_1973,Hodge_1961,Appleby_Duffy_Ogden_1987}. For ranks up to 8, Ref.~\cite{Kearsley_1975} lists a complete set of linearly independent isotropic tensors. For two-, three- and four-body interactions, we list in Table~\ref{Table.7} these independent isotropic tensors, the most general isotropic Hamiltonian and necessary and sufficient conditions for a given interaction tensor to have no isotropic component, which we call \textit{anisotropy conditions}. 

In the case where we are interested in a $K<N$ interaction between $N$ spins, the isotropic tensors, and therefore the anisotropy conditions, are equivalent to those obtained in $K$ interactions between $K$ spins. The only difference is that the isotropic tensors must be multiplied by $N-K$ identity operators. The anisotropy conditions for the $K$-body terms of $K$ spins define $\binom{N}{K}$ anisotropy conditions for the $K$-body terms of $N$ spins, each of which is associated with the choice of $K$ spins interacting among the $N$ spins.

\section{Comparison of the $\mathrm{TEDD}$ sequence with state-of-the-art sequences}
\label{Ap.Comp.}
In this Appendix, we briefly compare the $\mathrm{TEDD}$ sequence and its time-antisymmetric variant $\mathrm{TT}^{\dagger}$ with the DROID-60 and yxx24 sequences~\cite{Lukin_2020,Cappellaro_2022}. We consider the Hamiltonian~\eqref{eq:Dis.Dip.} for a system of $N=3$ interacting spin-$1/2$, where we define $\delta \equiv \norm{H_{\mathrm{dis}}}$ and $\Delta \equiv  \norm{H_{\mathrm{dd}}}$ as the two parameters describing the strength of the disorder and dipolar Hamiltonian respectively. We consider the finite-pulse regime where each sequence is applied without any waiting time between the pulses. Each rotation pulse is implemented in a time $\tau_p = \theta/\chi$, where $\theta$ is the angle of rotation and $\chi$ the amplitude of the pulse. Sequences are compared by calculating the average distance $D$ between the identity and the noisy evolution for a wide range of parameters in the $(\delta/\chi, \Delta/\chi)$ parameter space, for a sample of 30 randomly generated frequencies $\qty{\delta_i}$ and $\qty{\Delta_i}$. We add axis-misspecification errors by faultily implementing each rotation axis as $\mathbf{n}_{\epsilon_{AM}} = \sqrt{1-2\epsilon_{AM}^2}\mathbf{n} + \epsilon_{AM}\mathbf{n}^1_{\perp}+\epsilon_{AM}\mathbf{n}^2_{\perp}$, where $\mathbf{n}$ is the target rotation axis associated to the pulse, $\qty{\mathbf{n},\mathbf{n}^{1}_{\perp},\mathbf{n}^{2}_{\perp}}$ forms a set of three orthogonal axes and $\epsilon_{AM}$ corresponds to the amplitude of the axis-misspecification errors.

 \par
\begin{figure*}
    \centering
    \includegraphics[width=\linewidth]{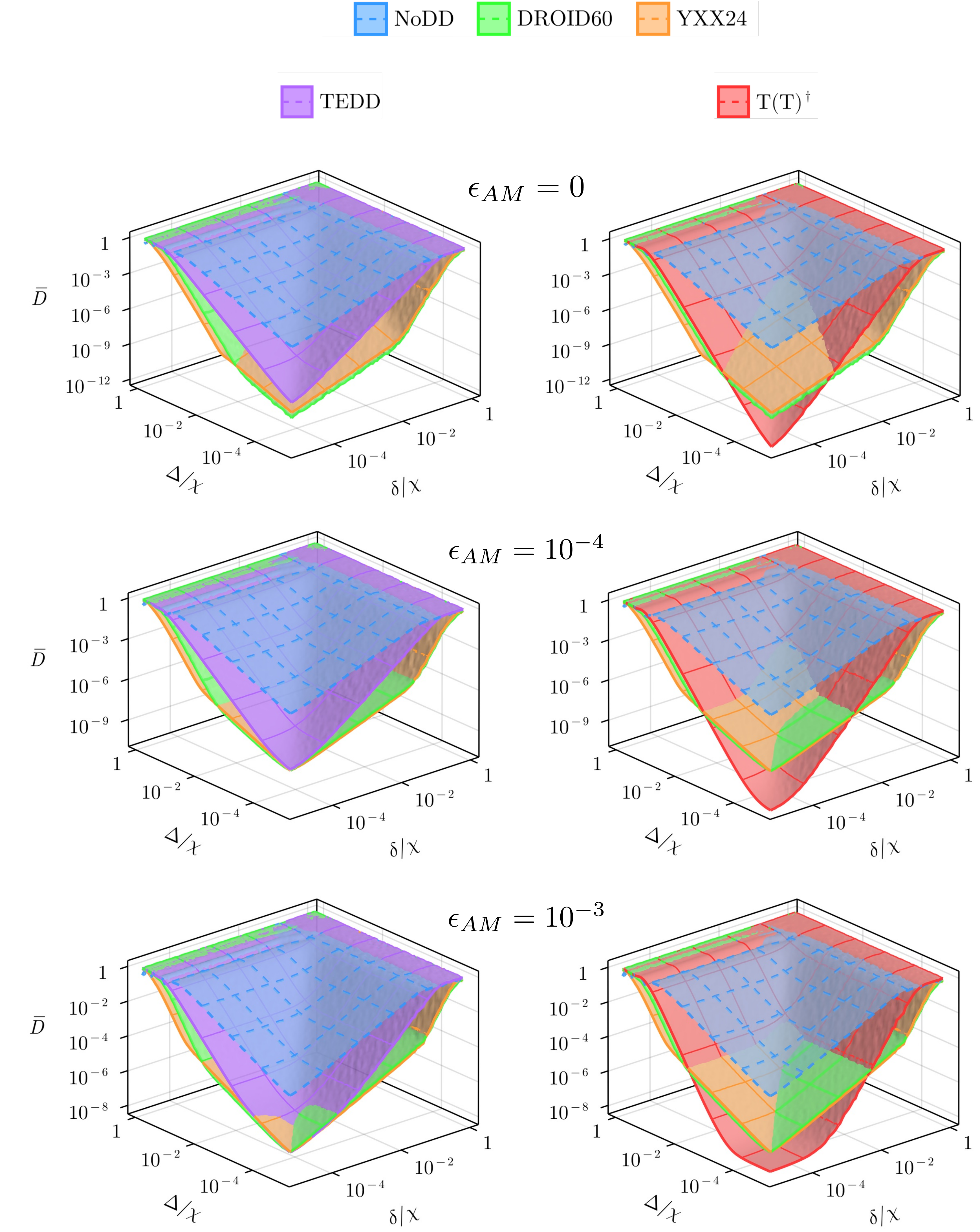}
    \caption{Distance between the identity and the noisy propagator, for a system of 3 interacting spin-$1/2$ of Hamiltonian~\eqref{eq:Dis.Dip.} subjected to different DD sequences. The dimensionless number $\delta/\chi$ (resp.\ $\Delta/\chi$) corresponds to the ratio between the amplitude of the disorder (resp.\ dipolar) Hamiltonian and the Hamiltonian implementing the rotation pulses. $\mathrm{NoDD}$ corresponds to a free evolution of a total duration equal to that of the shortest sequence (yxx24).}
    \label{fig:comparison}
\end{figure*}

 The results for $\epsilon_{AM}=0$ (top left panel in Fig.~\ref{fig:comparison}) show that $\mathrm{TEDD}$ is greatly outperformed in the absence of perturbation as DROID-60 and yxx24 both cancel some high-order terms while being robust to first-order to finite-duration errors. The antisymmetric $\mathrm{TT}^{\dagger}$ sequence (top right panel in Fig.~\ref{fig:comparison}), on the other hand, outperforms both state-of-the-art sequences for small enough decoherence rates, as it decouples both Hamiltonian to second order even in the finite-pulse regime.

 \par The results for $\epsilon_{AM}>0$ (center and bottom panels in Fig.~\ref{fig:comparison}) show that $\mathrm{TEDD}$ starts to match the performance of its competitors in a regime where the amplitude of the axis-misspecification errors $\epsilon_{AM}$ becomes non-negligible compared to $\delta/\chi$ and $\Delta/\chi$. Moreover, the addition of axis-misspecifcation errors further extends the region in the parameter space where $\mathrm{TT}^{\dagger}$ outperforms both DROID-60 and yxx24. 

 \par In conclusion, due to the versatility and robustness of the $\mathrm{TEDD}$ sequence, small perturbations are enough to significantly reduce the gap between the performance of the state-of-the-art sequences and our Platonic sequences. Furthermore, with a simple time-reflection symmetry, we have constructed a variant of $\mathrm{TEDD}$ (denoted $\mathrm{TT}^{\dagger}$) that outperforms both DROID-60 and yxx24 in certain parameter regimes. The addition of a small pulse imperfection ($\epsilon_{AM}=10^{-4}$) is enough to significantly extend the parameter regime in which $\mathrm{TT}^{\dagger}$ performs best. 

\end{document}